\documentclass[aps,prx,twocolumn,groupedaddress,amsmath,amssymb, nofootinbib,floatfix]{revtex4-2}

\usepackage{graphicx}
\usepackage{physics,booktabs,soul,xcolor, xr}
\usepackage{amsmath,amssymb} 
\usepackage{hyperref}
\hypersetup{
    colorlinks,
    linkcolor={red!50!black},
    citecolor={blue!50!black},
    urlcolor={blue!80!black}
}
\usepackage{algorithm}
\usepackage[normalem]{ulem}
\usepackage{amsfonts}
\usepackage{mathtools}
\usepackage{tikz}
\usetikzlibrary{quantikz2}
\DeclareMathOperator{\E}{\mathbb{E}}

\NewDocumentCommand{\hta}{o}{\hat{a}\IfValueT{#1}{_\mathrm{#1}}}
\NewDocumentCommand{\w}{o}{\omega\IfValueT{#1}{_\mathrm{#1}}}
\NewDocumentCommand{\alfa}{o}{\alpha\IfValueT{#1}{_\mathrm{#1}}}
\NewDocumentCommand{\kap}{t^ o o}{
    \kappa
    \IfBooleanTF{#1}{
        ^\mathrm{#2}\IfValueT{#3}{_\mathrm{#3}}
        }
        {\IfValueT{#2}{_\mathrm{#2}}}
}
\makeatletter
\RenewDocumentCommand\eqref{s m}{
  \IfBooleanTF#1%
  {\textup{\tagform@{\ref*{#2}}}}
  {\textup{\tagform@{\ref{#2}}}}
}
\makeatother

\newcommand*{\htb}{\hat{b}}

\newcommand*{\nth}{n_\mathrm{th}}
\NewDocumentCommand{\htn}{o}{\hat{n}\IfValueT{#1}{_\mathrm{#1}}}

\newcommand*{\K}{\hat{\mathbb{K}}}
\newcommand*{\prre}{p_\mathrm{re}}

\newcommand*{\ketL}[1]{\ket{#1}_\mathrm{L}}

\definecolor{phononlight}{HTML}{00DC72}
\definecolor{w2light}{HTML}{FFE065}

\tikzstyle{node} = [rectangle, rounded corners, 
    minimum width=2cm, minimum height=1cm,
    text centered, draw=black, align=center]

\graphicspath{{Figures/}}

\begin{document}

\title{Optomechanical resource for fault-tolerant quantum computing}
\author{Margaret Pavlovich}
\email{margaret.pavlovich@yale.edu}
\author{Peter Rakich}
\author{Shruti Puri}
\affiliation{Yale Quantum Institute, Yale University, New Haven, CT, USA, 06511}
\affiliation{Department of Applied Physics, Yale University, New Haven, CT, USA, 06511}
\date{July 7, 2025}

\begin{abstract}
    Fusion-based quantum computing with dual-rail qubits is a leading candidate for scalable quantum computing using linear optics. This paradigm requires single photons which are entangled into small resource states before being fed into a fusion network. The most common sources for single optical photons and for small entangled states are probabilistic and heralded. The realization of a single reliable deterministic source requires many redundant probabilistic sources and a complex optical network for rerouting and retiming probabilistic outputs. In this work, we show how optomechanics enables reliable production of resources for photonic quantum computing without the redundancy of the all-optical approach. This is achieved by using acoustic modes as caches of quantum resources, ranging from single-particle states to small entangled states, with on-demand read-out. The advantages of acoustic modes as optical quantum memories, compared to other technologies, include their intrinsically long lifetimes and that they are solid state, highly tailorable, and insensitive to electromagnetic noise. We show how the resource states can be prepared directly in the acoustic modes using optical controls. This is still probabilistic and heralded, as in the all-optical approach, but the acoustic modes act as a quantum memory which is integrated into the production of the states. The quantum states may be deterministically transferred from acoustic modes to optical modes, on demand, with another optical drive. 
\end{abstract}

\maketitle

Linear optics has long been studied as a platform for quantum information processing~\cite{knillSchemeEfficientQuantum2001,gimeno-segoviaPracticalLinearOptical2015,liResourceCostsFaultTolerant2015,rudolphWhyAmOptimistic2017,slussarenkoPhotonicQuantumInformation2019,walmsleyLightQuantumComputing2023}. 
In this platform, a qubit is encoded in the presence of a single photon in one of two optical modes that are differentiated by, for example, polarization, spatial mode, or time bin.
Since single optical photons generally do not interact with each other, they offer a path to qubits with enhanced coherence. However, one cannot implement deterministic entangling gates.
To circumvent this limitation, the leading approach for linear optical quantum computing (LOQC) relies on measurement-based quantum computing in which computation is carried out by photon-counting measurements on probabilistically generated small entangled resource states.
In this work, we show how optomechanical (OM) systems can be leveraged to produce single optical photons and photonic entangled states as a basis for fault-tolerant quantum computing. 

It is experimentally challenging to realize single photon sources that satisfy all the criteria necessary for practical and efficient LOQC.
An ideal source should produce exactly one photon, and the photons it produces sequentially should be indistinguishable from each other \cite{oxborrowSinglephotonSources2005}. 
An array of such sources should produce indistinguishable single photons. 
Finally, we prefer the photons to be produced at telecom wavelengths ($\sim1.5~\mu$m) to take advantage of existing photonic technologies. 

There are two broad approaches to making a single optical photon source. 
The first is to excite some matter (e.g., a quantum dot or single atom) which will deterministically emit a single photon when it spontaneously decays. 
While much experimental progress has been made, quantum emitter-type sources still suffer from poor indistinguishability and inadequate efficiency for LOQC \cite{aharonovichSolidstateSinglephotonEmitters2016,arakawaProgressQuantumdotSingle2020,caoTelecomWavelengthSingle2019}. 
The second approach is to probabilistically produce a photon pair through a nonlinear optical process (e.g., spontaneous parametric down-conversion) and use one of the photons to ``herald'' the other. 
This method can produce very high-quality single photons, but it is necessarily probabilistic. Furthermore, the photon-pair generation probability 
must be kept low in order to avoid producing more than one pair. 
Because the production of the single photon is heralded, many probabilistic sources may be multiplexed together to effect one ``deterministic'' source~\cite{meyer-scottSinglephotonSourcesApproaching2020,moodyChipscaleNonlinearPhotonics2020,wangIntegratedPhotonpairSources2021}. 
The drawback of multiplexing is that it requires actively routing a single photon, using optical switches which ideally should be fast and low loss. 
However, schemes for low-loss switching in classical optical communications are not suitable for quantum applications 
\cite{liResourceCostsFaultTolerant2015,mendozaActiveTemporalSpatial2016}.

Single photons must then be combined into entangled states to be used as input to a measurement-based quantum computer.
As originally proposed in 2001 \cite{raussendorfOneWayQuantumComputer2001}, measurement-based quantum computing uses large entangled states and single-qubit measurements. 
Since then, it has been shown that fault-tolerant quantum computing is achievable using collections of small ($\sim10$ qubits) entangled states and two-qubit measurements~\cite{liResourceCostsFaultTolerant2015,bartolucciFusion2023}. While producing small entangled states is easier than large entangled states, it is still a challenge. 

As with single photons, there are two broad approaches for producing entangled optical photons.
The first uses quantum emitters. One may either use a single quantum emitter whose successive emissions are entangled \cite{schonSequentialGenerationEntangled2005,lindnerProposalPulsedOnDemand2009} or pre-entangle several quantum emitters so that the photons they emit are also entangled \cite{economouOpticallyGenerated2dimensional2010}. Recent experiments have implemented the former strategy in single atom \cite{thomasEfficientGenerationEntangled2022} and semiconductor quantum dot \cite{coganDeterministicGenerationIndistinguishable2023} systems. 
Unfortunately, entangled-photon sources using quantum emitters suffer from poor indistinguishability, decoherence, and inadequate efficiency for LOQC. 
The second approach to producing entangled optical photons uses single-photon sources, beam splitters, and single-photon detectors to project and herald entangled states probabilistically \cite{varnavaHowGoodMust2008,liResourceCostsFaultTolerant2015}. This approach has been demonstrated experimentally, \cite{chenHeraldedThreephotonEntanglement2024}, but has the same shortcoming as probabilistic single-photon generation: one must use optical multiplexing or quantum memories (effectively temporal multiplexing) to make a reliable source out of many probabilistic sources. 

This paper shows how optomechanics can provide single optical photons and photonic entangled states, the crucial ingredients enabling fault-tolerant optical quantum computing. The unique properties of acoustic modes allow this proposal to overcome the challenges described above. 
We choose to use mechanical or acoustic resonators for their low loss and dephasing rates, which is partially attributable to their insensitivity to electromagnetic noise \cite{safavi-naeiniElectromagneticallyInducedTransparency2011}. Moreover, they are tailorable and flexible, meaning they can be designed to operate at almost any mechanical and optical frequency \cite{shinTailorableStimulatedBrillouin2013}.
For an overview of optomechanics and its applications in quantum technology, see \cite{aspelmeyerCavityOptomechanics2014,safavi-naeiniControllingPhononsPhotons2019,barzanjehOptomechanicsQuantumTechnologies2022}. 
Our proposal brings together several techniques which have been experimentally demonstrated: 
laser-cooling of mechanical resonators to their ground states even in a relatively warm environment \cite{chanLaserCoolingNanomechanical2011}, optical heralding of a single phonon \cite{gallandHeraldedSinglePhononPreparation2014}, and phonons as quantum memories \cite{wallucksQuantumMemoryTelecom2020}.
While other work has proposed similar methods to prepare exotic quantum states in acoustic resonators \cite{shepherdMultiphononFockState2024}, we believe this is the first examination of the application of optomechanics to fault-tolerant linear optical quantum computing.  

Our paper is organized as follows:
In Sec.~\ref{sec:background}, we introduce our model of the OM system---its Hamiltonian and parameters---and suggest methods for building such a system. In Sec.~\ref{sec:photon-source}, we describe how the system can be used to produce single photons. We quantify the fidelity of the single photons and show that many single photons may be efficiently produced simultaneously. Finally, in Sec.~\ref{sec:entanglement-source}, we show how entangled states may be prepared in the acoustic modes and read out on demand to enable linear optical quantum computing.

\section{Theory of the Optomechanical Setup}\label{sec:background}

\begin{figure}[htb]
    \centering
    \includegraphics{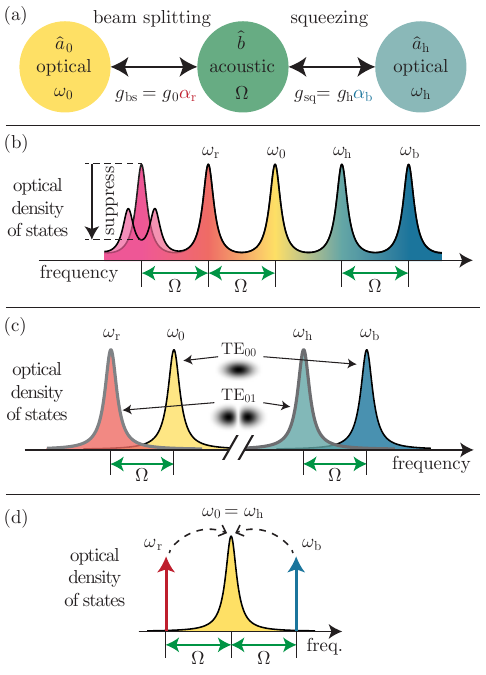}
    \caption{\textbf{Proposed optomechanical system.} 
    (a) Coupled mode diagram. The output optical mode, $\hat{a}_0$ (yellow circle), and herald optical mode, $\hat{a}_\mathrm{h}$ (teal circle) independently couple to quantum acoustic mode $\hat{b}$ (green circle), with strongly-driven classical mode amplitudes $\alfa[r]$ and $\alfa[b]$ controlling the effective coupling strengths. (b-c) Possible arrangements of the optical mode frequencies, using the same color-coding as (a). The frequency differences between mode pairs \{$\hat{a}_\mathrm{r}$, $\hat{a}_0$\} and \{$\hat{a}_\mathrm{h}$, $\hat{a}_\mathrm{b}$\} must be $\Omega$, the acoustic frequency, but other differences are unconstrained. (b) If there exists an optical mode at $\w[r]-\Omega$, it must be suppressed, e.g., by coupling to an auxiliary cavity, to inhibit undesired interactions. 
    (c) The gray or black outlines represent different mode characters, e.g., transverse mode profile (TE$_{00}$ and TE$_{01}$ labels) or propagation direction. (d) The proposed scheme would also work in a conventional cavity-OM system featuring sideband scattering. Additionally, the output and herald roles may be played by the same optical mode.}
    \label{fig:modes}
\end{figure}

In this section, we describe the OM Hamiltonian at the center of our proposal. 
Light and sound interact via radiation pressure and the photo-elastic effect (see Sec.~\ref*{supp-sec:om}). 
This paper focuses on triply-resonant\footnote{We say the interaction is triply-resonant if the three waves involved are each resonant with their own modes, rather than one of the optical fields being a side-band of the other.} interactions in which an acoustic mode, $\htb$, couples to a pair of optical modes which are separated by the phonon frequency $\Omega$.\footnote{Engineering such a mode arrangement can be achieved by the techniques described in \cite{otterstromSiliconBrillouinLaser2018,diamandiQuantumOptomechanicalControl2024}.} 
For conceptual clarity, we suppose that our system has two such pairs of optical modes, though this is not necessary for the proposal.
One pair of optical modes is used during the preparation phase and consists of the ``blue'' mode, $\hta[b]$, and the ``herald'' mode, $\hta[h]$: $\w[b]-\w[h]=\Omega$. The other pair is used during the retrieval phase and consists of the ``output'' mode, $\hta[0]$, and the ``red'' mode, $\hta[r]$: $\w[0]-\w[r]=\Omega$. 
The total Hamiltonian is 
\begin{align}
    \hat{H}&=\hat{H}_\mathrm{O}+\hat{H}_\mathrm{M}+\hat{H}_\mathrm{OM}\\
    \hat{H}_\mathrm{O} &= 
        \sum_{i\in\{\mathrm{r},\;0,\;\mathrm{h},\;\mathrm{b}\}}  \omega_i \hat{a}^\dag_i\hat{a}_i,\;\;\; 
        \hat{H}_\mathrm{M}= \Omega \hat{b}^\dag\hat{b}, \\
    \hat{H}_\mathrm{OM}& =  g_0 \left( \hat{a}_0 \hat{a}^\dag_\mathrm{r} \hat{b}^{\dag} + \hat{a}^\dag_0 \hat{a}_\mathrm{r} \hat{b} \right)
        + g_\mathrm{h} \left( \hat{a}_\mathrm{b} \hat{a}^\dag_\mathrm{h} \hat{b}^\dag + \hat{a}^\dag_\mathrm{b} \hat{a}_\mathrm{h} \hat{b} \right)\label{eq:om-ham},
\end{align}
taking $\hbar=1$ (see Fig.~\ref{fig:modes}a).
The vacuum coupling constants $g_i$ depend on material properties and the spatial overlap between the optical and acoustic modes (see Sec.~\ref*{supp-sec:om}). 

The optical modes couple to an input/output waveguide at rates $\kappa_i^\mathrm{ex}\gg \kappa_i^\mathrm{int}$, where  $\kappa_i^\mathrm{int}$ are the internal decay rates. The total loss rate for each optical mode is $\kappa_i= \kappa_i^\mathrm{int}+\kappa_i^\mathrm{ex}$. The linewidth of the acoustic mode, $\Gamma$, is many orders of magnitude smaller than the total optical loss. A higher ratio $\kappa/\Gamma\gg 1$ allows for more quantum resources to be prepared in parallel (Sec.~\ref{sec:parallelization}). 
The acoustic frequency should be at least a few times larger than the total optical loss rates, $\Omega\gg\kappa_i$, in order for the different optical modes to be well-resolved and the OM interaction to be selective. 
See Tab.~\ref{tab:pams} for the values of these parameters used for numerical simulations. 

Each term of the OM Hamiltonian, Eq.~\eqref{eq:om-ham}, has no effect if both optical modes involved in the interaction are in vacuum. 
Without any external drives $\hat{H}_\mathrm{OM}$ is effectively turned off because we may safely ignore the high-frequency optical modes' thermal populations. 

When the system is optically driven at $\w[r]=\w[0]-\Omega$, the red optical mode, $\hta[r]$, is populated. None of the other optical modes are appreciably populated because they are off-resonant with the drive. 
By replacing the red optical mode operator with its mean value, $\alfa[r]=\langle\hta[r]\rangle$, we linearize the OM Hamiltonian, Eq.~\eqref{eq:om-ham}, to obtain a beam-splitting interaction between the output optical mode, $\hta[0]$, and the acoustic mode, $\htb$:
\begin{align}\label{eq:bs}
    \hat{H}_\mathrm{OM}\rightarrow 
    g_0( \alfa[r]^*\hta[0] \htb^\dag + \alfa[r]\hta[0]^\dag \htb )
    =\hat{H}_\mathrm{bs}.
\end{align} 
Here, $|\alfa[r]|^2\gg 1$ is the average number of photons in mode $\hta[r]$ induced by the pump and the quantum fluctuations in this strongly pumped mode are ignored. This interaction transfers excitations between the acoustic mode and the output optical mode. 
One may also view the beam-splitting interaction as anti-Stokes scattering---a pump photon gaining energy by coherently absorbing a phonon and up-converting into anti-Stokes photon. 

On the other hand, when the blue mode, $\hta[b]$, is excited at $\w[b]=\w[h]+\Omega$, the second term in $\hat{H}_\mathrm{OM}$ is activated. 
We can thus linearize the Hamiltonian into a two-mode squeezing interaction: 
\begin{align}\label{eq:sq}
    \hat{H}_\mathrm{OM}\rightarrow 
    g_\mathrm{h}( \alfa[b]\hta[h]^\dag \hat{b}^\dag + \alfa[b]^*\hta[h] \hat{b} )
    =\hat{H}_\mathrm{sq},
\end{align}
which generates photon-phonon entanglement. $\alfa[b]$ is defined analogously to $\alfa[r]$. 
One may also view the squeezing interaction as Stokes scattering---a pump photon red-shifting into a Stokes photon by emitting a partner phonon. 

In simplifying the OM Hamiltonian from Eq.~\eqref{eq:om-ham} to Eqs.~\eqref{eq:bs}--\eqref{eq:sq}, we assume that the strongly driven modes are unaffected by their interactions with the quantum modes (stiff pump approximation). This requires
$|\alfa[r]|\gg\frac{2g_0}{\kap[r]}|\langle\hta[0]\htb\rangle|$ and
$|\alfa[b]|\gg\frac{2g_\mathrm{h}}{\kap[b]}|\langle\hta[h]\htb\rangle|$. 
The classical mode amplitudes are given by 
$|\alfa[r]|^2=(4\kap^[ex][r]/\kap[r]^2)(P_\mathrm{r}/\hbar\w[r])$ and 
$|\alfa[b]|^2=(4\kap^[ex][b]/\kap[b]^2) (P_\mathrm{b}/\hbar\w[b])$, 
where $P_\mathrm{r}$ and $P_\mathrm{b}$ are input drive powers. 

Eqs.~\eqref{eq:bs}--\eqref{eq:sq} show that even with small ($\sim$~Hz) single-photon couplings, strong pumps at the appropriate frequencies can amplify the photon-phonon interaction to useful strengths. 
In our proposal, the squeezing interaction is used for heralded generation of single phonons which are later converted to photons via the beam-splitting interaction. 

\begin{table}[htb]
    \centering
    \begin{tabular}{lcrlrl}
        \toprule
        Name                & Symbol & \multicolumn{2}{c}{Target} & \multicolumn{2}{c}{Near-term}\\ \midrule
        Optical frequency   & $\omega_0/2\pi$    & 193  & THz   & 193  & THz \\
        Internal optical loss & $\kap^[int]/2\pi$&   1  & MHz   &   1  & MHz \\
        Total optical loss  & $\kappa/2\pi$      &   1  & GHz   &  50  & MHz \\
        Acoustic frequency  & $\Omega/2\pi$      &  10  & GHz   &  13  & GHz \\
        Acoustic loss       & $\Gamma/2\pi$      &  10  &  Hz   &  50  &  Hz \\
        Coupling rate       & $g_0/2\pi$         & 100  & kHz   & {1}  &  {k}Hz \\
        \bottomrule
    \end{tabular}
    \caption{
    Parameters used for numerical results. Each optical mode is taken to have the same loss rate for simulations ($\kap^[int]_i=\kap^[int]$, $\kappa_i=\kappa$), and OM coupling rates for the two interactions are taken as equal ($g_\mathrm{h}=g_0$). 
    We include two sets of parameters: the target parameters, which combine the best properties of various experimental systems; and the near-term parameters, which are feasible for a bulk acousto-optic system \cite{diamandiQuantumOptomechanicalControl2024,luoLifetimelimitedGigahertzfrequencyMechanical2025}. 
    The high target acoustic Q-factor and coupling rate have been achieved in microresonators \cite{renTwodimensionalOptomechanicalCrystal2020}, and the target optical Q-factor is within reach for microresonators \cite{liuHighyieldWaferscaleFabrication2021}.
    }\label{tab:pams}
\end{table}

\subsection{System Design}\label{subsec:design}

While it is possible to engineer $\hat{H}_\mathrm{OM}$ in a variety of systems \cite{aspelmeyerCavityOptomechanics2014}, we outline some possible implementations and design considerations below. 

Realizing the proposed OM scheme requires careful engineering of the optical mode spectrum. 
For example, an optical mode at $\w[r]-\Omega$ would cause particular problems by introducing an unwanted two-mode squeezing interaction between this extraneous mode and the acoustic mode when the red mode is driven. 
This issue appears if the OM interaction involves regularly-spaced optical modes, such as those of a Fabry-Pérot cavity or ring resonator (Fig.~\ref{fig:modes}b). 
This undesired coupling may be avoided by shifting or splitting the extraneous mode using hybridization \cite{kharelHighfrequencyCavityOptomechanics2019,blackOpticalparametricOscillationPhotoniccrystal2022,liuIntegratedPhotonicMolecule2024}. 

Another way to avoid deleterious coupling to an optical mode at or near $\w[r]-\Omega$ is to engineer the OM interaction to couple optical modes of different character (Fig.~\ref{fig:modes}c).
For example, the two coupled optical modes could have different propagation directions or transverse profiles. 
Such inter-modal couplings are more selective and inhibit cascaded interactions \cite{kittlausNonreciprocalInterbandBrillouin2018,chenOptomechanicalRingResonator2023}.
To achieve this type of interaction, the OM resonator must be designed to support an acoustic mode which couples dissimilar optical modes.

Finally, while we focus on the triply-resonant interaction, the fundamental squeezing and beam-splitter interactions can also be obtained with one optical mode coupled to a phonon mode via the more-common cavity-OM interaction $g\hta^\dag\hta(\htb^\dag+\htb)$, using detuned drives (sidebands) instead of distinct optical modes for control (Fig.~\ref{fig:modes}d)\cite{aspelmeyerCavityOptomechanics2014}. 
As this interaction type does not require careful engineering of the optical resonance spacing, such systems may be easier to fabricate. 
This type of interaction is often explored in nano-mechanical systems, whose small mode volumes enable large vacuum coupling rates. Triply-resonant interactions are much more difficult to engineer in nano-scale systems because their small size limits the number of optical modes and tends to make the frequency spacing between optical modes far greater than any acoustic frequency. 
However, nano-mechanical systems suffer from optical absorption and heating as well as decoherence due to high surface participation \cite{meenehanPulsedExcitationDynamics2015,maccabeNanoacousticResonatorUltralong2020,fiaschiOptomechanicalQuantumTeleportation2021,clelandStudyingPhononCoherence2023,mayorTwodimensionalOptomechanicalCrystal2024}. 
Optical heating limits the possible drive power and therefore achievable effective coupling rates in nanoscale mechanical systems. 
On the other hand, macroscopic systems are more resistant to dephasing and to heating by optical absorption, and they allow for triply-resonant interactions~\cite{diamandiQuantumOptomechanicalControl2024,luoLifetimelimitedGigahertzfrequencyMechanical2025,doelemanBrillouinOptomechanicsQuantum2023}. Their robustness against parasitic laser heating enables macroscopic resonators to be strongly driven to enhance the effective coupling rate even with a lower vacuum coupling rate. 
For these reasons, we choose to analyze triply-resonant OM interactions in this work.

With this overview of the OM setup, in the following section we analyze its performance as a cached single photon source.
In Sec.~\ref{sec:entanglement-source}, we extend our analysis to a cached entanglement source. 

\section{Cached single photon source}\label{sec:photon-source}

There are two key steps to fashion an on-demand single-photon source out of an OM system. 
The first step is to use the squeezing interaction, Eq.~\eqref{eq:sq}, to generate entangled photon-phonon excitations. The detection of the photon in the output waveguide heralds the presence of a single phonon in the resonator with a high probability. We call this the single-phonon preparation phase. Note that for high-fidelity generation of a single phonon, it is necessary to ensure that the modes of the OM system are near vacuum (i.e., ground state) before squeezing is activated. 
After a successful herald, the phonon survives in the acoustic mode due to its long lifetime. When a single photon is required, the beam-splitting interaction, Eq.~\eqref{eq:bs}, is driven, converting the phonon to a photon which can travel through the waveguide to a subsequent optical network for further processing.

The utility of this system is apparent when attempting to produce many single photons. In LOQC, one requires a single photon in each of many modes. Prototypically, each mode is a traveling wave packet in its own optical waveguide. 
To this end, each waveguide can be coupled to its own identical OM resonator, in which single phonons are prepared in parallel. As will be shown, it may be nearly guaranteed that each resonator will be successfully populated by one phonon before any appreciable deterioration. Then, the resonators may be read out simultaneously, behaving as many single photon sources. We analyze the system in the order of operation: initialization (Sec.~\ref{sec:init}), preparation (Sec.~\ref{sec:prep}), and retrieval (Sec.~\ref{sec:beamsplit}). Then we consider how well the system parallelizes (Sec.~\ref{sec:parallelization}) and find the optimal operating point for various system parameters (Sec.~\ref{sec:sp-tradeoff}).
Finally, we estimate how large an OM coupling rate is required to produce single photons with fidelities high enough for LOQC (Sec.~\ref{sec:min-g0}).

\subsection{Initialization of optomechanical resonator to ground state} 
\label{sec:init}

Before attempting to prepare a single phonon, we require that all relevant modes are set to their ground state, i.e., to vacuum. 
Particularly after failed heralds, the optical and acoustic modes may be left in an unknown state, and we need a state-independent way to reset them to vacuum. 
Moreover, this reset must be quick because it contributes to the heralding cycle duration. A long heralding cycle inhibits parallelization, as will be shown in Sec.~\ref{sec:parallelization}. 
At a minimum, the reset time must be much faster than the phonon decay time, $1/\Gamma$.

The optical modes have large decay rates ($\kappa/2\pi\sim1$~GHz), and are naturally cold due to their high frequency, and thus these modes quickly equilibrate to near vacuum. Consequently, no active protocol is necessary to reset the optical modes, and we can simply wait for a time $O(1/\kappa)\ll (1/\Gamma)$ for these modes to equilibrate. 

Unlike the optical modes, we cannot wait for the acoustic mode to reach thermal equilibrium due to its long lifetime. Thus, it is necessary to actively laser-cool the acoustic mode by driving the red optical mode at frequency $\w[r]=\w[0]-\Omega$ to activate the beam-splitting interaction, $\hat{H}_\mathrm{bs}$ [Eq.~\eqref{eq:bs}]. 
Once the beam-splitting interaction is activated, the acoustic mode decays through output optical mode ($\hta[0]$) at rate
\begin{equation}\label{eq:Gamma-opt}
    \Gamma_\mathrm{opt}=\frac{4g_0^2
    |\alfa[r]|^2}{\kappa_0}
\end{equation}
to the steady-state phonon population 
\begin{equation}\label{eq:cooled-pop}
    n_\mathrm{ph} \rightarrow \frac{\nth}{(C_0|\alfa[r]|^2+1)}. 
\end{equation} 
Here, $|\alfa[r]|^2=(4\kap^[ex][r]/\kappa_\mathrm{r}^2)(P_\mathrm{r}/\hbar\omega_\mathrm{r})$ is the population of the driven red optical mode and $\nth$ is the phonon population at thermal equilibrium (i.e., in the absence of laser cooling). 
In terms of the cooperativity $C_0=4g_0^2/\kappa_0\Gamma$, the effective optically-induced phonon damping is $\Gamma_\mathrm{opt}=\Gamma C_0|\alfa[r]|^2$. 
Eqs.~\eqref{eq:Gamma-opt}--\eqref{eq:cooled-pop} are valid in the Born-Markov (i.e., weak coupling) limit: $\kappa_0,\; \kappa_\mathrm{r} \gg\Gamma,\; 2g_0 |\alfa[r]|$ (Sec.~\ref*{supp-sec:cooling-eff}), but it is not experimentally necessary to operate in this limit. One may cool even faster and to identically zero phonon population in the strong coupling regime---see Sec.~\ref{sec:beamsplit}. 

We define the ground state occupancy of the acoustic mode after cooling to be the \emph{initialization fidelity}, $F_\mathrm{init}$.
When $n_\mathrm{ph}\ll 1$, then $F_\mathrm{init}\approx 1-n_\mathrm{ph}$. 
Because of the acoustic mode's slow thermalization, it remains in the ground state for a long time without further cooling, removing the need to operate at millikelvin temperatures. 

From Tab.~\ref{tab:pams}, we see that $C_0$ is order unity for the target parameters and that the thermal phonon population is $\nth\sim 3.7$ at $2~\text{K}$ given $\Omega=2\pi\times10~\text{GHz}$. 
Thus, the above equations show that $\Gamma_\mathrm{opt}\sim 2\pi\times0.2~\text{GHz}$ and $n_\mathrm{ph}\rightarrow 2\times 10^{-7} $ even with a reasonable pump power of 1~mW, guaranteeing a fast, high-fidelity initialization to vacuum. 
We note that this level of cooling is fundamentally harder to achieve with sideband cooling in a conventional cavity-OM system, making the triply-resonant approach more attractive~\cite{marquardtQuantumTheoryCavityAssisted2007}. 
The above analysis of the fidelity of vacuum state preparation does not include system imperfections such as the spurious interactions with other optical resonator modes. Such interactions can be minimized by following the design principles outlined in Sec.~\ref{subsec:design}, and the potential effects are outlined in Sec.~\ref*{supp-sec:imperfections}.

For each set of parameters (Tab.~\ref{tab:pams}), we consider two initial phonon populations from which we need to cool. 
The first possible initial phonon population is the thermal population at 2~K, which is $n_\mathrm{ph}=3.7$ (2.7) for the target (near-term) set of parameters. 
The second possible initial phonon population is that which results from a failed attempt to generate a single phonon, i.e., when the squeezing interaction is driven but no herald photon is detected (see Sec.~\ref{sec:prep}). We take $n_\mathrm{ph}=0.06$ (0.1) to be the representative no-herald-detection residual phonon population when using the target (near-term) set of parameters.\footnote{The no-herald residual phonon population depends on the single-phonon heralding probability and the optical efficiency (see Sec.~\ref*{supp-sec:sp-herald-state}). These are the highest residual phonon populations given the ranges shown in Fig.~\ref{fig:herald-fid}.} 
For the target parameters, initialization fidelity $0.999$ ($n_\mathrm{ph}\approx 0.001$) is reached using a tanh-shaped beam-splitting drive pulse with power 1~mW and duration 5.2 or 2.8~ns when cooling from $n_\mathrm{ph}=3.7$ or $0.06$, respectively. 
For the near-term parameters, initialization fidelity $0.99$ ($n_\mathrm{ph}\approx 0.01$) is reached using a pulse with power {1}~mW and duration 96 or 43~ns when cooling from $n_\mathrm{ph}=2.7$ or $0.1$, respectively. 

\subsection{Preparation of a single phonon}\label{sec:prep}

Once the modes are initialized in vacuum, we are ready to prepare a single phonon. We use the photon-phonon squeezing interaction, $\hat{H}_\mathrm{sq}$, Eq.~\eqref{eq:sq}. 
The squeezing interaction generates photons and phonons in pairs  (see Sec.~\ref*{supp-sec:pairs}); 
therefore, counting the number of herald photons which have been generated indicates how many phonons have been generated.
We activate the squeezing interaction by driving the blue mode at $\w[h]+\Omega$ for a short time and send the herald mode's output to a photodetector, which we assume can distinguish between 0, 1, and more than 1 photons.  
We monitor the photodetector while the squeezing interaction is driven, plus a short period ($\sim\kap[h]^{-1}$) after. 
If exactly one photon is detected during this window, we \emph{herald} a single phonon in the resonator. Equivalently, the detection of a single herald photon projects the acoustic mode into the single-phonon state. 
How well the detection of a single herald photon corresponds to the acoustic resonator containing a single phonon is the \emph{single-phonon heralding fidelity}, $F_\mathrm{h,sp}$, which we calculate below. 
For the analytical calculations in this section, we neglect acoustic loss and assume the acoustic resonator starts in vacuum at $t=0$. 

Under continuous driving, the dynamics are governed by the Hamiltonian Eq.~\eqref{eq:sq}, and the phonon population grows as 
\begin{equation}\label{eq:sq-phonpop}
    n_\mathrm{ph}  
    = e^{-\kappa_\mathrm{h} t/2} 
        \left( \cosh\left({\tilde{g}_\mathrm{h} t}\right)  
            + \frac{\kappa_\mathrm{h}}{4\tilde{g}_\mathrm{h}} \sinh \left({\tilde{g}_\mathrm{h} t}\right) \right)^2 - 1. 
\end{equation}
Here, $\tilde{g}_\mathrm{h}=\sqrt{(g_\mathrm{h}\alfa[b])^2+(\kappa_\mathrm{h}/4)^2}$ is the effective coupling rate, and we have let $\alfa[b]=|\alfa[b]|$ (see Sec.~\ref*{supp-sec:squeezing-analytics}). 
For short times ($g_\mathrm{h}\alfa[b]t\ll1$), the population grows quadratically: $n_\mathrm{ph}\approx (g_\mathrm{h}\alfa[b]t)^2$. 

Two-mode squeezing may be understood as heating, and the variance in the expected phonon number is $\mathrm{var}(\htn[ph])=n_\mathrm{ph}(n_\mathrm{ph}+1)$, which is consistent with a thermal (i.e., geometric) distribution (Sec.~\ref*{supp-sec:squeezing-analytics}). Simulations confirm that the number of photon-phonon pairs generated follows this distribution. 
Hence, if the expected number of pairs generated is $\bar{n}$, then the probability of generating exactly one pair is $p_1=\bar{n}/(1+\bar{n})^2$, the probability of generating no pairs is $p_0=1/(1+\bar{n})$, and the probability of generating $k$ pairs is $p_k=p_0(p_1/p_0)^k$.  

The {single-phonon heralding fidelity} is the figure of merit for single-phonon preparation: 
$F_\mathrm{h,sp}\coloneq\langle1|\rho_{\mathrm{h}}|1\rangle$, where $\rho_{\mathrm{h}}$ is the density matrix of the acoustic mode given a successful herald (i.e., the detection of a single herald phonon). 
This overlap can alternatively be understood as the classical probability that the acoustic resonator contains a single phonon conditional on a successful herald, allowing us to use Bayes' theorem (see Sec.~\ref*{supp-sec:bayes}):
\begin{align}\label{eq:bayes}
    F_\mathrm{h,sp} &= P(\text{single phonon} ~|~ \text{herald}) \nonumber\\
    &= P(\text{herald} ~|~ \text{single phonon}) \frac{P(\text{single phonon})}{P(\text{herald})}.
\end{align}
\begin{itemize}
\item $P(\text{herald} ~|~ \text{single phonon})$ is the probability that the detector registers a single herald photon given that resonator is in the single-phonon state. Because we neglect acoustic loss and thermal effects, a single phonon being in the resonator means that a single photon-phonon pair was generated. 
Therefore, $P(\text{herald} ~|~ \text{single phonon})$ is the probability that, given a single herald photon is generated, it is detected.
In terms of the system's physical parameters, 
$P(\text{herald} | \text{single phonon})=\eta=\eta_\mathrm{h}^\mathrm{ex}\eta_\mathrm{d}$, where $\eta_\mathrm{h}^\mathrm{ex}=\kap^[ex][h]/\kap$ is the extraction efficiency of the optical herald mode and $\eta_\mathrm{d}$ is the detection system efficiency, including optical losses between the resonator and waveguide. 
\item $P(\text{single phonon})$ is the unconditioned probability that the acoustic mode contains a single phonon. Thus, $P(\text{single phonon})=p_\mathrm{1}$, the probability of generating a single photon-phonon pair. 
\item $P(\text{herald})=p_\mathrm{h,sp}$ is the \emph{single-phonon heralding probability}, i.e., the total probability of a herald irrespective of the actual number of pairs generated.
We must consider all the different ways a ``successful'' herald can occur. 
These are: exactly one photon-phonon pair is generated, and its photon is successfully detected; more than one photon-phonon pair is generated, but only one photon is detected; or no photon-phonon pairs are generated and the detector registers a dark count.\footnote{A dark count is when a photodetector registers a single photon despite no incident photons and is caused by thermal and quantum fluctuations.}
Considering only the most significant contributions, we have 
\begin{equation}\label{eq:herald-prob}
    p_\mathrm{h,sp}\approx\eta p_1 + 2\eta(1-\eta)p_1^2/p_0+p_\mathrm{d}p_0,
\end{equation}
where $p_\mathrm{d}$ is the dark count probability. 
\end{itemize}

\begin{figure}[htb]
    \centering
    \includegraphics{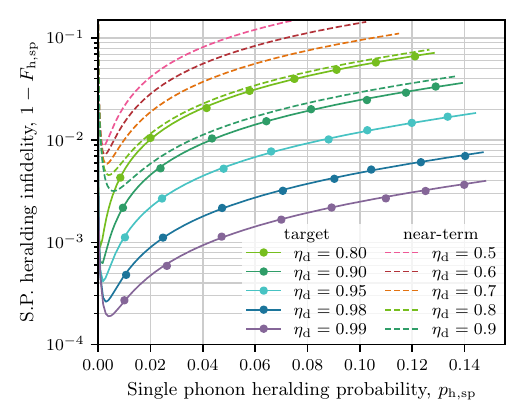}
    \caption{\textbf{Trade-off between single-phonon heralding fidelity and probability.}
    The single-phonon (SP) heralding fidelity, $F_\mathrm{h,sp}$ [Eq.~\eqref{eq:herald-simp}], and heralding probability, $p_\mathrm{h,sp}$, [Eq.~\eqref{eq:herald-prob}], are inversely correlated for most of their ranges. 
    Both heralding fidelity and probability are functions of the single-pair generation probability ($p_1$), itself approximately proportional to the total energy of the blue squeezing drive. Detection efficiency ($\eta_\mathrm{d}$) includes filtering, coupling, etc. Higher detection efficiency and lower heralding probability improve fidelity. 
    Solid and dashed lines show analytical results using the target and near-term parameters, respectively. Dots show results of QuTiP \cite{qutip,qutip2} simulations using the target parameters. Simulation uncertainty is commensurate with marker size. 
    Dark count probability ($p_\mathrm{d}$) is $4\times10^{-7}$ (target) or $9.5\times 10^{-6}$ (near-term). Herald extraction efficiency ($\eta_\mathrm{h}^\mathrm{ex}$) is $99.9\%$ (target) or $98\%$ (near-term).
    Simulations are made under the stiff-pump approximation. 
    Simulations include phonon loss, but assume perfect initialization to vacuum and 0 temperature. Imperfect initialization and finite temperature are considered in Sec.~\ref{sec:sp-tradeoff}. 
    }
    \label{fig:herald-fid}
\end{figure}

All together, in the operating limit $p_\mathrm{d}\ll p_1\ll p_0$, the single-phonon heralding fidelity is given by 
\begin{equation}\label{eq:herald-simp}
    1-F_\mathrm{h,sp} \approx 2p_1(1-\eta)+\frac{p_\mathrm{d}}{p_1 \eta}.
\end{equation}
The first term in Eq.~\eqref{eq:herald-simp} captures the infidelity arising from the possibility of multiple photon-phonon pairs being generated, but only one photon being detected, which leads to the possibility that the acoustic mode has 2 or more phonons after heralding. 
Therefore, the single-pair generation probability ($p_1$) should not be made too large, in order to avoid unwittingly generating more than one photon-phonon pair. This effect causes the infidelity to increase with increased heralding probability (see Fig.~\ref{fig:herald-fid}).
Increasing the likelihood that every generated photon is detected ($\eta\rightarrow1$) mitigates this risk. 
The second term in Eq.~\eqref{eq:herald-simp} captures the infidelity arising from detector dark counts, which leads to the acoustic mode possibly having no phonons after heralding. 
If the pair-generation probability is too low ($p_1\sim \sqrt{p_\mathrm{d}/(2\eta(1-\eta))}$), it becomes likely that a ``detection'' is actually a dark count, increasing the infidelity at the lowest heralding probabilities (see Fig.~\ref{fig:herald-fid}). 
To mitigate this effect, the herald detection integration time should be kept short. This requires both a high interaction strength, permitting the interaction to be kept short; and a high external coupling rate for the optical modes, so that the optical modes can be populated and depopulated quickly. 
For details of this calculation and full phonon state after heralding, see Sec~\ref*{supp-sec:sp-herald-fid}--\ref*{supp-sec:sp-herald-state}. 

Using the target set of parameters (see Tab.~\ref{tab:pams}), we suppose that the beam-splitting interaction is driven by a \mbox{1-ns} long tanh-shaped pulse with power up to 175~$\mu$W. We use a 4~ns detection window, which, combined with dark count rate 100 c.p.s. \cite{IDQSNSPD}, gives a dark count probability of $p_\mathrm{d}=4\times 10^{-7}$. 
With the near-term set of parameters, a beam-splitting drive duration of 25~ns and power up to {200~$\mu$W} gives an appropriate range of single-phonon heralding probabilities. Supposing a detector integration time of 95~ns gives dark count probability $p_\mathrm{d}=9.5\times 10^{-6}$.

If no herald photon is detected, the acoustic mode is left in a state which approximates a thermal state with population $(1-{F_\mathrm{h,sp}})/2$ (see Sec.~\ref*{supp-sec:sp-herald-state}). It is important to cool the acoustic mode back to the ground state (Sec.~\ref{sec:init}) before attempting to generate a single phonon again to avoid cumulative error. 

Once a single phonon is prepared, it may be retrieved as a single photon when required. 

\subsection{Retrieval of a single photon}\label{sec:beamsplit}

After a single phonon is prepared in an OM resonator, 
driving the red mode at frequency $\w[0]-\Omega$ activates the beam-splitting interaction, $H_\mathrm{bs}$, Eq.~\eqref{eq:bs}, and converts the single phonon into a single photon.
In contrast to the initialization of the acoustic mode to vacuum (Sec.~\ref{sec:init}), we are here interested in the quantum dynamics of the output optical mode. 
Under continuous driving, the dynamics are governed by the Hamiltonian Eq.~\eqref{eq:bs}, and the mode population swaps from the acoustic to the optical mode according to
\begin{align}
    n_\mathrm{ph}  
    &= e^{-\kappa_0t/2} \left( \cosh\left(\zeta t\right) 
        + \frac{\kappa_0}{4\zeta}\sinh\left(\zeta t\right) \right)^2  n_\mathrm{ph}(0), \label{eq:bs-phonon}\\
    n_0
    &= \frac{(g_0 \alfa[r])^2}{\zeta^2} e^{-\kappa_0t/2} 
        \sinh^2\left(\zeta t\right) n_\mathrm{ph}(0). \label{eq:bs-photon}
\end{align}
Here, $\zeta=\sqrt{(\kappa_0/4)^2-(g_0\alfa[r])^2}$ is the effective coupling rate, $ n_\mathrm{ph}(0)$ is the initial phonon population, and we have neglected acoustic loss and heating (see Sec.~\ref*{supp-sec:bs-analytics}). 
This expression assumes that we operate in the weak-coupling limit, ${g_0\alfa[r] < \kappa_0/4}$, i.e., ${P_\mathrm{r}<1.3}$~mW ($P_\mathrm{r}<1.6$~mW) given the target (near-term) parameters in Tab.~\ref{tab:pams}. 

The retrieved single-photon wave packet will have a temporal envelope determined by the beam-splitting drive pulse.\footnote{The shape of this envelope does not matter because our scheme does not require converting photons back into phonons.} We take the envelope to extend from $t=0$, when the beam-splitting drive is turned on, until $t=T_\mathrm{re}$, a few optical lifetimes after the beam-splitting drive is turned off. 
We define the \emph{retrieval efficiency} as the total number of photons extracted from the resonator during this time, relative to the initial number of phonons: 
\begin{equation}\label{eq:retrieval-eff}
    \eta_\mathrm{re} \coloneq 
    \frac{\kap^[ex][0]\int_0^{T_{\mathrm{re}}} n_0(t) ~dt}
            {n_\mathrm{ph}(0) },
\end{equation}
The retrieval efficiency is the figure of merit for the single photon retrieval step. 
Even if a single phonon is prepared perfectly, a low retrieval efficiency will lead to the optical mode possibly missing photons. 
The retrieval efficiency can be improved by increasing the strength and duration of the beam-splitting drive. 
In the weak-coupling limit, the excitation cannot swap back from the photon into the phonon because the photon escapes from the cavity too fast. 
Specifically, ${g_0\alfa[r] < \kappa_0/4}$ implies $\Gamma_\mathrm{opt}<\kappa_0/4$, where $\Gamma_\mathrm{opt}$ is the optically-induced phonon loss rate, Eq.~\eqref{eq:Gamma-opt}. 
In the weak-coupling regime, the retrieval efficiency may not exceed $\frac{\kappa_0^\mathrm{ex}}{\kappa_0}\frac{C}{C+1}$, where $C=C_0|\alfa[r]|^2$ is the effective cooperativity. 

The retrieval efficiency is ultimately limited by the output-mode extraction efficiency: $\eta_\mathrm{re}\leq\eta_0^\mathrm{ex}=\kap^[ex][0]/\kappa_0$. 
Using the target parameters, this limit is $\eta_0^\mathrm{ex}=99.9\%$, so we aim to achieve retrieval efficiency $\eta_\mathrm{re}=99.8\%$, which is reached by a 21~ns-long, 0.25~mW beam-splitting drive; a 10~ns-long, 0.5~mW drive; or a 4.4~ns-long, 1~mW drive. 
With the near-term parameters, the limit is $\eta_0^\mathrm{ex}=98\%$, so we aim to achieve retrieval efficiency $\eta_\mathrm{re}=97\%$, which is reached by 
a 444~ns-long, 0.2~mW drive; or a 79~ns-long 1~mW pump drive. 

Therefore, it is possible to efficiently convert single phonons into single photons. Combined with the ability to prepare high-fidelity single phonons, this shows how an OM resonator can be used as a single-photon source. 

\subsection{Parallel preparation of many single phonons}\label{sec:parallelization}

In this section, we show that it is possible to produce many single optical photons by generating single phonons in parallel and then converting them to photons simultaneously. 
Because the preparation of a single phonon is probabilistic, some single phonons will be prepared earlier than others and thus have to idle.
The idling time depends on the number of single phonons to be prepared and the heralding probability per cycle. 
How fast a heralded single phonon degrades while idling depends on the acoustic mode's intrinsic loss and its equilibrium temperature.

\begin{figure}[htb]
    \centering
    \includegraphics{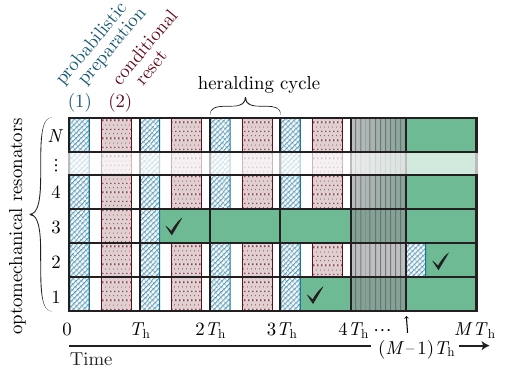}
    \caption{
    \textbf{Parallel preparation of $N$ single phonons divided into heralding cycles.}
    Each heralding cycle consists of two steps: (1) (blue hatch) attempt to prepare a single phonon in every OM resonator not already so prepared, and (2) (red dots) re-initialize modes to the ground state where preparation does not herald success. Successful heralds are indicated by a check mark, and the solid green fill signifies that the resonator is prepared in the single-phonon state.
    Each OM resonator is independent, and has probability $p_\mathrm{h,sp}$ to herald a single phonon during each cycle (see Sec.~\ref{sec:prep}). 
    The cycle during which resonator $i$ heralds is $m_i$, which are independent random variables. 
    In this illustration, $m_1=4$, $m_2=M$, $m_3=2$, etc. 
    The entire preparation process finishes after cycle $M=\max{\{m_i\}}$, which is a random variable with cumulative distribution function given by Eq.~\eqref{eq:M-cdf}.}
    \label{fig:heralding-cycle}
\end{figure}

The parallelized single-phonon preparation process is divided into heralding cycles. 
Each heralding cycle has duration $T_\mathrm{h}$ and consists of two steps (see Fig.~\ref{fig:heralding-cycle}). 
The first step is to attempt to prepare every OM resonator in the single-phonon state (Sec.~\ref{sec:prep}), except those resonators which were successfully prepared in this state during a prior cycle. 
Physically, this step consists of driving the OM squeezing interaction and detecting any herald photons. 
The second step of the heralding cycle is to reinitialize modes which did not herald a single phonon by driving the OM beam-splitting interaction (Sec.~\ref{sec:init}). 
In Sec.~\ref{sec:prep}, we used a 4~ns (95~ns) detector integration time for the preparation step, and in Sec.~\ref{sec:init}, we showed that a 2.8~ns (43~ns) cooling pulse is sufficient to reinitialize unheralded modes to the ground state, using the target (near-term) set of parameters. 
Adding these together and allowing time for classical feedback propagation and for optical modes to ring down when not driven, we take a heralding cycle duration of $T_\mathrm{h}=10$~ns (200~ns) for the target (near-term) parameters.

The cycle during which resonator $i$ heralds is $m_i$, which are independent and identically distributed random variables. 
The expected cycle at which any individual resonator heralds a single phonon is $\overline{m}=1/p_\mathrm{h,sp}$, where $p_\mathrm{h,sp}$ is the individual single-phonon heralding probability per generation attempt [Eq.~\eqref{eq:herald-prob}]. 
The cycle at which the final resonator heralds success, $M=\max{\{m_i\}}$, is also a random variable, which has cumulative (in $M$) distribution function\footnote{Intuition: $(1-p_\mathrm{h,sp})$ is the probability a resonator doesn't herald. $(1-p_\mathrm{h,sp})^M$ is the probability that it fails to herald $M$ times in a row. $\left(1-(1-p_\mathrm{h,sp})^M\right)$ is the probability that a resonator does not fail $M$ times in a row---that is, if you give it $M$ chances, it succeeds at some point between rounds 1 and $M$. Finally, $\left(1-(1-p_\mathrm{h,sp})^M\right)^N$ is the probability all $N$ resonators succeed by round $M$.}
\begin{equation}\label{eq:M-cdf}
    \mathcal{P}(M;N,p_\mathrm{h,sp})=\left(1-(1-p_\mathrm{h,sp})^M\right)^N.
\end{equation}
That is, we will succeed at preparing a single phonon in all $N$ resonators by heralding cycle $M$ (inclusive) with probability $\mathcal{P}(M;N,p_\mathrm{h,sp})$. Note $M$ is the random variable, and $N$ and $p_\mathrm{h,sp}$ are parameters. The expected cycle on which the final resonator heralds is 
\begin{equation}\label{eq:Mbar}
    \overline{M}(N,p_\mathrm{h,sp})=\sum_{M=1}^\infty M \times \mathsf{P}(M;N,p_\mathrm{h,sp}) 
\end{equation}
where $\mathsf{P}(M;N,p_\mathrm{h,sp})=\mathcal{P}(M;N,p_\mathrm{h,sp}) - \mathcal{P}(M-1;N,p_\mathrm{h,sp})$ is the probability that the final resonator will herald \emph{on} the $M$th cycle.
Therefore, each individual phonon idles $(\overline{M}-\overline{m})$ cycles  on average between when it is heralded and when all resonators have heralded.
Note that $\overline{m}$, $\overline{M}$, and $(\overline{M}-\overline{m})$ all decrease when the heralding probability, $p_\mathrm{h,sp}$, is increased. 

Given phonon loss rate $\Gamma$, time per heralding cycle $T_\mathrm{h}$, and equilibrium thermal phonon population $\nth$, each acoustic mode's single-phonon fidelity decays by a factor of  
\begin{equation}\label{eq:fid-idle}
    F_\mathrm{idle}= \exp\left[-\left(3\nth+1\right) \Gamma \left(\overline{M}-\overline{m}\right)T_\mathrm{h} \right]
\end{equation}
on average
\footnote{This is a conservative estimate: 
$\exp\left[-\Gamma_\mathrm{eff} \E\left\{{M}-{m}\right\}T_\mathrm{h}\right] \leq\E\left\{\exp\left[-\Gamma_\mathrm{eff} \left({M}-{m}\right)T_\mathrm{h} \right]\right\}$
by the arithmetic mean--geometric mean inequality for non-negative numbers. Here $\E\{\bullet\}$ is the classical expected value, i.e., the weighted arithmetic mean.}
before the entire set is prepared. We call $F_\mathrm{idle}$ the \emph{idling fidelity}.
As before, the fidelity is defined as the square-overlap of the acoustic mode state with the ideal single-phonon state. 
Above,
$(\overline{M}-\overline{m})T_\mathrm{h}$ is the expected amount of time each phonon idles. 
The factor $(3\nth+1)\Gamma$ in Eq.~\eqref{eq:fid-idle} has two contributions: the rate of decay from the single-phonon state to vacuum, $(\nth+1)\Gamma$; and the rate of excitation from the single-phonon state to the two-phonon state, $2\nth\Gamma$.

\subsubsection{Photon (in)distinguishability}

A crucial figure of merit of any single photon source is the indistinguishability of single photons from nominally identical sources. 
The indistinguishability of two photons is typically quantified by the Hong-Ou-Mandel visibility. Photons may be distinguished by their frequency, time of arrival, linewidth/temporal duration, and/or spatial mode. Frequency and linewidth are our main concerns as the others are primarily technical challenges.

The frequency of the single photon produced by the beam-splitting interaction is simply the sum of optical beam-splitting drive frequency and the phonon frequency. Differences in phonon frequency among the different resonators may be compensated by adjusting the drive frequency. On the other hand, if the frequency of the mode chosen for generating the single photon, $\omega_0$, differs appreciably across the resonators, we may be forced to drive the beam-splitting interaction off-resonantly in order to produce a single photon at the correct frequency. 

The spectrum of the single photon produced by the beam-splitting interaction is determined by the convolution of the spectra of the optical drive and the acoustic mode, and by the width of the optical resonance. 
Differences in the intrinsic acoustic mode linewidths are unlikely to matter as such linewidths will still be many orders of magnitude smaller than the optical linewidths. Small differences in the optical mode linewidths can be compensated for by adjusting the shape of the retrieval drive, but larger differences will degrade Hong-Ou-Mandel visibility. 

\subsection{Total fidelity of parallelized single photon source}
\label{sec:sp-tradeoff}

\begin{table*}[htb]
    \centering
    \begin{tabular}{lllc}
        \toprule
        Figure of merit & Factor/input & Scaling \\\midrule
        Initialization fidelity, 
            & Strength and duration, $T_\mathrm{init}$, of re-initialization drive & $+$ \\
        \quad$F_\mathrm{init}$ (Sec.~\ref{sec:init}) 
            & Thermal phonon population, $\nth$ & $-$ \\ 
            & Single-phonon heralding fidelity, $F_\mathrm{h,sp}$ & $+$ \\ 
            & \quad via residual phonon population after non-herald \\\midrule[0.1pt]
        Single-phonon heralding fidelity,  
            & Single-phonon heralding probability, $p_\mathrm{h,sp}$ & $-$ \\
        \quad$F_\mathrm{h,sp}$ (Sec.~\ref{sec:prep}) 
            & Detection probability, $\eta_\mathrm{h}^\mathrm{ex}\eta_\mathrm{d}$ & + \\ \midrule[0.1pt]
        Idling fidelity, 
            & Single-phonon heralding probability, $p_\mathrm{h,sp}$ & $+$ \\
        \quad$F_\mathrm{idle}$ (Sec.~\ref{sec:parallelization}) 
            &\quad via number of cycles idled, $\overline{M}-\overline{m}$ & \\
            & Thermal phonon population, $\nth$ & $-$ \\
            & Duration of re-initialization drive, $T_\mathrm{init}$, & $-$ \\
            &\quad via heralding cycle duration, $T_\mathrm{h}$ & \\
            & Number of systems to prepare, $N$ & $-$ \\
            &\quad via number of cycles idled, $\overline{M}-\overline{m}$ & \\ \midrule[0.1pt]
        Retrieval efficiency, 
            & Strength and duration of retrieval drive & $+$ \\
        \quad$\eta_\mathrm{re}$ (Sec.~\ref{sec:beamsplit}) 
            & Extraction efficiency, $\eta_\mathrm{0}^{\mathrm{ex}}$ & + \\ 
        \bottomrule
    \end{tabular}
    \caption{\textbf{Contributions to total fidelity, Eq.~\eqref{eq:total-fid}}, when preparing $N$ single photons in parallel using the OM scheme. 
    When single-phonon (SP) heralding probability is increased, the SP heralding fidelity decreases (negative scaling, $-$), while idling fidelity increases  (positive scaling, $+$).
    These opposite dependences lead to a trade-off between SP heralding fidelity and idling fidelity, which is optimized in Fig.~\ref{fig:total-fid}.
    Furthermore, the initialization and idling fidelity trade off against each other via the duration of the re-initialization drive. This trade-off is addressed in Sec.~\ref{sec:min-g0}.
    }
    \label{tab:total-fid}
\end{table*}

In the previous sections, we outlined the scheme for generating synchronous single photons using optomechanics and calculated the fidelity associated with each step. Here, we combine all these contributions and consider their trade-offs.
The average final single photon fidelity is the product of the initialization fidelity, heralding fidelity, idling fidelity, and retrieval efficiency,
\begin{equation}\label{eq:total-fid}
    F_\mathrm{tot} = F_\mathrm{init}F_\mathrm{h,sp}F_\mathrm{idle}\eta_\mathrm{re},
\end{equation}
because the steps are independent. See Tab.~\ref{tab:total-fid} for a summary of how each figure of merit depends on OM system properties and drive strengths.

\begin{figure}[htb]
    \centering
    \includegraphics{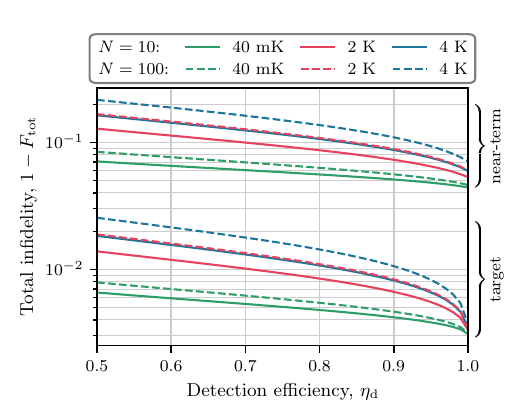}
    \caption{\textbf{Total fidelity for generating $N$ single photons},
    $F_\mathrm{tot}$ [Eq.~\eqref{eq:total-fid}], maximized by adjusting single-phonon (SP) heralding probability to balance the trade-off between SP heralding fidelity, $F_\mathrm{h,sp}$ [Eq.~\eqref{eq:bayes}--\eqref{eq:herald-simp}], and idling fidelity, $F_\mathrm{idle}$, [Eq.~\eqref{eq:fid-idle}]; see Tab.~\ref{tab:total-fid}. 
    Total fidelity is numerically optimized using analytical expressions for component fidelities.
    The achievable fidelity depends on phonon bath temperature (line colors), detection efficiency ($\eta_\mathrm{d}$, $x$-axis), the number of single photons to be prepared ($N$, line dashing), and which parameter set we use (indicated by braces). 
    We conservatively take the initialization fidelity to be $F_\mathrm{init}=99.9\%$ ($99\%$), the retrieval efficiency to be $\eta_\mathrm{re}=99.8\%$ ($97\%$), and the heralding cycle duration to be $T_\mathrm{h}=10$~ns (200~ns) for the target (near-term) parameters. 
    We see that the proposed scheme can produce high-fidelity single photons even with relatively high bath temperatures when using the target parameter set. 
    Using the near-term parameters, the total fidelity exceeds 90\% when using a cold bath or high-efficiency photodetector, though it is limited by the higher phonon loss, longer heralding cycle time, and lower extraction efficiency.
    }
    \label{fig:total-fid}
\end{figure}

There is a trade-off between single-phonon heralding fidelity and idling fidelity. 
When the single-phonon heralding probability is increased, the heralding fidelity decreases (where dark counts are negligible) due to the likelihood of unwittingly generating two phonons (Sec.~\ref{sec:prep}). 
On the other hand, if the single-phonon heralding probability is too low, it will take many rounds for all $N$ OM systems to herald a single phonon, leaving more time for acoustic modes which heralded earlier to thermalize (Sec.~\ref{sec:parallelization}). 
The total fidelity may be maximized by adjusting the single-phonon heralding probability to balance this trade-off (Fig.~\ref{fig:total-fid}). 

Considering only the target parameters, we see that the proposed OM scheme allows one to produce many single photons, each with $99\%$ fidelity, even with modest detection efficiency and relatively high phonon bath temperatures.
The single-photon infidelity can be decomposed into a probability that the optical mode has no photons and the probability that the mode has more than one photon. 
At the lowest bath temperature ($40$~mK), the single-photon infidelity is dominated by the erroneous vacuum component, so this infidelity can be treated as single-photon loss for the purpose of LOQC. 
Thus, this scheme can provide single photons for fault-tolerant LOQC, which has a photon loss threshold of $0.2\%$ \cite{sahayTailoringFusionbasedError2023} to $10\%$ \cite{songEncodedFusionBasedQuantumComputation2024}, depending on the size of the input states and complexity of the fusion operation.
At higher temperatures, the infidelity is dominated by extra photons, and the effect of extra photons in fusion-based quantum error correction is a largely open question requiring future research. 

When we consider the near-term parameters, the single-photon infidelity is considerably higher and is always dominated by the multi-photon component. In this case, the OM scheme does not produce single photons suitable for quantum computing. 
However, this parameter set shows that a proof-of-concept experiment is possible in the near term. 
One factor which limits the total fidelity is the extraction efficiency $\eta_0^\mathrm{ex}=0.98$,  which limits the retrieval efficiency $\eta_\mathrm{re}\leq 0.98$.
Additionally, parallelization is inhibited by the higher acoustic loss, $\Gamma$, and the longer heralding cycle duration, $T_\mathrm{h}$. 
The extraction efficiency can be improved by increasing the external optical coupling, $\kap^[ex][0]$, assuming the internal optical loss, $\kap^[int][0]$ cannot be reduced. However, increasing the total optical loss reduces the resonant enhancement of the OM interactions, necessitating a higher classical drive strength or longer duration to maintain the same initialization fidelity between each heralding attempt. 
If the drive strength is limited, this increases the heralding cycle duration, hindering parallelization. 
On the other hand, lower total optical loss necessitates more time for optical modes to populate and depopulate, which can also increase the heralding cycle duration. 
Thus, there is a trade-off to consider when choosing the external coupling rate of the optical modes. We chose $\kap^[ex]\approx\kappa=2\pi\times50$~MHz for the near-term parameters to give an initialization fidelity of $99\%$ when cooling from 0.1 phonons using a beam-splitting drive with strength 1~mW and duration $\sim10/\kappa$.
{In the following section, we numerically optimize the external coupling rate (and other parameters) to maximize the total parallelized single-photon fidelity, with necessary approximations to make the calculations tractable. }

\subsection{Minimum necessary OM coupling rate}\label{sec:min-g0}

In this section, we find the system parameters necessary to reach 99\% total single-photon fidelity, $F_\mathrm{tot}$ [Eq.~\eqref{eq:total-fid}], using bilevel optimization. 
The lower-level task is to maximize the total fidelity ($F_\mathrm{tot}$) given the OM coupling rate ($g_0$), detection efficiency ($\eta_\mathrm{d}$), thermal phonon population ($\nth$), and number of OM systems ($N$).
The free parameters are the external optical coupling rate ($\kap^[ex]$), photon-phonon pair generation probability ($p_1$), and the re-initialization drive duration ($T_\mathrm{init}$). 
These parameters can have different impacts on each figure of merit, as shown in the previous section and in Tab.~\ref{tab:total-fid}, leading to trade-offs.
In all cases, we limit the classical drive strengths to 1~mW, and we take the near-term values for internal optical loss ($\kap^[int]$) and acoustic loss ($\Gamma$). 
The upper-level optimization task is to find the OM coupling rate ($g_0$) where the maximized total fidelity crosses 99\% for a given $\eta_\mathrm{d}$, $\nth$, $N$. 
The results of the upper-level optimization are the minimum OM coupling rates required to reach 99\% total fidelity, and are shown in Fig.~\ref{fig:min-g0}. The corresponding $\kap^[ex]$, $p_1$, and $T_\mathrm{init}$ are shown in supplementary Fig.~\ref*{supp-fig:min-g0-pams}.

From Fig.~\ref{fig:min-g0}, we see that our proposed OM scheme can produce single photons suitable for LOQC with OM coupling rates only slightly higher than our near-term coupling rate, $g_0/2\pi=1$~kHz (Tab.~\ref{tab:pams}). 
In contrast to Sec.~\ref{sec:sp-tradeoff}, we took an ambitious approach to the heralding cycle, with minimal allowance for classical feedback and minimal drive down-time, as described below. 
This allows for faster heralding cycles and more effective parallelization, yielding more hopeful results than those in Sec.~\ref{sec:sp-tradeoff} and Fig.~\ref{fig:total-fid}.

\begin{figure}[htb]
    \centering
    \includegraphics{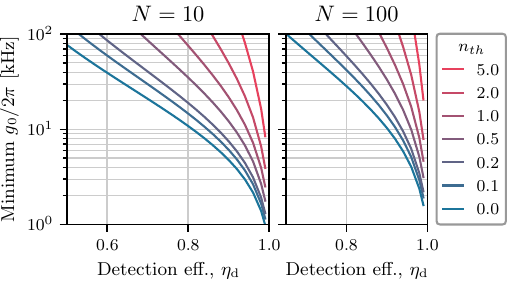}
    \caption{\textbf{Minimum OM coupling, $g_0$, to reach 99\% total fidelity, $F_\mathrm{tot}$} [Eq.~\eqref{eq:total-fid}], depending on detection efficiency, $\eta_\mathrm{d}$; thermal phonon population, $\nth$; and number of OM systems to prepare in parallel, $N$. 
    For this calculation, the squeezing and beam-splitting drive powers are taken to be no more than 1~mW, while the internal optical loss, $\kap^[int]$, and acoustic loss, $\Gamma$, are given their near-term values in Tab.~\ref{tab:pams}. 
    The detector dark count rate is 100~c.p.s.
    The external optical coupling rate, $\kap^[ex]$, and photon-phonon pair generation probability, $p_1$ are free parameters. 
    The squeezing drive duration and the re-initialization drive duration, $T_\mathrm{init}$, are constrained by the optical cavity rise and fall time but otherwise free. These drive durations determine the heralding cycle duration, $T_\mathrm{h}$.
    For this calculation, the retrieval efficiency, $\eta_\mathrm{re}$ [Eq.~\eqref{eq:retrieval-eff}], takes its long-time weak-coupling limit. 
    For comparison, recall that the near-term and target OM coupling rates used throughout this paper are $g_0/2\pi=1$~kHz and 100~kHz, respectively. 
    }
    \label{fig:min-g0}
\end{figure}

We now describe how we calculated the total single-photon fidelity ($F_\mathrm{tot}=F_\mathrm{init}F_\mathrm{h,sp}F_\mathrm{idle}\eta_\mathrm{re}$) as a function of the parameters $g_0$, $\eta_\mathrm{d}$, $\nth$, $N$, $\kap^[ex]$, $p_1$, and $T_\mathrm{init}$. 

The SP heralding fidelity ($F_\mathrm{h,sp}$, Eq.~\eqref{eq:herald-simp}, ~\eqref*{supp-eq:Fhsp}) depends primarily on the photon-phonon pair generation probability ($p_1$), the detection efficiency ($\eta_\mathrm{d}$), and optical external coupling rate ($\kap^[ex]$). 
The optical extraction efficiency ($\eta^\mathrm{ex}=\kap^[ex]/\kappa$, where $\kappa=\kap^[int]+\kap^[ex]$) combines with the detection efficiency to give the total efficiency, $\eta=\eta^\mathrm{ex}\eta_\mathrm{d}$. 
The SP heralding fidelity also depends on the detector dark count probability, which we estimate using the squeezing drive duration, the optical cavity response time $2\pi/\kappa$, and a conservative dark count rate of 100~c.p.s. 
The calculation of the SP heralding probability [Eq.~\eqref{eq:herald-prob},~\eqref*{supp-eq:ph}] uses these same parameters. 
See Sec.~\ref*{supp-sec:min-g0} for details. 

The idling fidelity ($F_\mathrm{idle}$, Eq.~\eqref{eq:fid-idle}) depends on the thermal phonon population ($\nth$), the number of OM systems ($N$), the SP heralding probability, and the heralding cycle duration. The heralding cycle duration is the sum of the squeezing drive duration and the re-initialization drive duration ($T_\mathrm{init}$), with an additional $8\times2\pi/\kappa$ to allow for drive rise/fall time and 5~ns to allow for classical feedback. 

The retrieval efficiency ($\eta_\mathrm{re}$, Eq.~\eqref{eq:retrieval-eff}) depends on the OM coupling rate ($g_0$) and optical external coupling rate ($\kap^[ex]$).  
We take the long-time, weak-coupling limit: $\eta_\mathrm{re}\rightarrow \eta^\mathrm{ex}\frac{C}{C+1}$. Recall $C=4g_0^2|\alpha_\mathrm{r}|^2/\kappa\Gamma$ is the effective cooperativity.

The initialization fidelity ($F_\mathrm{init}$) depends on all given parameters except $N$. 
The phonon population added during a heralding cycle in which no herald photon is detected depends on the SP heralding fidelity, the heralding cycle duration, and the thermal phonon population. 
The phonon population subtracted during the re-initialization drive depends on the OM coupling rate ($g_0$), the optical external coupling rate ($\kap^[ex]$), the re-initialization drive duration ($T_\mathrm{init}$), and the thermal phonon population ($\nth$).
The balance of these heating and cooling processes determine the initialization fidelity. For details see Sec.~\ref*{supp-sec:min-g0}.

These four factors determine the total single-photon fidelity's dependence on the given parameters, allowing us to perform the bilevel optimization. Our results are presented in Fig.~\ref{fig:min-g0}.

In summary, OM systems may practically be used to generate single photons for fault-tolerant LOQC (Fig.~\ref{fig:total-fid}--\ref{fig:min-g0}). 
Parallel production of many single photons for this purpose requires a high OM coupling rate to enable fast operation, and a low temperature to avoid acoustic mode heating while idling. 
Further investigation into the impact of excess photons on quantum error-correcting codes will determine whether higher operating temperatures are tolerable. 
The proposed OM single-photon source is inspired by all-optical photon-pair sources (e.g., based on spontaneous parametric down-conversion) which may be multiplexed to reliably produce single photons. 
The advantage of the scheme presented herein, relative to optically-multiplexed photon-pair sources, lies in the separation of the preparation and retrieval phases. Instead of requiring fast routing of heralded single photons, the single phonon is automatically preserved in place and can be read out on demand. 
The long lifetimes of the acoustic modes enable a high degree of multiplexing, and ultimately single-photon fidelity, which is not achievable in all-optical systems (see Sec.~\ref*{supp-sec:sp-comparison}).

\section{Entanglement resource}\label{sec:entanglement-source}

Recall that our goal with this paper is to show that harnessing optomechanics can remove barriers to fault-tolerant LOQC. In the previous section, we showed how OM systems can be used as on-demand single photon sources for LOQC. 
In this section, we extend our protocol to prepare small entangled states, another key ingredient for LOQC~\cite{bartolucciFusion2023}. 

Our proposal shows how entangled states, namely GHZ states, can be prepared in acoustic modes without direct phonon-phonon interactions. 
The principal technique is to realize a tunable OM beam splitter by adjusting the strength and duration of the red beam-splitting drive, rather than fully converting phonons into photons as we did in Sec.~\ref{sec:beamsplit}. 
This tunable OM beam-splitting interaction is applied to acoustic modes which have been prepared in the single-phonon state according to Sec.~\ref{sec:prep}. 
The optical modes are sent through a network of optical beam splitters and finally to single-photon detectors. A successful detection pattern heralds an entangled state in the dual-rail basis of the phonons. 
These states can be cached in the acoustic modes for a long time due to the high acoustic lifetime, and converted to the optical domain on demand. 
This approach reduces the redundancy and routing requirements for photonic quantum computing relying on probabilistically generated entangled states. 

In this section, we analyze the OM system's utility as a source of entangled optical photons. First, we estimate the lifetime of phononic dual-rail qubits (Sec.~\ref{sec:dual-rail}). Then, we show how a GHZ state may be prepared in the acoustic resonators and calculate the fidelity of such preparation (Sec.~\ref{sec:ghz-state-prep}). Finally, we show that the probability of successfully preparing a GHZ state may be improved by iteratively building the state (Sec.~\ref{sec:bleeding}). 

\subsection{Phononic dual-rail qubit lifetimes}\label{sec:dual-rail}

The dual-rail qubit is a popular logical encoding in quantum computing for its robustness against loss of single excitations, which is the dominant source of errors in linear optics~\cite{kokLinearOpticalQuantum2007}. 
Here we consider a dual rail qubit encoded in two acoustic resonator modes. The logical states are defined by location of a single excitation: 
$\ketL{0} \coloneq \ket{10}$ and $\ketL{1} \coloneq \ket{01}$. 

\begin{figure}[htb]
    \centering
    \includegraphics{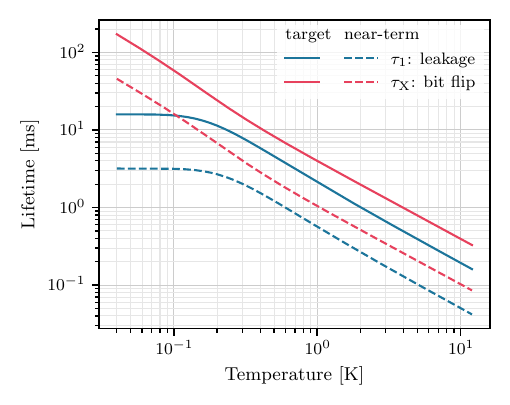}
    \caption{
    \textbf{Phononic dual-rail qubit lifetimes,} using target or near-term parameters (Tab.~\ref{tab:pams}). Resonators are assumed to be identical. 
    The leakage lifetime, defined as the inverse of the rate at which the qubit leaves the dual-rail subspace, is $\tau_1=((4\nth+1)\Gamma)^{-1}$, where $\nth$ is the thermal phonon population. Upon detection, leakage errors are converted into erasure errors. 
    $\tau_\mathrm{X}$ is the bit-flip (or $X$-error) lifetime, defined as the time at which the probability of a bit flip reaches $1/2e$ within the dual-rail subspace. It is roughly inversely proportional to the phonon bath temperature $T_\mathrm{b}$ (see Sec.~\ref*{supp-sec:dual-rail-lifetimes} for expression.)
    Because of the acoustic mode's low loss, millisecond-scale lifetimes are achievable even at single-digit Kelvin temperatures. However, the error bias---the tendency to have  only one kind of error---decreases as the temperature increases. Unbiased errors are more difficult to handle in quantum error-correcting codes~\cite{gottesmanStabilizerCodesQuantum1997}.
    }
    \label{fig:lifetimes}
\end{figure}

Single-phonon loss takes both logical states to the vacuum state $\ket{00}$, which is no longer in the logical basis. This is an example of a \emph{leakage error}. 
The rate at which the phononic qubit leaves the dual-rail subspace is $\tau_1^{-1}=(4\nth+1)\Gamma$, which defines to be the leakage lifetime, $\tau_1$ (Fig.~\ref{fig:lifetimes}). 
The contributions to the leakage rate are 
$(\nth+1)\Gamma$ for cooling from $\ket{1}$ to $\ket{0}$, 
$\nth\Gamma$ for heating from $\ket{0}$ to $\ket{1}$, and 
$2\nth\Gamma$ for heating from $\ket{1}$ to $\ket{2}$ (see Sec.~\ref*{supp-sec:dual-rail-lifetimes}). 
The acoustic dual-rail qubits are to be used in a measurement-based quantum computing scheme; therefore, every  qubit is eventually converted to the optical domain and detected. This measurement will reveal whether the system had the correct number of excitations. 
This converts a leakage error into an \emph{erasure error}---an error whose location can be identified, though the quantum information cannot be recovered~\cite{grasslCodesQuantumErasure1997}. 
Erasure errors are easier to accommodate in quantum error correction compared to computational errors such as bit or phase flips~\cite{gottesmanStabilizerCodesQuantum1997}. 

Bit and phase flips are known as \emph{Pauli errors} because they can be represented as the application of the Pauli $X$ or $Z$ operators on the logical state of the qubit, respectively. 
If the dual-rail qubit modes are at zero temperature, bit flips ($X$ errors) cannot occur. 
However, if the two modes have different decay rates, phase information can be lost even in the absence of heating. For example, if the second mode has higher loss than the first, both $\ketL{+}=(\ketL{0}+\ketL{1})/\sqrt{2}$ and $\ketL{-}=(\ketL{0}-\ketL{1})/\sqrt{2}$ will tend towards $\ketL{0}$ (Sec.~\ref*{supp-sec:dual-rail-lifetimes}). Therefore, if the qubit prepared in $\ketL{+}$ or $\ketL{-}$ is idled and then measured in the $X$ basis, it will sometimes return the incorrect result---an apparent phase flip ($Z$ error). 
At finite temperature, bit and phase flips are both possible (Fig.~\ref{fig:lifetimes}, Sec.~\ref*{supp-sec:dual-rail-lifetimes}). 

These error lifetimes determine how long entangled states can be stored in the acoustic modes before being retrieved. 

\subsection{Phononic entangled state preparation}\label{sec:ghz-state-prep}

In this section we show how the OM systems can be used to prepare and cache small entangled states in the acoustic modes for later use. 
These small entangled states can then be converted from the acoustic to the optical domain as described in Sec.~\ref{sec:beamsplit}, and used as inputs for LOQC. 

\begin{figure}[tb]
    \centering
    \begin{quantikz}[row sep={18pt, between origins}, column sep={14 pt}]
        \lstick[4,label style={rotate=90, anchor=south, yshift=2pt}]{acoustic modes\\
        $\ket{\psi_\mathrm{in}}\rightarrow\ket{1111}$} 
            \midstick{$[1,0]$}
            &[-4pt] \octrl{4}\wire[d][1]["p_\mathrm{re}"{yshift=2pt}]{a}
            \gategroup[4,steps=8,style={fill=phononlight!30, draw=none, xshift=-3pt,inner ysep=2pt, inner xsep=5pt}, background]{} 
            \gategroup[8,steps=4, style={thin,dashed,inner xsep=2pt},
                label style={label position=above,anchor=south,yshift=-0.1cm}]{\parbox{3cm}{OM BS interaction}} &&&&&&&&[-10pt] 
            \rstick[4, label style={anchor=north, rotate=90,yshift=-2pt}]{Bell state?}\\
            \midstick{$[1,1]$} && \octrl{4} &&&&&&& \\
            \midstick{$[2,0]$} &&& \octrl{4} &&&&&& \\
            \midstick{$[2,1]$} &&&& \octrl{4} &&&&& \\[4pt]
        \lstick[4,label style={rotate=90, anchor=south, yshift=2pt}]{optical modes\\
        $\ket{\psi_\mathrm{in}}\rightarrow\ket{0000}$} 
            \midstick{$[1,0]$} & \ocontrol{}        
            \gategroup[4,steps=8,style={fill=w2light!30, draw=none, xshift=-3pt, inner ysep=2pt, inner xsep=5pt}, background]{} 
            &&&& \ctrl{1}
            \gategroup[4,steps=3, 
                style={thin,dashed,inner xsep=2pt}, 
                label style={label position=below,anchor=north,yshift=-0.2cm, xshift=0pt}]{optical BSs} 
            && \ctrl{3} & \meterD{}
            \gategroup[4,steps=1, 
                style={thin,dashed,inner xsep=2pt,xshift=0pt}, 
                label style={label position=below,anchor=north,yshift=-0.2cm, xshift=0pt}]{SPDs} \\
            \midstick{$[1,1]$} && \ocontrol{} &&& \control{} & \ctrl{1} & & \meterD{} \\
            \midstick{$[2,0]$} &&& \ocontrol{} && \ctrl{1}  &  \control{}  && \meterD{} \\
            \midstick{$[2,1]$} &&&& \ocontrol{} & \control{} && \control{} & \meterD{} 
    \end{quantikz}
    \caption{
    \textbf{Linearized optomechanical circuit for preparing Bell state in phononic memory. }Certain detection patterns---e.g., single photons detected at $[1,1]$ and $[2,1]$ only---herald a Bell state in the acoustic modes. 
    Each horizontal wire represents an acoustic (green background) or optical (yellow background) mode, vertical lines with solid endpoints represent 50/50 beam splitters, and vertical lines with open endpoints represent tunable beam splitters (tuning parameter $p_\mathrm{re}$, Eq.~\eqref{eq:P_re}).
    BS=beam-splitting/beam splitter, SPD=single-photon detector.
    See Fig.~\ref*{supp-fig:bell-circuit} for circuit in conventional optical notation.
    }\label{fig:Bell-circuit}
\end{figure}
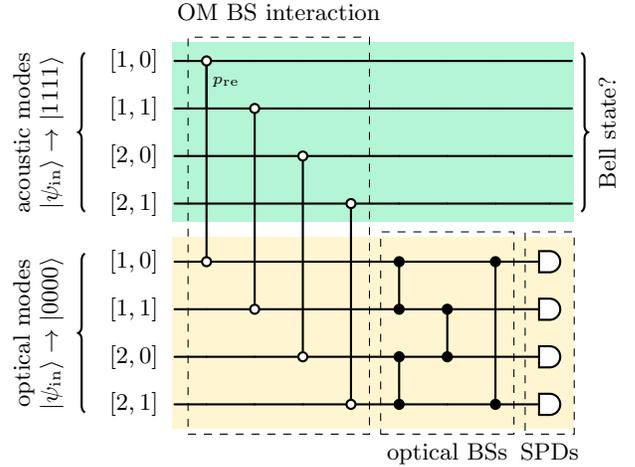

We show that it is possible to probabilistically produce one of the two $X$-basis GHZ states of $n$ phononic dual-rail qubits: $|n\text{-GHZ}^{\pm}\rangle=(\ketL{+}^{\otimes n}\pm\ketL{-}^{\otimes n})/\sqrt{2}$, where $\ketL{+}=(\ketL{0}+\ketL{1})/\sqrt{2}$ and $\ketL{-}=(\ketL{0}-\ketL{1})/\sqrt{2}$ are the eigenstates of $X$ for each phononic dual-rail qubit. 
Both whether we prepare a GHZ state and which of the two GHZ states is prepared is random, but heralded. 
The procedure is adapted from \cite{varnavaHowGoodMust2008,liResourceCostsFaultTolerant2015}. 
Our scheme is identical to that for preparing a GHZ state in optical photons, with the single-photon sources replaced by single phonons and the initial set of optical beam splitters replaced by the OM beam-splitting interaction. 
Figure~\ref{fig:Bell-circuit} illustrates the scheme for $n=2$, i.e., a Bell state. 
For describing our protocol, we group the acoustic modes into pairs and label the pairs by $i\in\{1,2,\ldots,n\}$. Within each pair, the modes are labeled by $q\in\{0,1\}$. Therefore, each acoustic mode has a unique label $[i,q]$. 
We refer to ``mode pairs'' rather than ``qubits'' because the modes are not in valid qubit states until the end of the protocol. 
For example, mode $[2,1]$ will represent the ``1'' rail of the second qubit at the end of a successful GHZ-state preparation. 
Each acoustic mode is coupled to an optical mode which shares the same label, $[i,q]$. We will always specify whether we are referring to the acoustic or optical mode.  

The GHZ-state preparation protocols relies on partially converting phonons into photons. 
Partial conversion is possible simply by reducing the beam-splitting drive strength and/or duration from that used to retrieve full photons in Sec.~\ref{sec:beamsplit}. 
Given a particular beam-splitting drive strength and duration, the probability that a single phonon is converted into a single photon is the \emph{retrieval probability}, 
\begin{equation}\label{eq:P_re}
    p_\mathrm{re} \coloneq 
    \frac{\kap[0]\int_0^{T_{\mathrm{re}}}  n_0(t) ~dt}
    {n_\mathrm{ph}(0) }
    \approx 1 - \frac{n_\mathrm{ph}(T_\mathrm{re})}{n_\mathrm{ph}(0)},
\end{equation}
with the approximation holding when phonon loss is negligible. 
For a drive of constant strength, the phonon and photon populations are given by Eqs.~\eqref{eq:bs-phonon}--\eqref{eq:bs-photon}. 
As in Eq.~\eqref{eq:retrieval-eff}, the beam-splitting drive is turned on at $t=0$ and the retrieval window ends at $t=T_\mathrm{re}$, which is a few photon lifetimes after the beam-splitting drive is turned off. 
The retrieval probability may be understood as an effective beam-splitting ratio or beam splitter transmissivity. 
The retrieval probability, Eq.~\eqref{eq:P_re}, differs from the retrieval efficiency, Eq.~\eqref{eq:retrieval-eff}, by a factor of the extraction efficiency: $p_\mathrm{re}=\eta_\mathrm{re}/\eta_0^\mathrm{ex}$.\footnote{Recall $\eta_0^\mathrm{ex}=\kap^[ex][0]/\kappa_0$.} We make this choice because in this section, the state of the acoustic mode after beam-splitting is important, whereas in Sec.~\ref{sec:beamsplit}, only the photon extracted to the waveguide was relevant. For the purposes of generating entanglement in the phonons, an optical photon absorbed by the internal loss of the OM cavity still removes the phonon from the acoustic mode, so we count the phonon as having been retrieved.

To prepare the $n$-GHZ entangled state, we start with $2n$ single phonons (prepared as described in Sec.~\ref{sec:parallelization}) and drive the OM beam-splitting interaction such that we expect no more than half of them to convert into photons ($p_\mathrm{re}\leq 0.5$), as illustrated in Fig.~\ref{fig:Bell-circuit} for $n=2$. 
Following the retrieval, there are two rounds of optical beam-splitting, which effectively connect the optical modes in a ring. 
First, each pair of optical modes $[i,0]$ and $[i,1]$, where $i\in\{1,2,\dots,n\}$, is put through a 50/50 beam splitter. 
Then, each pair of optical modes $[i,1]$ and $[(i \bmod n)+1,0]$ is put through a 50/50 beam splitter. 
Finally, the optical modes are sent to single-photon detectors. 

A successful herald requires that for each pair of detectors $[i,1]$ and $[(i\bmod n)+1,0]$, one detector counts a single photon and the other detector counts no photons, for a total of $n$ single-photon detections. 
If the detection outcome is one of the $2^n$ patterns satisfying this condition, we herald a GHZ state in the acoustic resonators. 
For example, when preparing a $2$-GHZ (i.e., Bell) state, successful heralds are single detections on and only on ([1,1] and [2,1]), ([1,1] and [1,0]), ([2,0] and [2,1]), or ([2,0] and [1,0]).\footnote{
    Detections on ([1,0] and [2,1]) and ([1,1] and [2,0]) produce Bell states, but in the wrong dual-rail basis. One may transform these into the correct states by swapping modes [1,1] and [2,0], but we discount this possibility and insist our Bell state be in the predefined basis.\label{fn:bell-other}
}
Notice that we start with $2n$ phonons and detect $n$ photons to get a state of $2n-n=n$ phonons. 
The herald detection pattern is described by a set of parity bits $\{s_i\}$, where $s_i=1$ ($s_i=-1$) if, for detector pair $[i,1]$ and $[(i\bmod n)+1,0]$, the single photon was detected at the first (second) detector of the pair.
Neglecting loss, the full herald detection event is described by the Kraus operator
\begin{equation}\label{eq:KGHZ}
    \K_\mathrm{GHZ}^{\{s_i\}} = \left( \frac{p_\mathrm{re}}{2} \right)^{n/2}
    (1-p_\mathrm{re})^{\htn[ph]/2} 
   \prod_{i=1}^n
       \left(\htb_{[i,+]}-s_i\htb_{[i+1,-]}\right),
\end{equation}
where 
$\htn[ph]=\sum_{i,q}\htb_{[i,q]}^\dag\htb_{[i,q]}$ is the total number of phonons, 
and $\htb_{[i ,\pm]}=(\htb_{[i,0]}\pm\htb_{[i,1]})/\sqrt{2}$ subtract a phonon from the positive or negative superposition of acoustic mode pair $i$ (see Sec.~\ref*{supp-sec:ghz-state-prep}). 
The total parity of the detection pattern, $\bar{s}\coloneq\prod_{i=1}^n s_i$, determines which of the two possible GHZ states is prepared. If $\bar{s}=1$ ($\bar{s}=-1$), the state prepared in the acoustic modes is $\ket{n\text{-GHZ}^+}$ ($\ket{n\text{-GHZ}^-}$). 
The GHZ state can be preserved in the acoustic modes as long as the phononic qubits remain error-free (Sec.~\ref{sec:dual-rail}), and converted from acoustic phonons to optical photons on demand. 

\subsubsection{Heralding probability and fidelity}
\label{sec:ghz-fid}

In this section, we calculate the fidelity of GHZ-state preparation and the probability of success following similar methods as in Sec.~\ref{sec:prep}. 
For this analytical calculation, we assume the single phonons have been perfectly prepared, and we neglect acoustic loss and heating. The imperfections we consider are optical loss/detection inefficiency and detector dark counts, which are expected to dominate in experiments. 
The independent variable in these calculations is the retrieval probability, $p_\mathrm{re}$, playing analogous role to the single-pair generation probability, $p_1$, in Sec.~\ref{sec:prep}.  

The \emph{GHZ-state heralding fidelity}, $F_\mathrm{h,ghz}$, is the square-overlap of the actual heralded state with the true $n$-GHZ state, i.e., 
$F_\mathrm{h,ghz}\coloneq\langle n\text{-GHZ}|\rho_\mathrm{h}|n\text{-GHZ}\rangle$, where $\rho_\mathrm{h}$ is the state of the $2n$ acoustic modes after heralding. 
Equivalently, the heralding fidelity is the classical probability of having a GHZ state given a successful herald, so we can use Bayes' theorem, as we did in Sec.~\ref{sec:prep}, Eq.~\eqref{eq:bayes}:
\begin{align}\label{eq:bayes-ghz}
    F_\mathrm{h,ghz} &= P(\text{GHZ state} ~|~ \text{herald}) \nonumber\\
    &= P(\text{herald} ~|~ \text{GHZ state}) \frac{P(\text{GHZ state})}{P(\text{herald})}.
\end{align}
\begin{itemize}
    \item $P(\text{herald} ~|~ \text{GHZ state})$ is the probability of heralding success given that the acoustic modes are truly in a GHZ state. Because we neglect acoustic imperfections, it is simply related to the detection efficiency: 
    $P(\text{herald} ~|~ \text{GHZ state})=\eta^n=(\eta_0^\mathrm{ex}\eta_\mathrm{d})^n$. 
    \item $P(\text{GHZ state})$ is the unconditioned probability of preparing a GHZ state, i.e., the square-overlap between a GHZ state and the state of the acoustic modes after the OM beam-splitting interaction, without considering the optical detection pattern. 
    We have $P(\text{GHZ state})=2p_\mathrm{re}^n(1-p_\mathrm{re})^n$, which may be understood intuitively because $n$ phonons must be converted into photons, $n$ must remain as phonons, and there are 2 GHZ states that can be generated, i.e., the positive and negative superpositions (see Sec.~\ref*{supp-sec:ghz-overlap}). 
    \item $P(\text{herald})=p_\mathrm{h,ghz}$ is the \emph{GHZ-state heralding probability}: the probability of getting a GHZ herald detection pattern irrespective of the actual state of the acoustic modes. Considering both detector inefficiency and the possibility of dark counts, we have
    \begin{equation}\label{eq:ph-ghz}
        p_\mathrm{h,ghz}=2(\eta p_\mathrm{re})^n(1-\eta p_\mathrm{re})^n 
            \left(1+ n^2p_\mathrm{d}\frac{1-\eta p_\mathrm{re}}{\eta p_\mathrm{re}}\right),
    \end{equation}
    where $p_\mathrm{d}$ is the detector dark count probability. Notice that the first part of the above expression is the same as $P(\text{GHZ state})$ with $p_\mathrm{re}\rightarrow\eta p_\mathrm{re}$. Intuitively, this is because the photons must be retrieved \emph{and} detected, not only retrieved (see Sec.~\ref*{supp-sec:lossy-det-general}). The final term in Eq.~\eqref{eq:ph-ghz} comes from the possibility of a dark count leading to a herald if the true detection pattern was one short of a herald pattern (see Sec.~\ref*{supp-sec:herald-minus-one}).
\end{itemize}

\begin{figure}[tb]
    \centering
    \includegraphics{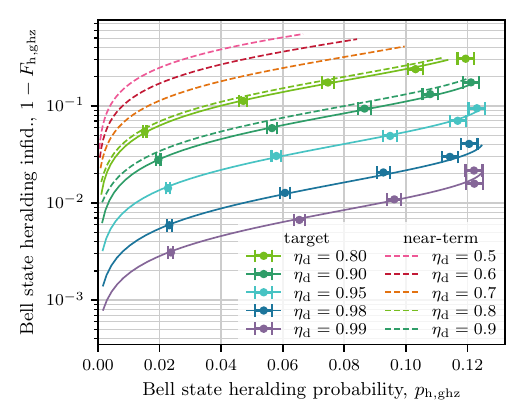}
    \caption{\textbf{Trade-off between phononic Bell state heralding fidelity and probability,}
    prepared using the circuit in Fig.~\ref{fig:Bell-circuit}. Lines show analytical expressions, Eq.~\eqref{eq:ph-ghz} and \eqref{eq:herald-fid-ghz}, and dots with error bars show results of QuTiP simulations. Both heralding probability and heralding fidelity are functions of retrieval probability, $p_\mathrm{re}$ [Eq.~\eqref{eq:P_re}], which is adjusted in simulation via the OM beam-splitting drive strength. 
    The dark count probability is $4.4\times10^{-7}$ ($1\times10^{-5}$) using the target (near-term) set of parameters. }
    \label{fig:Bell-Fh}
\end{figure}

All together, the GHZ-state heralding fidelity, Eq.~\eqref{eq:bayes-ghz}, is
\begin{equation}\label{eq:herald-fid-ghz}
    F_\mathrm{h,ghz} = \left(\frac{1-p_\mathrm{re}}{1-\eta p_\mathrm{re}}\right)^n
        \left( 1 + n^2p_\mathrm{d}\frac{1-\eta p_\mathrm{re}}{\eta p_\mathrm{re}} \right)^{-1}
\end{equation}
in the presence of optical loss and detector dark counts (Fig.~\ref{fig:Bell-Fh}). We have assumed that the dark count probability is low enough that the probability of multiple dark counts is negligible ($np_\mathrm{d}\ll \eta p_\mathrm{re}$).
We see that the GHZ-state heralding fidelity is generally improved by lowering the retrieval probability. 
This is because a higher retrieval probability increases the likelihood that more than $n$ phonons were converted into photons, but some detections were missed, leaving the acoustic modes with fewer than $n$ phonons in total. 
This trade-off between heralding probability and fidelity is analogous to the result in Sec.~\ref{sec:prep} that the single-phonon heralding fidelity improves with decreasing single-pair generation probability.  
If the retrieval probability is too low ($p_\mathrm{re}\sim \sqrt{2np_\mathrm{d}/(\eta(1-\eta))}$), detector dark counts start to impact the heralding fidelity. 
When dark counts are negligible, heralding a GHZ state using inefficient detectors is identical to heralding with perfect detectors and imposing uniform, $p_\mathrm{re}$-dependent, loss on the output modes afterwards.

Eq.~\eqref{eq:ph-ghz} and \eqref{eq:herald-fid-ghz} are validated with QuTiP simulations of Bell state generation circuit (Fig.~\ref{fig:Bell-circuit}), with results shown in Fig.~\ref{fig:Bell-Fh}. 
The simulation concretely implements the OM beam splitter using its Hamiltonian, Eq.~\eqref{eq:bs}, with a time-dependent drive (see Sec.~\ref*{supp-sec:ghz-herald-sim}). 
Optical beam-splitting is implemented via a scattering matrix transformation of the detected optical modes. 
As described previously, there are four detection patterns which herald a Bell state in the chosen dual-rail basis. Another two detection patterns herald Bell states in a different dual-rail basis (see footnote~\ref{fn:bell-other}). 
For this simulation, all six detection patterns were counted as heralding events in order to collect more statistics and reduce simulation uncertainty. Including these extra detection events increases the heralding probability by $3/2$ and does not affect the heralding fidelity. The simulated heralding probability is rescaled by $2/3$ in Fig.~\ref{fig:Bell-Fh} to match Eq.~\eqref{eq:ph-ghz}.
Using the target parameters, we take a beam-splitting drive with duration 1.18~ns, which gives retrieval probability $p_\mathrm{re}=0.5$ at 0.5~mW drive strength. The detector integration time is 4.4~ns, giving dark count probability $p_\mathrm{d}=4.4\times10^{-7}$. 
With the near-term parameters, a beam-splitting drive duration of 70~ns and strength of 0.2~mW gives retrieval probability $p_\mathrm{re}=0.5$. We assume a 100~ns detector integration time, giving dark count probability $p_\mathrm{d}=1\times10^{-5}$.

\subsection{Increasing GHZ-state heralding probability by iterative ``bleeding''}\label{sec:bleeding}

The process described in the previous section takes a single shot at probabilistically preparing a phononic GHZ state---there is one round of beam-splitting and detection, after which the process is judged to have either succeeded or failed. 
The single-shot GHZ-state heralding probability is maximized when the retrieval probability is $p_\mathrm{re}=0.5$, giving $p_\mathrm{h,ghz}=1/2^{2n-1}$ [Eq.~\eqref{eq:ph-ghz}]. This probability may be improved by using the iterative ``bleeding'' strategy, adapted from \cite{bartolucciCreationEntangledPhotonic2021}.
We neglect loss and detector dark counts unless otherwise stated. 

\begin{figure}[tb]
    \centering 
    \begin{quantikz}[row sep={14pt, between origins}, column sep={8 pt}]
            \lstick{$[1,0]$}
            &[-4pt] \octrl{4}\wire[d][1]["p_\mathrm{re}^{(1)}"{yshift=2pt}]{a}
            \gategroup[4,steps=17,style={fill=phononlight!30, draw=none, xshift=-2pt,inner ysep=1pt, inner xsep=0pt}, background]{}
            \gategroup[8,steps=8, label style={label position=above, yshift=-4pt}, style={draw=none}]{$\overbrace{\hphantom{\parbox{90pt}{foo}}}^{\textstyle\text{first iteration}}$}
            &&&&&&& 
            \slice{}            
            &&[-4pt] \octrl{4}\wire[d][1]["p_\mathrm{re}^{(2)}"{yshift=2pt}]{a}
            \gategroup[8,steps=8, label style={label position=above, yshift=-4pt}, style={draw=none}]{$\overbrace{\hphantom{\parbox{90pt}{foo}}}^{\textstyle\text{second iteration}}$}
            &&&&&&&&[-8pt]
            \\
            \lstick{$[1,1]$} && \octrl{4} &&&&&& &&& \octrl{4} &&&&&&& \\
            \lstick{$[2,0]$} &&& \octrl{4} &&&&& &&&& \octrl{4} &&&&&& \\
            \lstick{$[2,1]$} &&&& \octrl{4} &&&& &&&&& \octrl{4} &&&&& \\[4pt]
            \lstick{$[1,0]$} 
            & \ocontrol{}        
            \gategroup[4,steps=8,style={fill=w2light!30, draw=none, xshift=-2pt, inner ysep=1pt, inner xsep=0pt}, background]{} 
            &&&& \ctrl{1} && \ctrl{3} & \meterD{} & \wireoverride{n} 
            & \ocontrol{}        
            \gategroup[4,steps=8,style={fill=w2light!30, draw=none, xshift=-2pt, inner ysep=1pt, inner xsep=0pt}, background]{} 
            &&&& \ctrl{1} && \ctrl{3} & \meterD{} \\
            \lstick{$[1,1]$} 
            && \ocontrol{} &&& \control{} & \ctrl{1} & & \meterD[fill=w2light]{} & \wireoverride{n} 
            && \ocontrol{} &&& \control{} & \ctrl{1} & & \meterD{} \\
            \lstick{$[2,0]$} 
            &&& \ocontrol{} && \ctrl{1}  &  \control{}  && \meterD{} & \wireoverride{n} 
            &&& \ocontrol{} && \ctrl{1}  &  \control{}  && \meterD{} \\
            \lstick{$[2,1]$} 
            &&&& \ocontrol{} & \control{} && \control{} & \meterD{} & \wireoverride{n}
            &&&& \ocontrol{} & \control{} && \control{} & \meterD[fill=w2light]{}
    \end{quantikz}    
    \caption{
    \textbf{Iterated application of the Bell-state preparation protocol} to increase heralding probability. 
    Figure uses same notation as Fig.~\ref{fig:Bell-circuit}.
    If one or fewer photons is detected in the first iteration of the Bell-state preparation protocol, the protocol is repeated. Detections from the first and second iterations are combined to determine if a Bell state was heralded. 
    The same acoustic modes (green background) participate in both iterations, but each iteration has a separate set of spatio-temporal optical modes (yellow background), which always start in vacuum. 
    The retrieval probability used for each iteration, $p_\mathrm{re}^{(k)}$, may be adjusted depending on how many photons have been detected so far. 
    If a single photon is detected at $[1,1]$ [Eq.~\eqref{eq:K0100}] in the first iteration of the protocol (represented by filled-in detector), the state of the acoustic modes at the location of the red dashed line is $\ket{\psi'}$~[Eq.~\eqref{eq:Bell-intermediate-state}]. 
    If, after this, a photon is then detected at $[2,1]$ [Eq.~\eqref{eq:K0001}] in the second iteration, the acoustic modes are prepared in the positive Bell state. 
    This iterative strategy increases the probability of successfully preparing a Bell state compared to the single-shot strategy. 
    }\label{fig:bleeding}
\end{figure}
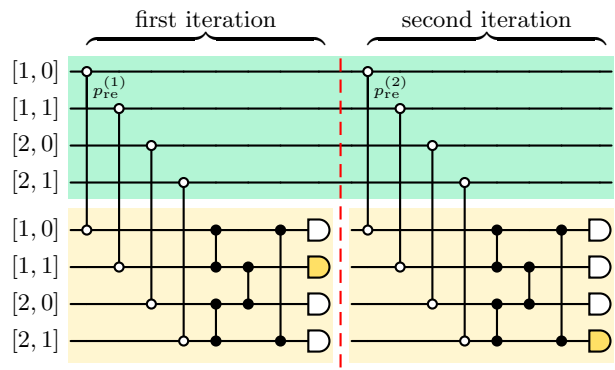

We begin with an illustrative example: suppose that upon running the Bell-state preparation protocol (Fig.~\ref{fig:Bell-circuit}), a single photon is detected at $[1,1]$ and nowhere else. 
As shown in Sec.~\ref*{supp-sec:ghz-bleeding}, this detection event can be represented by the Kraus operator\footnote{This Kraus operator is labeled $\K_{12}^{\mathrm{F},+}$ in Sec.~\ref*{supp-sec:ghz-bleeding}.} 
\begin{equation}\label{eq:K0100}
    \K_{0100}=\sqrt{\frac{p_\mathrm{re}}{2}}(1-p_\mathrm{re})^{\htn[ph]/2}
    \left(\htb_{[1,+]}-\htb_{[2,-]}\right),
\end{equation}
leaving the acoustic modes in the state\footnote{This state is labeled $|{\pi_2^+}\rangle$ in Sec.~\ref*{supp-sec:ghz-bleeding}.} 
\begin{equation}\label{eq:Bell-intermediate-state}
    \K_{0100}\ket{\psi_\mathrm{in}}
    \propto\ket{\psi'}=\frac{1}{\sqrt{2}}\left(\ketL{+}\ket{11}+\ket{11}\ketL{-}\right). 
\end{equation}
The acoustic mode pairs are entangled, but only one mode pair can be in a qubit state at a time because we have only subtracted one phonon from the initial four-phonon state $\ket{\psi_\mathrm{in}}=\ket{1111}$. 
At this point, we can repeat the Bell-state preparation protocol on the acoustic modes, as illustrated in  Fig.~\ref{fig:bleeding}, in the hope of detecting one more photon on an appropriate detector to project the acoustic modes onto a Bell state. 
Suppose that in this second round, a single photon is detected at $[2,1]$ and nowhere else. This corresponds to the Kraus operator\footnote{This Kraus operator is labeled $\K_{21}^{\mathrm{F,+}}$ in Sec.~\ref*{supp-sec:ghz-bleeding}.}
\begin{equation}\label{eq:K0001}
    \K_{0001}=\sqrt{\frac{p_\mathrm{re}}{2}}(1-p_\mathrm{re})^{\htn[ph]/2}
    \left(\htb_{[2,+]}-\htb_{[1,-]}\right),
\end{equation}
which prepares the acoustic modes in a Bell state: $\K_{0001}\ket{\psi'}\propto\left(\ketL{++}+\ketL{--}\right)/\sqrt{2}$.
Using this strategy of repeating the protocol as needed, we can boost the probability of preparing a Bell state from $1/8=12.5\%$ to $18.3\%$ when allowing up to two iterations 
(see Sec.~\ref*{supp-sec:2-round-bell}).
In this example, we showed that sequential detections on detectors $[1,1]$ and $[2,1]$ (i.e., iterative preparation) produces the same Bell state as simultaneous detections on these detectors (i.e., single-shot preparation, Sec.~\ref{sec:ghz-state-prep}). 
That is, $\K_{0001}\K_{0100}\ket{1111}\propto \K_{0101}\ket{1111}$, where $\K_{0101}=\K_\mathrm{GHZ}^{\{1,1\}}$ [Eq.~\eqref{eq:KGHZ}].

This iterative strategy can be generalized to preparing $n$-GHZ states, $n\geq2$ (see Sec.~\ref*{supp-sec:ghz-bleeding}). 
The iterative strategy works because the which-path information (i.e., which acoustic modes gave up their phonons) is erased by the optical beam-splitting regardless of when a photon is detected.
Mathematically, the terms in the product in $\K_\mathrm{GHZ}$ [Eq.~\eqref{eq:KGHZ}] commute with each other, showing that the order of detections is immaterial. 

The iterative strategy to prepare an $n$-GHZ state begins the same as the single-shot strategy:
we start with $2n$ single phonons and drive OM beam-splitting between each acoustic mode and its corresponding optical mode. Then, the optical modes are beam-split among each other and sent to single-photon detectors. 
If $n$ photons are detected in any of the $2^n$ herald detection patterns, the acoustic modes are projected into an $n$-GHZ state (Sec.~\ref{sec:ghz-state-prep}). 
The single-shot strategy ends here, but the iterative strategy continues.
If fewer than $n$ photons are detected, but the locations of the detections comprise a subset of a herald detection pattern, 
then the acoustic mode pairs are left in an entangled state which is not (yet) a GHZ state. 
This entangled state can be probabilistically transformed into a GHZ state by iterating the GHZ-state preparation protocol.  
If the new detections complete a herald detection pattern when added to the previous detections, the state preparation is successful. 
The only change made in later iterations of the GHZ-state preparation protocol is that if there 
have already been single photons detected at each of detector pairs $\{[i-1,1],[i,0]\}$ and $\{[i,1],[i+1,0]\}$, 
then acoustic mode pair $i$ is not retrieved, i.e., the OM beam-splitting interaction is not driven for those modes. 

The iterative strategy of GHZ state preparation increases the probability of success because it allows one to decide whether to repeat the protocol based on the number and location of single-photon detections thus far. 
One may also adjust the retrieval probability based on the cumulative detections. 
Bartolucci et al. named the iterative strategy for GHZ state preparation ``bleeding'' because the original single phonons (photons, in their case) are slowly ``bled out'' from their modes, rather than being subjected to a strong beam splitter \cite{bartolucciCreationEntangledPhotonic2021}. 
We believe the iterative bleeding strategy is particularly well-suited to the OM system because the acoustic modes are stationary and naturally allow repeated interactions. 

\subsubsection{Heralding fidelity}\label{sec:bleeding-fid}

The heralding fidelity of a Bell state prepared in $2$ iterations of the preparation protocol is given by the previously calculated formula, Eq.~\eqref{eq:herald-fid-ghz}, replacing the retrieval prob, $\prre$, with effective retrieval probability 
\begin{equation}\label{eq:p-re-eff}
    \prre^\mathrm{eff} = 1 - \left(1-\prre^{(1)}\right)\left(1-\prre^{(2)}\right),
\end{equation}
where $\prre^{(1)}$ ($\prre^{(2)}$) is the retrieval probability used in the first (second) iteration (see Fig.~\ref{fig:bleeding}). 
Intuitively, this is because the acoustic modes cannot distinguish between two weak beam-splitting drives and one longer or stronger beam-splitting drive. 
The dark count probability must also be scaled to account for the effectively longer detection window: $p_\mathrm{d}\rightarrow 1-(1-p_\mathrm{d})^2 \approx 2p_\mathrm{d}$. 
The increased effective retrieval probability and increased dark count probability lower the GHZ-state heralding fidelity as the number of iterations increases, holding the retrieval probability per round constant. 
This relation is validated by QuTiP simulations. 

In practice, the retrieval probability per round is not constant, and high heralding fidelity may be maintained even while heralding probability is enhanced using bleeding. 
For example, with single-shot preparation, the Bell-state heralding fidelity is $F_\mathrm{h,ghz} = 0.959$ when the heralding probability is maximized, taking optical efficiency to be $\eta=\eta_0^\mathrm{ex}\eta_\mathrm{d}=0.999\times0.98$. 
Allowing up to two rounds of bleeding, we find that the average heralding fidelity is largely unaffected, $F_\mathrm{h,ghz}= 0.959$, when maximizing the heralding probability. This heralding fidelity is calculated by averaging together the fidelities associated with each way the system can herald success, weighted by their relative probabilities. 
Specifically, 
$59\%$ of heralds occur in the first round, producing a Bell state with fidelity $0.976$; 
$30\%$ of heralds have one detection in the first round and one detection in the second round, producing a Bell state with fidelity $0.944$; and
$11\%$ of heralds have no detections in the first round and a full herald detection pattern in the second round, producing a Bell state with fidelity $0.914$ (see Sec.~\ref*{supp-sec:bleeding-fid}).
The average heralding fidelity is not decreased even while the heralding probability is increased from $12.5\%$ to $18.3\%$
because using the iterative strategy allows for a lower retrieval probability on the first iteration, and most heralds occur in the first iteration. This offsets the decreased fidelity in the worst case. 
The average heralding fidelity is a relevant metric when preparing many small entangled states for fusion-based quantum computing. 

For more than two bleeding stages, or $n$-GHZ states with $n>2$, such a formula for the heralding fidelity is not readily available because the no-jump back-action (i.e., the effect of non-detection) does not commute with detection events. 
Nonetheless, we may intuitively expect the heralding fidelity to be negatively impacted by excessive cumulative retrieval probability. 
Crucially, the iterative strategy of GHZ-state preparation allows errors to accumulate. This is in contrast with the iterative strategy of single-phonon preparation outlined in Sec.~\ref{sec:parallelization}, because the iterative preparation of GHZ states cannot include a reset. Despite this, we showed by example that high average heralding fidelity may be maintained even when heralding probability is increased by the bleeding strategy. 
This is because bleeding allows one to use lower retrieval probabilities in the first iteration of the state preparation protocol.

\subsubsection{Asymptotic heralding probability}\label{sec:asym-succ-prob}

The probability of eventually heralding an $n$-GHZ state is maximized in the asymptotic limit where the retrieval probability per iteration tends to zero ($\prre\rightarrow 0$) and the number of iterations tends to infinity. 
This maximizes the heralding probability because it guarantees that the correct number of photons will be detected, because no more than one phonon will be converted to a photon per iteration. 
In this limit, the only way that the GHZ state preparation can fail is if the locations of the detections do not match any of the $2^n$ heralding detection patterns.   
The heralding probability in the asymptotic limit serves as a benchmark for the benefit of bleeding compared to single-shot preparation of GHZ states, though it is not a practical operational limit. 
Operating in the $\prre\rightarrow 0$ limit would magnify the effect of detector dark counts and of phonon loss. 
Therefore, we do not consider detection imperfections when calculating the asymptotic heralding probabilities.

The asymptotic heralding probability must be calculated separately for each size of GHZ state, $n$, because the detection probabilities are order-dependent. 
The asymptotic heralding probabilities for $n=2$ through 15 were calculated in \texttt{Mathematica} (see Sec.~\ref*{supp-sec:asym-succ-prob}) and 
found empirically to follow $p_\text{asym}(n)\approx0.759/2.24^{n-1}$. This is far better than the maximum heralding probability of $p_\mathrm{h,ghz}=0.5/4^{n-1}$ [Eq.~\eqref{eq:ph-ghz}] for single-shot preparation.\footnote{
The asymptotic success probability in \cite{bartolucciCreationEntangledPhotonic2021} is $1/2^{n-1}$. Ours is worse because of the limitation that the acoustic modes cannot be directly beam-split with each other.
}

\subsubsection{Rounds to success with single-phonon reset}\label{sec:ghz-adaptive}

We now combine our analysis of the iterative preparation of $n$-GHZ states with our analysis of parallelized single-phonon preparation (Sec.~\ref{sec:photon-source}). 

Recall that the iterative GHZ-state preparation succeeds when $n$ single photons have cumulatively been detected in a heralding pattern (see Sec.~\ref{sec:ghz-state-prep}). 
The other endpoint of GHZ-state preparation is failure, which occurs when too many photons are detected, or if photons are detected in a pattern inconsistent with a heralding pattern. 
The $n$-GHZ state preparation protocol reaches an endpoint (success or failure) in fewer iterations when the retrieval probability per iteration is increased. 
On the other hand, the probability that this endpoint is success decreases as the retrieval probability is increased, because a higher retrieval probability increases the likelihood of too many photons being detected. 
Therefore, there is a trade-off between how quickly the protocol reaches an endpoint and the likelihood that that endpoint is success. 

If the GHZ-state preparation fails, the acoustic modes are reset to the single-phonon state (Sec.~\ref{sec:init}--\ref{sec:parallelization}), and the GHZ-state preparation protocol is restarted. 
The number of single-phonon heralding cycles (Fig.~\ref{fig:heralding-cycle}) it takes to prepare the $2n$ single phonons is the \emph{single-phonon reset time}, given by $\overline{M}(2n,p_\mathrm{h,sp})$ [Eq.~\eqref{eq:Mbar}].  

The total number of rounds to prepare an $n$-GHZ state includes the iterations of the GHZ-state preparation protocol as well as the heralding cycles to reset the acoustic modes.
We calculate this expected total number of rounds as a function of the retrieval probabilities used in the iterative GHZ-state preparation and the single-phonon reset time (see Sec.~\ref*{supp-sec:ghz-rounds} for details). 
Each iteration of the GHZ-state preparation protocol is expected to be of comparable duration to the single-phonon heralding cycle because it is similarly limited by the response time of the driven optical modes and by classical feedback. 
Therefore, the total number of rounds serves as a proxy for the time to prepare an $n$-GHZ state.

\begin{figure}[tb]
    \centering
    \includegraphics{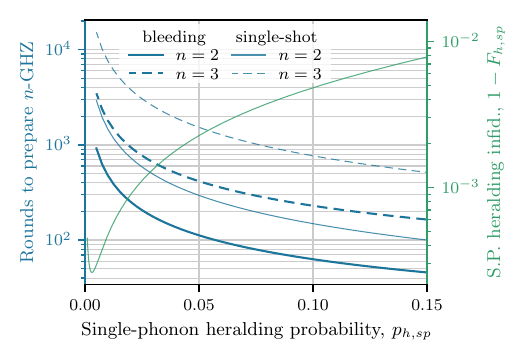}
    \caption{\textbf{Adaptive bleeding compared to single-shot strategy for preparing entangled states.} 
    The expected number of rounds to prepare an $n$-GHZ state (blue lines, left $y$-axis) is minimized when using the bleeding strategy for GHZ-state preparation (thick lines) by adjusting the retrieval probability at each step.
    The optimal retrieval probabilities depend on the number of rounds it takes to reset the acoustic modes to the single-phonon (SP) state after failure, which depends on the SP heralding probability ($p_\mathrm{h,sp}$, $x$-axis).
    For the single-shot strategy (thin lines), the optimal retrieval probability is always $1/2$.
    The single-phonon heralding probability also determines the single-phonon heralding fidelity, $F_\mathrm{h,sp}$ [Eq.~\eqref{eq:herald-simp}] (green line, right $y$-axis, taking $\eta=\eta_\mathrm{h}^\mathrm{ex}\eta_\mathrm{d}=0.999\times0.98$). Lower-fidelity single phonons will lead to lower-fidelity GHZ states. 
    When preparing many GHZ states in parallel, earlier-prepared states decay while idling and waiting for later-prepared states, yielding a fidelity trade-off analogous to that shown in Sec.~\ref{sec:parallelization}--\ref{sec:sp-tradeoff} for parallelized single-phonon preparation.
    }
    \label{fig:rounds-to-ghz}
\end{figure}

If the single-phonon reset time is high ($p_\mathrm{h,sp}$ is low), it is advantageous to use low retrieval probability in order to increase the likelihood that the GHZ-state preparation endpoint is success, and avoid wasting time resetting the acoustic modes after failure. 
On the other hand, if the reset time is low ($p_\mathrm{h,sp}$ is high), it is advantageous to have a higher retrieval probability and risk GHZ-state preparation failure in order to reach an endpoint faster.  
We minimize the expected total number of rounds to prepare a phononic $n$-GHZ state by adjusting the retrieval probabilities used in the iterative GHZ-state preparation protocol (Fig.~\ref{fig:rounds-to-ghz}).

Recall from Sec.~\ref{sec:prep} that the single-phonon heralding fidelity decreases when the single-phonon heralding probability is increased (Fig.~\ref{fig:herald-fid}, green line in Fig.~\ref{fig:rounds-to-ghz}). 
This will impact the final fidelity of the prepared GHZ state, which remains to be quantified. 
The purpose of minimizing the average number of rounds to prepare a GHZ state is to enable parallel preparation of many small entangled states in separate sets of acoustic modes.
Entangled states prepared earlier may incur errors, as shown in Sec.~\ref{sec:dual-rail}, while idling and waiting for other acoustic GHZ states to be prepared. 
Therefore, there exists a trade-off when preparing many GHZ states between the fidelity of the initial single phonons and errors incurred while the GHZ states idle, analogous to the trade-off between single-phonon heralding fidelity and idling fidelity outlined in Sec.~\ref{sec:sp-tradeoff}. 
Optimizing this trade-off is left for future work. 

\section{Conclusion}

In this paper, we provided an in-depth analysis on the use of OM systems (Sec.~\ref{sec:background}) to provide resources---single optical photons (Sec.~\ref{sec:photon-source}) and small entangled states (Sec.~\ref{sec:entanglement-source})---for fault-tolerant linear optical quantum computing.
Our proposal avoids two main challenges of previous strategies, namely, the need to reroute and retime quantum optical states in the all-optical approach, and difficulty collecting photons and maintaining indistinguishability in matter-based approaches. 
The fundamental strategy of our proposal is to probabilistically prepare the quantum resource in many copies of the OM system in parallel, cache the states in the acoustic modes, and retrieve the states as needed. 
The acoustic modes have very long lifetimes relative to the optical modes. 
This disparity enables many attempts at preparing the desired state in one acoustic mode before an idling, previously prepared state in a different acoustic mode decays. 
We calculated that the single photons produced by this scheme would have high enough fidelity for quantum computing when the OM coupling rate is high enough (e.g. $100$~kHz) to enable fast operation and the phonons have negligible thermal population (see Tab.~\ref{tab:pams}, Fig.~\ref{fig:total-fid}, and Fig.~\ref{fig:min-g0}). 
Finally, we showed that entangled states can be probabilistically prepared in the acoustic modes using linear optics. We calculated that the fidelities of such states  easily exceed 99\% (see Fig.~\ref{fig:Bell-Fh}), and are primarily determined by the strength of the induced photon-phonon interaction and the detection system efficiency. 
Additionally, we showed that the probability of successfully preparing an entangled state may be enhanced by using an iterative strategy, without harming the average state fidelity. This iterative strategy is particularly well-suited to OM systems as the acoustic modes are stationary and naturally admit repeated interactions. 

Our studies show which experimental issues are most important to address.
The first is the tendency of OM systems to heat when strongly driven. 
Strong optical drives are necessary to increase the effective OM interaction strength. 
The heating problem may be mitigated by careful choice of system material and design to reduce optical absorption and improve thermal anchoring \cite{diamandiQuantumOptomechanicalControl2024,brubakerOptomechanicalGroundStateCooling2022,mayorHighPhotonphononPair2025}.
The second challenge is achieving very low optical loss inside the resonator. This is necessary to have high extraction efficiency while keeping the total optical linewidth low enough to avoid intramodal scattering. 
Furthermore, to separate a single photon from the strong pump requires optical filters with very high rejection and minimal insertion loss. 
Finally, fabricating identical devices is challenging in any situation, and optomechanics is no exception. While the acoustic frequency does not matter per se, the output optical frequencies must be identical to produce indistinguishable photons. This challenge may be addressed by in-situ tuning of optical resonances.
These challenges can be overcome, and we believe our proposed OM scheme is a practical alternative for producing quantum optical resources. 

We hope this proposal will provide motivation for experimental progress in quantum optomechanics, and foster further exploration of these systems' usefulness for quantum computing. 
While we focus on optomechanics, the strategies described in this work are broadly applicable to emissive-type quantum optical memories, such as atom clouds or NV centers \cite{barrettEfficientHighfidelityQuantum2005,leiQuantumOpticalMemory2023}. 
These OM techniques may be extended in the future to produce cat states \cite{dodonovEvenOddCoherent1974,lundConditionalProductionSuperpositions2004,ourjoumtsevGenerationOpticalSchrodinger2007,chenGenerationHeraldedOptical2024,zhaoFastRobustCat2023} or GKP states \cite{gottesmanEncodingQubitOscillator2001,konnoLogicalStatesFaulttolerant2024} in acoustic modes. 

Finally, our proposal invites inquiry into the design of quantum error-correcting codes for linear optics. 
The impact of imperfect single photons, especially the possibility of biphotons, on quantum error correction remains largely an open question. This impact will affect the requirements on any single photon source intended for linear optical quantum computing. 
Furthermore, our proposed OM system lends itself to parallel, asynchronous operation. 
An error-correcting code designed to use resources as they become available, rather than on a fixed clock, may be better able to harness the potential of the quantum OM system.

{\emph{Acknowledgements:}}
We thank Hilel Hagai Diamandi, David Mason, Freek Ruesink and Kaavya Sahay for fruitful discussion. 
We thank to Hilel Hagai Diamandi, Steven M. Girvin, Kaavya Sahay, Daniel K. Weiss for help editing the manuscript. 
This material is based upon work supported by the National Science Foundation (NSF) under Grant No. 2137740. Additional support was also provided by the Air Force Office of Scientific Research (AFOSR) and the Office of Naval Research (ONR) under award No. FA9550-23-1-0338 and the National Science Foundation (NSF) under QLCI Award No. OMA 2016244.
Any opinions, findings, and conclusions or recommendations expressed in this material are those of the authors and do not necessarily reflect the views of the NSF, ONR, or AFOSR. 

{\emph{Competing Interests:}}
P.T.R is a founder and shareholder of Resonance Micro Technologies Inc.  

\bibliography{phonbib}

\begin{thebibliography}{69}%
\makeatletter
\providecommand \@ifxundefined [1]{%
 \@ifx{#1\undefined}
}%
\providecommand \@ifnum [1]{%
 \ifnum #1\expandafter \@firstoftwo
 \else \expandafter \@secondoftwo
 \fi
}%
\providecommand \@ifx [1]{%
 \ifx #1\expandafter \@firstoftwo
 \else \expandafter \@secondoftwo
 \fi
}%
\providecommand \natexlab [1]{#1}%
\providecommand \enquote  [1]{``#1''}%
\providecommand \bibnamefont  [1]{#1}%
\providecommand \bibfnamefont [1]{#1}%
\providecommand \citenamefont [1]{#1}%
\providecommand \href@noop [0]{\@secondoftwo}%
\providecommand \href [0]{\begingroup \@sanitize@url \@href}%
\providecommand \@href[1]{\@@startlink{#1}\@@href}%
\providecommand \@@href[1]{\endgroup#1\@@endlink}%
\providecommand \@sanitize@url [0]{\catcode `\\12\catcode `\$12\catcode `\&12\catcode `\#12\catcode `\^12\catcode `\_12\catcode `\%12\relax}%
\providecommand \@@startlink[1]{}%
\providecommand \@@endlink[0]{}%
\providecommand \url  [0]{\begingroup\@sanitize@url \@url }%
\providecommand \@url [1]{\endgroup\@href {#1}{\urlprefix }}%
\providecommand \urlprefix  [0]{URL }%
\providecommand \Eprint [0]{\href }%
\providecommand \doibase [0]{https://doi.org/}%
\providecommand \selectlanguage [0]{\@gobble}%
\providecommand \bibinfo  [0]{\@secondoftwo}%
\providecommand \bibfield  [0]{\@secondoftwo}%
\providecommand \translation [1]{[#1]}%
\providecommand \BibitemOpen [0]{}%
\providecommand \bibitemStop [0]{}%
\providecommand \bibitemNoStop [0]{.\EOS\space}%
\providecommand \EOS [0]{\spacefactor3000\relax}%
\providecommand \BibitemShut  [1]{\csname bibitem#1\endcsname}%
\let\auto@bib@innerbib\@empty
\bibitem [{\citenamefont {Knill}\ \emph {et~al.}(2001)\citenamefont {Knill}, \citenamefont {Laflamme},\ and\ \citenamefont {Milburn}}]{knillSchemeEfficientQuantum2001}%
  \BibitemOpen
  \bibfield  {author} {\bibinfo {author} {\bibfnamefont {E.}~\bibnamefont {Knill}}, \bibinfo {author} {\bibfnamefont {R.}~\bibnamefont {Laflamme}},\ and\ \bibinfo {author} {\bibfnamefont {G.~J.}\ \bibnamefont {Milburn}},\ }\bibfield  {title} {\bibinfo {title} {A scheme for efficient quantum computation with linear optics},\ }\href {https://doi.org/10.1038/35051009} {\bibfield  {journal} {\bibinfo  {journal} {Nature}\ }\textbf {\bibinfo {volume} {409}},\ \bibinfo {pages} {46} (\bibinfo {year} {2001})}\BibitemShut {NoStop}%
\bibitem [{\citenamefont {{Gimeno-Segovia}}(2015)}]{gimeno-segoviaPracticalLinearOptical2015}%
  \BibitemOpen
  \bibfield  {author} {\bibinfo {author} {\bibfnamefont {M.}~\bibnamefont {{Gimeno-Segovia}}},\ }\emph {\bibinfo {title} {Towards {{Practical Linear Optical Quantum Computing}}}},\ \href@noop {} {Ph.D. thesis},\ \bibinfo  {school} {Imperial College London} (\bibinfo {year} {2015})\BibitemShut {NoStop}%
\bibitem [{\citenamefont {Li}\ \emph {et~al.}(2015)\citenamefont {Li}, \citenamefont {Humphreys}, \citenamefont {Mendoza},\ and\ \citenamefont {Benjamin}}]{liResourceCostsFaultTolerant2015}%
  \BibitemOpen
  \bibfield  {author} {\bibinfo {author} {\bibfnamefont {Y.}~\bibnamefont {Li}}, \bibinfo {author} {\bibfnamefont {P.~C.}\ \bibnamefont {Humphreys}}, \bibinfo {author} {\bibfnamefont {G.~J.}\ \bibnamefont {Mendoza}},\ and\ \bibinfo {author} {\bibfnamefont {S.~C.}\ \bibnamefont {Benjamin}},\ }\bibfield  {title} {\bibinfo {title} {Resource {{Costs}} for {{Fault-Tolerant Linear Optical Quantum Computing}}},\ }\href {https://doi.org/10.1103/PhysRevX.5.041007} {\bibfield  {journal} {\bibinfo  {journal} {Physical Review X}\ }\textbf {\bibinfo {volume} {5}},\ \bibinfo {pages} {041007} (\bibinfo {year} {2015})}\BibitemShut {NoStop}%
\bibitem [{\citenamefont {Rudolph}(2017)}]{rudolphWhyAmOptimistic2017}%
  \BibitemOpen
  \bibfield  {author} {\bibinfo {author} {\bibfnamefont {T.}~\bibnamefont {Rudolph}},\ }\bibfield  {title} {\bibinfo {title} {Why {{I}} am optimistic about the silicon-photonic route to quantum computing},\ }\href {https://doi.org/10.1063/1.4976737} {\bibfield  {journal} {\bibinfo  {journal} {APL Photonics}\ }\textbf {\bibinfo {volume} {2}},\ \bibinfo {pages} {030901} (\bibinfo {year} {2017})}\BibitemShut {NoStop}%
\bibitem [{\citenamefont {Slussarenko}\ and\ \citenamefont {Pryde}(2019)}]{slussarenkoPhotonicQuantumInformation2019}%
  \BibitemOpen
  \bibfield  {author} {\bibinfo {author} {\bibfnamefont {S.}~\bibnamefont {Slussarenko}}\ and\ \bibinfo {author} {\bibfnamefont {G.~J.}\ \bibnamefont {Pryde}},\ }\bibfield  {title} {\bibinfo {title} {Photonic quantum information processing: {{A}} concise review},\ }\href {https://doi.org/10.1063/1.5115814} {\bibfield  {journal} {\bibinfo  {journal} {Applied Physics Reviews}\ }\textbf {\bibinfo {volume} {6}},\ \bibinfo {pages} {041303} (\bibinfo {year} {2019})}\BibitemShut {NoStop}%
\bibitem [{\citenamefont {Walmsley}(2023)}]{walmsleyLightQuantumComputing2023}%
  \BibitemOpen
  \bibfield  {author} {\bibinfo {author} {\bibfnamefont {I.}~\bibnamefont {Walmsley}},\ }\bibfield  {title} {\bibinfo {title} {Light in quantum computing and simulation: Perspective},\ }\href {https://doi.org/10.1364/OPTICAQ.507527} {\bibfield  {journal} {\bibinfo  {journal} {Optica Quantum}\ }\textbf {\bibinfo {volume} {1}},\ \bibinfo {pages} {35} (\bibinfo {year} {2023})}\BibitemShut {NoStop}%
\bibitem [{\citenamefont {Oxborrow}\ and\ \citenamefont {Sinclair}(2005)}]{oxborrowSinglephotonSources2005}%
  \BibitemOpen
  \bibfield  {author} {\bibinfo {author} {\bibfnamefont {M.}~\bibnamefont {Oxborrow}}\ and\ \bibinfo {author} {\bibfnamefont {A.~G.}\ \bibnamefont {Sinclair}},\ }\bibfield  {title} {\bibinfo {title} {Single-photon sources},\ }\href {https://doi.org/10.1080/00107510512331337936} {\bibfield  {journal} {\bibinfo  {journal} {Contemporary Physics}\ }\textbf {\bibinfo {volume} {46}},\ \bibinfo {pages} {173} (\bibinfo {year} {2005})}\BibitemShut {NoStop}%
\bibitem [{\citenamefont {Aharonovich}\ \emph {et~al.}(2016)\citenamefont {Aharonovich}, \citenamefont {Englund},\ and\ \citenamefont {Toth}}]{aharonovichSolidstateSinglephotonEmitters2016}%
  \BibitemOpen
  \bibfield  {author} {\bibinfo {author} {\bibfnamefont {I.}~\bibnamefont {Aharonovich}}, \bibinfo {author} {\bibfnamefont {D.}~\bibnamefont {Englund}},\ and\ \bibinfo {author} {\bibfnamefont {M.}~\bibnamefont {Toth}},\ }\bibfield  {title} {\bibinfo {title} {Solid-state single-photon emitters},\ }\href {https://doi.org/10.1038/nphoton.2016.186} {\bibfield  {journal} {\bibinfo  {journal} {Nature Photonics}\ }\textbf {\bibinfo {volume} {10}},\ \bibinfo {pages} {631} (\bibinfo {year} {2016})}\BibitemShut {NoStop}%
\bibitem [{\citenamefont {Arakawa}\ and\ \citenamefont {Holmes}(2020)}]{arakawaProgressQuantumdotSingle2020}%
  \BibitemOpen
  \bibfield  {author} {\bibinfo {author} {\bibfnamefont {Y.}~\bibnamefont {Arakawa}}\ and\ \bibinfo {author} {\bibfnamefont {M.~J.}\ \bibnamefont {Holmes}},\ }\bibfield  {title} {\bibinfo {title} {Progress in quantum-dot single photon sources for quantum information technologies: {{A}} broad spectrum overview},\ }\href {https://doi.org/10.1063/5.0010193} {\bibfield  {journal} {\bibinfo  {journal} {Applied Physics Reviews}\ }\textbf {\bibinfo {volume} {7}},\ \bibinfo {pages} {021309} (\bibinfo {year} {2020})}\BibitemShut {NoStop}%
\bibitem [{\citenamefont {Cao}\ \emph {et~al.}(2019)\citenamefont {Cao}, \citenamefont {Zopf},\ and\ \citenamefont {Ding}}]{caoTelecomWavelengthSingle2019}%
  \BibitemOpen
  \bibfield  {author} {\bibinfo {author} {\bibfnamefont {X.}~\bibnamefont {Cao}}, \bibinfo {author} {\bibfnamefont {M.}~\bibnamefont {Zopf}},\ and\ \bibinfo {author} {\bibfnamefont {F.}~\bibnamefont {Ding}},\ }\bibfield  {title} {\bibinfo {title} {Telecom wavelength single photon sources},\ }\href {https://doi.org/10.1088/1674-4926/40/7/071901} {\bibfield  {journal} {\bibinfo  {journal} {Journal of Semiconductors}\ }\textbf {\bibinfo {volume} {40}},\ \bibinfo {pages} {071901} (\bibinfo {year} {2019})}\BibitemShut {NoStop}%
\bibitem [{\citenamefont {{Meyer-Scott}}\ \emph {et~al.}(2020)\citenamefont {{Meyer-Scott}}, \citenamefont {Silberhorn},\ and\ \citenamefont {Migdall}}]{meyer-scottSinglephotonSourcesApproaching2020}%
  \BibitemOpen
  \bibfield  {author} {\bibinfo {author} {\bibfnamefont {E.}~\bibnamefont {{Meyer-Scott}}}, \bibinfo {author} {\bibfnamefont {C.}~\bibnamefont {Silberhorn}},\ and\ \bibinfo {author} {\bibfnamefont {A.}~\bibnamefont {Migdall}},\ }\bibfield  {title} {\bibinfo {title} {Single-photon sources: {{Approaching}} the ideal through multiplexing},\ }\href {https://doi.org/10.1063/5.0003320} {\bibfield  {journal} {\bibinfo  {journal} {Review of Scientific Instruments}\ }\textbf {\bibinfo {volume} {91}},\ \bibinfo {pages} {041101} (\bibinfo {year} {2020})}\BibitemShut {NoStop}%
\bibitem [{\citenamefont {Moody}\ \emph {et~al.}(2020)\citenamefont {Moody}, \citenamefont {Chang}, \citenamefont {Steiner},\ and\ \citenamefont {Bowers}}]{moodyChipscaleNonlinearPhotonics2020}%
  \BibitemOpen
  \bibfield  {author} {\bibinfo {author} {\bibfnamefont {G.}~\bibnamefont {Moody}}, \bibinfo {author} {\bibfnamefont {L.}~\bibnamefont {Chang}}, \bibinfo {author} {\bibfnamefont {T.~J.}\ \bibnamefont {Steiner}},\ and\ \bibinfo {author} {\bibfnamefont {J.~E.}\ \bibnamefont {Bowers}},\ }\bibfield  {title} {\bibinfo {title} {Chip-scale nonlinear photonics for quantum light generation},\ }\href {https://doi.org/10.1116/5.0020684} {\bibfield  {journal} {\bibinfo  {journal} {AVS Quantum Science}\ }\textbf {\bibinfo {volume} {2}},\ \bibinfo {pages} {041702} (\bibinfo {year} {2020})}\BibitemShut {NoStop}%
\bibitem [{\citenamefont {Wang}\ \emph {et~al.}(2021)\citenamefont {Wang}, \citenamefont {J{\"o}ns}, \citenamefont {{Klaus D. J{\"o}ns}},\ and\ \citenamefont {Sun}}]{wangIntegratedPhotonpairSources2021}%
  \BibitemOpen
  \bibfield  {author} {\bibinfo {author} {\bibfnamefont {Y.}~\bibnamefont {Wang}}, \bibinfo {author} {\bibfnamefont {K.~D.}\ \bibnamefont {J{\"o}ns}}, \bibinfo {author} {\bibnamefont {{Klaus D. J{\"o}ns}}},\ and\ \bibinfo {author} {\bibfnamefont {Z.}~\bibnamefont {Sun}},\ }\bibfield  {title} {\bibinfo {title} {Integrated photon-pair sources with nonlinear optics},\ }\href {https://doi.org/10.1063/5.0030258} {\bibfield  {journal} {\bibinfo  {journal} {Applied physics reviews}\ }\textbf {\bibinfo {volume} {8}},\ \bibinfo {pages} {011314} (\bibinfo {year} {2021})}\BibitemShut {NoStop}%
\bibitem [{\citenamefont {Mendoza}\ \emph {et~al.}(2016)\citenamefont {Mendoza}, \citenamefont {Santagati}, \citenamefont {Munns}, \citenamefont {Hemsley}, \citenamefont {Piekarek}, \citenamefont {{Mart{\'i}n-L{\'o}pez}}, \citenamefont {Marshall}, \citenamefont {Bonneau}, \citenamefont {Thompson},\ and\ \citenamefont {O'Brien}}]{mendozaActiveTemporalSpatial2016}%
  \BibitemOpen
  \bibfield  {author} {\bibinfo {author} {\bibfnamefont {G.~J.}\ \bibnamefont {Mendoza}}, \bibinfo {author} {\bibfnamefont {R.}~\bibnamefont {Santagati}}, \bibinfo {author} {\bibfnamefont {J.}~\bibnamefont {Munns}}, \bibinfo {author} {\bibfnamefont {E.}~\bibnamefont {Hemsley}}, \bibinfo {author} {\bibfnamefont {M.}~\bibnamefont {Piekarek}}, \bibinfo {author} {\bibfnamefont {E.}~\bibnamefont {{Mart{\'i}n-L{\'o}pez}}}, \bibinfo {author} {\bibfnamefont {G.~D.}\ \bibnamefont {Marshall}}, \bibinfo {author} {\bibfnamefont {D.}~\bibnamefont {Bonneau}}, \bibinfo {author} {\bibfnamefont {M.~G.}\ \bibnamefont {Thompson}},\ and\ \bibinfo {author} {\bibfnamefont {J.~L.}\ \bibnamefont {O'Brien}},\ }\bibfield  {title} {\bibinfo {title} {Active temporal and spatial multiplexing of photons},\ }\href {https://doi.org/10.1364/OPTICA.3.000127} {\bibfield  {journal} {\bibinfo  {journal} {Optica}\ }\textbf {\bibinfo {volume} {3}},\ \bibinfo {pages} {127} (\bibinfo {year} {2016})}\BibitemShut {NoStop}%
\bibitem [{\citenamefont {Raussendorf}\ and\ \citenamefont {Briegel}(2001)}]{raussendorfOneWayQuantumComputer2001}%
  \BibitemOpen
  \bibfield  {author} {\bibinfo {author} {\bibfnamefont {R.}~\bibnamefont {Raussendorf}}\ and\ \bibinfo {author} {\bibfnamefont {H.~J.}\ \bibnamefont {Briegel}},\ }\bibfield  {title} {\bibinfo {title} {A {{One-Way Quantum Computer}}},\ }\href {https://doi.org/10.1103/PhysRevLett.86.5188} {\bibfield  {journal} {\bibinfo  {journal} {Physical Review Letters}\ }\textbf {\bibinfo {volume} {86}},\ \bibinfo {pages} {5188} (\bibinfo {year} {2001})}\BibitemShut {NoStop}%
\bibitem [{\citenamefont {Bartolucci}\ \emph {et~al.}(2023)\citenamefont {Bartolucci}, \citenamefont {Birchall}, \citenamefont {Bomb{\'i}n}, \citenamefont {Cable}, \citenamefont {Dawson}, \citenamefont {{Gimeno-Segovia}}, \citenamefont {Johnston}, \citenamefont {Kieling}, \citenamefont {Nickerson}, \citenamefont {Pant}, \citenamefont {Pastawski}, \citenamefont {Rudolph},\ and\ \citenamefont {Sparrow}}]{bartolucciFusion2023}%
  \BibitemOpen
  \bibfield  {author} {\bibinfo {author} {\bibfnamefont {S.}~\bibnamefont {Bartolucci}}, \bibinfo {author} {\bibfnamefont {P.}~\bibnamefont {Birchall}}, \bibinfo {author} {\bibfnamefont {H.}~\bibnamefont {Bomb{\'i}n}}, \bibinfo {author} {\bibfnamefont {H.}~\bibnamefont {Cable}}, \bibinfo {author} {\bibfnamefont {C.}~\bibnamefont {Dawson}}, \bibinfo {author} {\bibfnamefont {M.}~\bibnamefont {{Gimeno-Segovia}}}, \bibinfo {author} {\bibfnamefont {E.}~\bibnamefont {Johnston}}, \bibinfo {author} {\bibfnamefont {K.}~\bibnamefont {Kieling}}, \bibinfo {author} {\bibfnamefont {N.}~\bibnamefont {Nickerson}}, \bibinfo {author} {\bibfnamefont {M.}~\bibnamefont {Pant}}, \bibinfo {author} {\bibfnamefont {F.}~\bibnamefont {Pastawski}}, \bibinfo {author} {\bibfnamefont {T.}~\bibnamefont {Rudolph}},\ and\ \bibinfo {author} {\bibfnamefont {C.}~\bibnamefont {Sparrow}},\ }\bibfield  {title} {\bibinfo {title} {Fusion-based quantum computation},\ }\href {https://doi.org/10.1038/s41467-023-36493-1} {\bibfield  {journal} {\bibinfo  {journal} {Nature Communications}\ }\textbf {\bibinfo {volume} {14}},\ \bibinfo {pages} {912} (\bibinfo {year} {2023})}\BibitemShut {NoStop}%
\bibitem [{\citenamefont {Sch{\"o}n}\ \emph {et~al.}(2005)\citenamefont {Sch{\"o}n}, \citenamefont {Solano}, \citenamefont {Verstraete}, \citenamefont {Cirac},\ and\ \citenamefont {Wolf}}]{schonSequentialGenerationEntangled2005}%
  \BibitemOpen
  \bibfield  {author} {\bibinfo {author} {\bibfnamefont {C.}~\bibnamefont {Sch{\"o}n}}, \bibinfo {author} {\bibfnamefont {E.}~\bibnamefont {Solano}}, \bibinfo {author} {\bibfnamefont {F.}~\bibnamefont {Verstraete}}, \bibinfo {author} {\bibfnamefont {J.~I.}\ \bibnamefont {Cirac}},\ and\ \bibinfo {author} {\bibfnamefont {M.~M.}\ \bibnamefont {Wolf}},\ }\bibfield  {title} {\bibinfo {title} {Sequential generation of entangled multiqubit states},\ }\href {https://doi.org/10.1103/PhysRevLett.95.110503} {\bibfield  {journal} {\bibinfo  {journal} {Physical Review Letters}\ }\textbf {\bibinfo {volume} {95}},\ \bibinfo {pages} {110503} (\bibinfo {year} {2005})}\BibitemShut {NoStop}%
\bibitem [{\citenamefont {Lindner}\ and\ \citenamefont {Rudolph}(2009)}]{lindnerProposalPulsedOnDemand2009}%
  \BibitemOpen
  \bibfield  {author} {\bibinfo {author} {\bibfnamefont {N.~H.}\ \bibnamefont {Lindner}}\ and\ \bibinfo {author} {\bibfnamefont {T.}~\bibnamefont {Rudolph}},\ }\bibfield  {title} {\bibinfo {title} {Proposal for {{Pulsed On-Demand Sources}} of {{Photonic Cluster State Strings}}},\ }\href {https://doi.org/10.1103/PhysRevLett.103.113602} {\bibfield  {journal} {\bibinfo  {journal} {Physical Review Letters}\ }\textbf {\bibinfo {volume} {103}},\ \bibinfo {pages} {113602} (\bibinfo {year} {2009})}\BibitemShut {NoStop}%
\bibitem [{\citenamefont {Economou}\ \emph {et~al.}(2010)\citenamefont {Economou}, \citenamefont {Lindner},\ and\ \citenamefont {Rudolph}}]{economouOpticallyGenerated2dimensional2010}%
  \BibitemOpen
  \bibfield  {author} {\bibinfo {author} {\bibfnamefont {S.~E.}\ \bibnamefont {Economou}}, \bibinfo {author} {\bibfnamefont {N.}~\bibnamefont {Lindner}},\ and\ \bibinfo {author} {\bibfnamefont {T.}~\bibnamefont {Rudolph}},\ }\bibfield  {title} {\bibinfo {title} {Optically generated 2-dimensional photonic cluster state from coupled quantum dots},\ }\href {https://doi.org/10.1103/PhysRevLett.105.093601} {\bibfield  {journal} {\bibinfo  {journal} {Physical Review Letters}\ }\textbf {\bibinfo {volume} {105}},\ \bibinfo {pages} {093601} (\bibinfo {year} {2010})}\BibitemShut {NoStop}%
\bibitem [{\citenamefont {Thomas}\ \emph {et~al.}(2022)\citenamefont {Thomas}, \citenamefont {Ruscio}, \citenamefont {Morin},\ and\ \citenamefont {Rempe}}]{thomasEfficientGenerationEntangled2022}%
  \BibitemOpen
  \bibfield  {author} {\bibinfo {author} {\bibfnamefont {P.}~\bibnamefont {Thomas}}, \bibinfo {author} {\bibfnamefont {L.}~\bibnamefont {Ruscio}}, \bibinfo {author} {\bibfnamefont {O.}~\bibnamefont {Morin}},\ and\ \bibinfo {author} {\bibfnamefont {G.}~\bibnamefont {Rempe}},\ }\bibfield  {title} {\bibinfo {title} {Efficient generation of entangled multiphoton graph states from a single atom},\ }\href {https://doi.org/10.1038/s41586-022-04987-5} {\bibfield  {journal} {\bibinfo  {journal} {Nature}\ }\textbf {\bibinfo {volume} {608}},\ \bibinfo {pages} {677} (\bibinfo {year} {2022})}\BibitemShut {NoStop}%
\bibitem [{\citenamefont {Cogan}\ \emph {et~al.}(2023)\citenamefont {Cogan}, \citenamefont {Su}, \citenamefont {Kenneth},\ and\ \citenamefont {Gershoni}}]{coganDeterministicGenerationIndistinguishable2023}%
  \BibitemOpen
  \bibfield  {author} {\bibinfo {author} {\bibfnamefont {D.}~\bibnamefont {Cogan}}, \bibinfo {author} {\bibfnamefont {Z.-E.}\ \bibnamefont {Su}}, \bibinfo {author} {\bibfnamefont {O.}~\bibnamefont {Kenneth}},\ and\ \bibinfo {author} {\bibfnamefont {D.}~\bibnamefont {Gershoni}},\ }\bibfield  {title} {\bibinfo {title} {Deterministic generation of indistinguishable photons in a cluster state},\ }\href {https://doi.org/10.1038/s41566-022-01152-2} {\bibfield  {journal} {\bibinfo  {journal} {Nature Photonics}\ }\textbf {\bibinfo {volume} {17}},\ \bibinfo {pages} {324} (\bibinfo {year} {2023})}\BibitemShut {NoStop}%
\bibitem [{\citenamefont {Varnava}\ \emph {et~al.}(2008)\citenamefont {Varnava}, \citenamefont {Browne},\ and\ \citenamefont {Rudolph}}]{varnavaHowGoodMust2008}%
  \BibitemOpen
  \bibfield  {author} {\bibinfo {author} {\bibfnamefont {M.}~\bibnamefont {Varnava}}, \bibinfo {author} {\bibfnamefont {D.~E.}\ \bibnamefont {Browne}},\ and\ \bibinfo {author} {\bibfnamefont {T.}~\bibnamefont {Rudolph}},\ }\bibfield  {title} {\bibinfo {title} {How {{Good Must Single Photon Sources}} and {{Detectors Be}} for {{Efficient Linear Optical Quantum Computation}}?},\ }\href {https://doi.org/10.1103/PhysRevLett.100.060502} {\bibfield  {journal} {\bibinfo  {journal} {Physical Review Letters}\ }\textbf {\bibinfo {volume} {100}},\ \bibinfo {pages} {060502} (\bibinfo {year} {2008})}\BibitemShut {NoStop}%
\bibitem [{\citenamefont {Chen}\ \emph {et~al.}(2024{\natexlab{a}})\citenamefont {Chen}, \citenamefont {Peng}, \citenamefont {Guo}, \citenamefont {Gu}, \citenamefont {Ding}, \citenamefont {Liu}, \citenamefont {Zhao}, \citenamefont {You}, \citenamefont {Qin}, \citenamefont {Wang}, \citenamefont {He}, \citenamefont {Renema}, \citenamefont {Huo}, \citenamefont {Wang}, \citenamefont {Lu},\ and\ \citenamefont {Pan}}]{chenHeraldedThreephotonEntanglement2024}%
  \BibitemOpen
  \bibfield  {author} {\bibinfo {author} {\bibfnamefont {S.}~\bibnamefont {Chen}}, \bibinfo {author} {\bibfnamefont {L.-C.}\ \bibnamefont {Peng}}, \bibinfo {author} {\bibfnamefont {Y.-P.}\ \bibnamefont {Guo}}, \bibinfo {author} {\bibfnamefont {X.-M.}\ \bibnamefont {Gu}}, \bibinfo {author} {\bibfnamefont {X.}~\bibnamefont {Ding}}, \bibinfo {author} {\bibfnamefont {R.-Z.}\ \bibnamefont {Liu}}, \bibinfo {author} {\bibfnamefont {J.-Y.}\ \bibnamefont {Zhao}}, \bibinfo {author} {\bibfnamefont {X.}~\bibnamefont {You}}, \bibinfo {author} {\bibfnamefont {J.}~\bibnamefont {Qin}}, \bibinfo {author} {\bibfnamefont {Y.-F.}\ \bibnamefont {Wang}}, \bibinfo {author} {\bibfnamefont {Y.-M.}\ \bibnamefont {He}}, \bibinfo {author} {\bibfnamefont {J.~J.}\ \bibnamefont {Renema}}, \bibinfo {author} {\bibfnamefont {Y.-H.}\ \bibnamefont {Huo}}, \bibinfo {author} {\bibfnamefont {H.}~\bibnamefont {Wang}}, \bibinfo {author} {\bibfnamefont {C.-Y.}\ \bibnamefont {Lu}},\ and\ \bibinfo {author} {\bibfnamefont {J.-W.}\ \bibnamefont {Pan}},\ }\bibfield  {title} {\bibinfo {title} {Heralded {{Three-Photon Entanglement}} from a {{Single-Photon Source}} on a {{Photonic Chip}}},\ }\href {https://doi.org/10.1103/PhysRevLett.132.130603} {\bibfield  {journal} {\bibinfo  {journal} {Physical Review Letters}\ }\textbf {\bibinfo {volume} {132}},\ \bibinfo {pages} {130603} (\bibinfo {year} {2024}{\natexlab{a}})}\BibitemShut {NoStop}%
\bibitem [{\citenamefont {{Safavi-Naeini}}\ \emph {et~al.}(2011)\citenamefont {{Safavi-Naeini}}, \citenamefont {Alegre}, \citenamefont {Chan}, \citenamefont {Eichenfield}, \citenamefont {Winger}, \citenamefont {Lin}, \citenamefont {Hill}, \citenamefont {Chang},\ and\ \citenamefont {Painter}}]{safavi-naeiniElectromagneticallyInducedTransparency2011}%
  \BibitemOpen
  \bibfield  {author} {\bibinfo {author} {\bibfnamefont {A.~H.}\ \bibnamefont {{Safavi-Naeini}}}, \bibinfo {author} {\bibfnamefont {T.~P.~M.}\ \bibnamefont {Alegre}}, \bibinfo {author} {\bibfnamefont {J.}~\bibnamefont {Chan}}, \bibinfo {author} {\bibfnamefont {M.}~\bibnamefont {Eichenfield}}, \bibinfo {author} {\bibfnamefont {M.}~\bibnamefont {Winger}}, \bibinfo {author} {\bibfnamefont {Q.}~\bibnamefont {Lin}}, \bibinfo {author} {\bibfnamefont {J.~T.}\ \bibnamefont {Hill}}, \bibinfo {author} {\bibfnamefont {D.~E.}\ \bibnamefont {Chang}},\ and\ \bibinfo {author} {\bibfnamefont {O.}~\bibnamefont {Painter}},\ }\bibfield  {title} {\bibinfo {title} {Electromagnetically induced transparency and slow light with optomechanics},\ }\href {https://doi.org/10.1038/nature09933} {\bibfield  {journal} {\bibinfo  {journal} {Nature}\ }\textbf {\bibinfo {volume} {472}},\ \bibinfo {pages} {69} (\bibinfo {year} {2011})}\BibitemShut {NoStop}%
\bibitem [{\citenamefont {Shin}\ \emph {et~al.}(2013)\citenamefont {Shin}, \citenamefont {Qiu}, \citenamefont {Jarecki}, \citenamefont {Cox}, \citenamefont {Olsson}, \citenamefont {Starbuck}, \citenamefont {Wang},\ and\ \citenamefont {Rakich}}]{shinTailorableStimulatedBrillouin2013}%
  \BibitemOpen
  \bibfield  {author} {\bibinfo {author} {\bibfnamefont {H.}~\bibnamefont {Shin}}, \bibinfo {author} {\bibfnamefont {W.}~\bibnamefont {Qiu}}, \bibinfo {author} {\bibfnamefont {R.}~\bibnamefont {Jarecki}}, \bibinfo {author} {\bibfnamefont {J.~A.}\ \bibnamefont {Cox}}, \bibinfo {author} {\bibfnamefont {R.~H.}\ \bibnamefont {Olsson}}, \bibinfo {author} {\bibfnamefont {A.}~\bibnamefont {Starbuck}}, \bibinfo {author} {\bibfnamefont {Z.}~\bibnamefont {Wang}},\ and\ \bibinfo {author} {\bibfnamefont {P.~T.}\ \bibnamefont {Rakich}},\ }\bibfield  {title} {\bibinfo {title} {Tailorable stimulated {{Brillouin}} scattering in nanoscale silicon waveguides},\ }\href {https://doi.org/10.1038/ncomms2943} {\bibfield  {journal} {\bibinfo  {journal} {Nature Communications}\ }\textbf {\bibinfo {volume} {4}},\ \bibinfo {pages} {1} (\bibinfo {year} {2013})}\BibitemShut {NoStop}%
\bibitem [{\citenamefont {Aspelmeyer}\ \emph {et~al.}(2014)\citenamefont {Aspelmeyer}, \citenamefont {Kippenberg},\ and\ \citenamefont {Marquardt}}]{aspelmeyerCavityOptomechanics2014}%
  \BibitemOpen
  \bibfield  {author} {\bibinfo {author} {\bibfnamefont {M.}~\bibnamefont {Aspelmeyer}}, \bibinfo {author} {\bibfnamefont {T.~J.}\ \bibnamefont {Kippenberg}},\ and\ \bibinfo {author} {\bibfnamefont {F.}~\bibnamefont {Marquardt}},\ }\bibfield  {title} {\bibinfo {title} {Cavity optomechanics},\ }\href {https://doi.org/10.1103/RevModPhys.86.1391} {\bibfield  {journal} {\bibinfo  {journal} {Reviews of Modern Physics}\ }\textbf {\bibinfo {volume} {86}},\ \bibinfo {pages} {1391} (\bibinfo {year} {2014})}\BibitemShut {NoStop}%
\bibitem [{\citenamefont {{Safavi-Naeini}}\ \emph {et~al.}(2019)\citenamefont {{Safavi-Naeini}}, \citenamefont {Baets},\ and\ \citenamefont {Laer}}]{safavi-naeiniControllingPhononsPhotons2019}%
  \BibitemOpen
  \bibfield  {author} {\bibinfo {author} {\bibfnamefont {A.~H.}\ \bibnamefont {{Safavi-Naeini}}}, \bibinfo {author} {\bibfnamefont {R.}~\bibnamefont {Baets}},\ and\ \bibinfo {author} {\bibfnamefont {R.~V.}\ \bibnamefont {Laer}},\ }\bibfield  {title} {\bibinfo {title} {Controlling phonons and photons at the wavelength scale: Integrated photonics meets integrated phononics},\ }\href {https://doi.org/10.1364/OPTICA.6.000213} {\bibfield  {journal} {\bibinfo  {journal} {Optica}\ }\textbf {\bibinfo {volume} {6}},\ \bibinfo {pages} {213} (\bibinfo {year} {2019})}\BibitemShut {NoStop}%
\bibitem [{\citenamefont {Barzanjeh}\ \emph {et~al.}(2022)\citenamefont {Barzanjeh}, \citenamefont {Xuereb}, \citenamefont {Gr{\"o}blacher}, \citenamefont {Paternostro}, \citenamefont {Regal},\ and\ \citenamefont {Weig}}]{barzanjehOptomechanicsQuantumTechnologies2022}%
  \BibitemOpen
  \bibfield  {author} {\bibinfo {author} {\bibfnamefont {S.}~\bibnamefont {Barzanjeh}}, \bibinfo {author} {\bibfnamefont {A.}~\bibnamefont {Xuereb}}, \bibinfo {author} {\bibfnamefont {S.}~\bibnamefont {Gr{\"o}blacher}}, \bibinfo {author} {\bibfnamefont {M.}~\bibnamefont {Paternostro}}, \bibinfo {author} {\bibfnamefont {C.~A.}\ \bibnamefont {Regal}},\ and\ \bibinfo {author} {\bibfnamefont {E.~M.}\ \bibnamefont {Weig}},\ }\bibfield  {title} {\bibinfo {title} {Optomechanics for quantum technologies},\ }\href {https://doi.org/10.1038/s41567-021-01402-0} {\bibfield  {journal} {\bibinfo  {journal} {Nature Physics}\ }\textbf {\bibinfo {volume} {18}},\ \bibinfo {pages} {15} (\bibinfo {year} {2022})}\BibitemShut {NoStop}%
\bibitem [{\citenamefont {Chan}\ \emph {et~al.}(2011)\citenamefont {Chan}, \citenamefont {Alegre}, \citenamefont {{Safavi-Naeini}}, \citenamefont {Hill}, \citenamefont {Krause}, \citenamefont {Gr{\"o}blacher}, \citenamefont {Aspelmeyer},\ and\ \citenamefont {Painter}}]{chanLaserCoolingNanomechanical2011}%
  \BibitemOpen
  \bibfield  {author} {\bibinfo {author} {\bibfnamefont {J.}~\bibnamefont {Chan}}, \bibinfo {author} {\bibfnamefont {T.~P.~M.}\ \bibnamefont {Alegre}}, \bibinfo {author} {\bibfnamefont {A.~H.}\ \bibnamefont {{Safavi-Naeini}}}, \bibinfo {author} {\bibfnamefont {J.~T.}\ \bibnamefont {Hill}}, \bibinfo {author} {\bibfnamefont {A.}~\bibnamefont {Krause}}, \bibinfo {author} {\bibfnamefont {S.}~\bibnamefont {Gr{\"o}blacher}}, \bibinfo {author} {\bibfnamefont {M.}~\bibnamefont {Aspelmeyer}},\ and\ \bibinfo {author} {\bibfnamefont {O.}~\bibnamefont {Painter}},\ }\bibfield  {title} {\bibinfo {title} {Laser cooling of a nanomechanical oscillator into its quantum ground state},\ }\href {https://doi.org/10.1038/nature10461} {\bibfield  {journal} {\bibinfo  {journal} {Nature}\ }\textbf {\bibinfo {volume} {478}},\ \bibinfo {pages} {89} (\bibinfo {year} {2011})}\BibitemShut {NoStop}%
\bibitem [{\citenamefont {Galland}\ \emph {et~al.}(2014)\citenamefont {Galland}, \citenamefont {Sangouard}, \citenamefont {Piro}, \citenamefont {Gisin},\ and\ \citenamefont {Kippenberg}}]{gallandHeraldedSinglePhononPreparation2014}%
  \BibitemOpen
  \bibfield  {author} {\bibinfo {author} {\bibfnamefont {C.}~\bibnamefont {Galland}}, \bibinfo {author} {\bibfnamefont {N.}~\bibnamefont {Sangouard}}, \bibinfo {author} {\bibfnamefont {N.}~\bibnamefont {Piro}}, \bibinfo {author} {\bibfnamefont {N.}~\bibnamefont {Gisin}},\ and\ \bibinfo {author} {\bibfnamefont {T.~J.}\ \bibnamefont {Kippenberg}},\ }\bibfield  {title} {\bibinfo {title} {Heralded {{Single-Phonon Preparation}}, {{Storage}}, and {{Readout}} in {{Cavity Optomechanics}}},\ }\href {https://doi.org/10.1103/PhysRevLett.112.143602} {\bibfield  {journal} {\bibinfo  {journal} {Physical Review Letters}\ }\textbf {\bibinfo {volume} {112}},\ \bibinfo {pages} {143602} (\bibinfo {year} {2014})}\BibitemShut {NoStop}%
\bibitem [{\citenamefont {Wallucks}\ \emph {et~al.}(2020)\citenamefont {Wallucks}, \citenamefont {Marinkovi{\'c}}, \citenamefont {Hensen}, \citenamefont {Stockill},\ and\ \citenamefont {Gr{\"o}blacher}}]{wallucksQuantumMemoryTelecom2020}%
  \BibitemOpen
  \bibfield  {author} {\bibinfo {author} {\bibfnamefont {A.}~\bibnamefont {Wallucks}}, \bibinfo {author} {\bibfnamefont {I.}~\bibnamefont {Marinkovi{\'c}}}, \bibinfo {author} {\bibfnamefont {B.}~\bibnamefont {Hensen}}, \bibinfo {author} {\bibfnamefont {R.}~\bibnamefont {Stockill}},\ and\ \bibinfo {author} {\bibfnamefont {S.}~\bibnamefont {Gr{\"o}blacher}},\ }\bibfield  {title} {\bibinfo {title} {A quantum memory at telecom wavelengths},\ }\href {https://doi.org/10.1038/s41567-020-0891-z} {\bibfield  {journal} {\bibinfo  {journal} {Nature Physics}\ }\textbf {\bibinfo {volume} {16}},\ \bibinfo {pages} {772} (\bibinfo {year} {2020})}\BibitemShut {NoStop}%
\bibitem [{\citenamefont {Shepherd}\ and\ \citenamefont {Behunin}(2024)}]{shepherdMultiphononFockState2024}%
  \BibitemOpen
  \bibfield  {author} {\bibinfo {author} {\bibfnamefont {A.~J.}\ \bibnamefont {Shepherd}}\ and\ \bibinfo {author} {\bibfnamefont {R.~O.}\ \bibnamefont {Behunin}},\ }\href {https://doi.org/10.48550/arXiv.2407.19120} {\bibinfo {title} {Multi-phonon {{Fock}} state heralding with single-photon detection}} (\bibinfo {year} {2024}),\ \Eprint {https://arxiv.org/abs/2407.19120} {arXiv:2407.19120} \BibitemShut {NoStop}%
\bibitem [{\citenamefont {Otterstrom}\ \emph {et~al.}(2018)\citenamefont {Otterstrom}, \citenamefont {Behunin}, \citenamefont {Kittlaus}, \citenamefont {Wang},\ and\ \citenamefont {Rakich}}]{otterstromSiliconBrillouinLaser2018}%
  \BibitemOpen
  \bibfield  {author} {\bibinfo {author} {\bibfnamefont {N.~T.}\ \bibnamefont {Otterstrom}}, \bibinfo {author} {\bibfnamefont {R.~O.}\ \bibnamefont {Behunin}}, \bibinfo {author} {\bibfnamefont {E.~A.}\ \bibnamefont {Kittlaus}}, \bibinfo {author} {\bibfnamefont {Z.}~\bibnamefont {Wang}},\ and\ \bibinfo {author} {\bibfnamefont {P.~T.}\ \bibnamefont {Rakich}},\ }\bibfield  {title} {\bibinfo {title} {A silicon {{Brillouin}} laser},\ }\href {https://doi.org/10.1126/science.aar6113} {\bibfield  {journal} {\bibinfo  {journal} {Science}\ }\textbf {\bibinfo {volume} {360}},\ \bibinfo {pages} {1113} (\bibinfo {year} {2018})}\BibitemShut {NoStop}%
\bibitem [{\citenamefont {Diamandi}\ \emph {et~al.}(2024)\citenamefont {Diamandi}, \citenamefont {Luo}, \citenamefont {Mason}, \citenamefont {Kanmaz}, \citenamefont {Ghosh}, \citenamefont {Pavlovich}, \citenamefont {Yoon}, \citenamefont {Behunin}, \citenamefont {Puri}, \citenamefont {Harris},\ and\ \citenamefont {Rakich}}]{diamandiQuantumOptomechanicalControl2024}%
  \BibitemOpen
  \bibfield  {author} {\bibinfo {author} {\bibfnamefont {H.~H.}\ \bibnamefont {Diamandi}}, \bibinfo {author} {\bibfnamefont {Y.}~\bibnamefont {Luo}}, \bibinfo {author} {\bibfnamefont {D.}~\bibnamefont {Mason}}, \bibinfo {author} {\bibfnamefont {T.~B.}\ \bibnamefont {Kanmaz}}, \bibinfo {author} {\bibfnamefont {S.}~\bibnamefont {Ghosh}}, \bibinfo {author} {\bibfnamefont {M.}~\bibnamefont {Pavlovich}}, \bibinfo {author} {\bibfnamefont {T.}~\bibnamefont {Yoon}}, \bibinfo {author} {\bibfnamefont {R.}~\bibnamefont {Behunin}}, \bibinfo {author} {\bibfnamefont {S.}~\bibnamefont {Puri}}, \bibinfo {author} {\bibfnamefont {J.~G.~E.}\ \bibnamefont {Harris}},\ and\ \bibinfo {author} {\bibfnamefont {P.~T.}\ \bibnamefont {Rakich}},\ }\href {https://doi.org/10.48550/arXiv.2410.18037} {\bibinfo {title} {Quantum optomechanical control of long-lived bulk acoustic phonons}} (\bibinfo {year} {2024}),\ \Eprint {https://arxiv.org/abs/2410.18037} {arXiv:2410.18037} \BibitemShut {NoStop}%
\bibitem [{\citenamefont {Luo}\ \emph {et~al.}(2025)\citenamefont {Luo}, \citenamefont {Diamandi}, \citenamefont {Li}, \citenamefont {Bi}, \citenamefont {Mason}, \citenamefont {Yoon}, \citenamefont {Guo}, \citenamefont {Tang}, \citenamefont {Behunin}, \citenamefont {Walker}, \citenamefont {Ahn},\ and\ \citenamefont {Rakich}}]{luoLifetimelimitedGigahertzfrequencyMechanical2025}%
  \BibitemOpen
  \bibfield  {author} {\bibinfo {author} {\bibfnamefont {Y.}~\bibnamefont {Luo}}, \bibinfo {author} {\bibfnamefont {H.~H.}\ \bibnamefont {Diamandi}}, \bibinfo {author} {\bibfnamefont {H.}~\bibnamefont {Li}}, \bibinfo {author} {\bibfnamefont {R.}~\bibnamefont {Bi}}, \bibinfo {author} {\bibfnamefont {D.}~\bibnamefont {Mason}}, \bibinfo {author} {\bibfnamefont {T.}~\bibnamefont {Yoon}}, \bibinfo {author} {\bibfnamefont {X.}~\bibnamefont {Guo}}, \bibinfo {author} {\bibfnamefont {H.}~\bibnamefont {Tang}}, \bibinfo {author} {\bibfnamefont {R.~O.}\ \bibnamefont {Behunin}}, \bibinfo {author} {\bibfnamefont {F.~J.}\ \bibnamefont {Walker}}, \bibinfo {author} {\bibfnamefont {C.}~\bibnamefont {Ahn}},\ and\ \bibinfo {author} {\bibfnamefont {P.~T.}\ \bibnamefont {Rakich}},\ }\href {https://doi.org/10.48550/arXiv.2504.07523} {\bibinfo {title} {Lifetime-limited {{Gigahertz-frequency Mechanical Oscillators}} with {{Millisecond Coherence Times}}}} (\bibinfo {year} {2025}),\ \Eprint {https://arxiv.org/abs/2504.07523} {arXiv:2504.07523} \BibitemShut {NoStop}%
\bibitem [{\citenamefont {Ren}\ \emph {et~al.}(2020)\citenamefont {Ren}, \citenamefont {Matheny}, \citenamefont {MacCabe}, \citenamefont {Luo}, \citenamefont {Pfeifer}, \citenamefont {Mirhosseini},\ and\ \citenamefont {Painter}}]{renTwodimensionalOptomechanicalCrystal2020}%
  \BibitemOpen
  \bibfield  {author} {\bibinfo {author} {\bibfnamefont {H.}~\bibnamefont {Ren}}, \bibinfo {author} {\bibfnamefont {M.~H.}\ \bibnamefont {Matheny}}, \bibinfo {author} {\bibfnamefont {G.~S.}\ \bibnamefont {MacCabe}}, \bibinfo {author} {\bibfnamefont {J.}~\bibnamefont {Luo}}, \bibinfo {author} {\bibfnamefont {H.}~\bibnamefont {Pfeifer}}, \bibinfo {author} {\bibfnamefont {M.}~\bibnamefont {Mirhosseini}},\ and\ \bibinfo {author} {\bibfnamefont {O.}~\bibnamefont {Painter}},\ }\bibfield  {title} {\bibinfo {title} {Two-dimensional optomechanical crystal cavity with high quantum cooperativity},\ }\href {https://doi.org/10.1038/s41467-020-17182-9} {\bibfield  {journal} {\bibinfo  {journal} {Nature Communications}\ }\textbf {\bibinfo {volume} {11}},\ \bibinfo {pages} {3373} (\bibinfo {year} {2020})}\BibitemShut {NoStop}%
\bibitem [{\citenamefont {Liu}\ \emph {et~al.}(2021)\citenamefont {Liu}, \citenamefont {Huang}, \citenamefont {Wang}, \citenamefont {He}, \citenamefont {Raja}, \citenamefont {Liu}, \citenamefont {Engelsen},\ and\ \citenamefont {Kippenberg}}]{liuHighyieldWaferscaleFabrication2021}%
  \BibitemOpen
  \bibfield  {author} {\bibinfo {author} {\bibfnamefont {J.}~\bibnamefont {Liu}}, \bibinfo {author} {\bibfnamefont {G.}~\bibnamefont {Huang}}, \bibinfo {author} {\bibfnamefont {R.~N.}\ \bibnamefont {Wang}}, \bibinfo {author} {\bibfnamefont {J.}~\bibnamefont {He}}, \bibinfo {author} {\bibfnamefont {A.~S.}\ \bibnamefont {Raja}}, \bibinfo {author} {\bibfnamefont {T.}~\bibnamefont {Liu}}, \bibinfo {author} {\bibfnamefont {N.~J.}\ \bibnamefont {Engelsen}},\ and\ \bibinfo {author} {\bibfnamefont {T.~J.}\ \bibnamefont {Kippenberg}},\ }\bibfield  {title} {\bibinfo {title} {High-yield, wafer-scale fabrication of ultralow-loss, dispersion-engineered silicon nitride photonic circuits},\ }\href {https://doi.org/10.1038/s41467-021-21973-z} {\bibfield  {journal} {\bibinfo  {journal} {Nature Communications}\ }\textbf {\bibinfo {volume} {12}},\ \bibinfo {pages} {2236} (\bibinfo {year} {2021})}\BibitemShut {NoStop}%
\bibitem [{\citenamefont {Kharel}\ \emph {et~al.}(2019)\citenamefont {Kharel}, \citenamefont {Harris}, \citenamefont {Kittlaus}, \citenamefont {Renninger}, \citenamefont {Otterstrom}, \citenamefont {Harris},\ and\ \citenamefont {Rakich}}]{kharelHighfrequencyCavityOptomechanics2019}%
  \BibitemOpen
  \bibfield  {author} {\bibinfo {author} {\bibfnamefont {P.}~\bibnamefont {Kharel}}, \bibinfo {author} {\bibfnamefont {G.~I.}\ \bibnamefont {Harris}}, \bibinfo {author} {\bibfnamefont {E.~A.}\ \bibnamefont {Kittlaus}}, \bibinfo {author} {\bibfnamefont {W.~H.}\ \bibnamefont {Renninger}}, \bibinfo {author} {\bibfnamefont {N.~T.}\ \bibnamefont {Otterstrom}}, \bibinfo {author} {\bibfnamefont {J.~G.~E.}\ \bibnamefont {Harris}},\ and\ \bibinfo {author} {\bibfnamefont {P.~T.}\ \bibnamefont {Rakich}},\ }\bibfield  {title} {\bibinfo {title} {High-frequency cavity optomechanics using bulk acoustic phonons},\ }\href {https://doi.org/10.1126/sciadv.aav0582} {\bibfield  {journal} {\bibinfo  {journal} {Science Advances}\ }\textbf {\bibinfo {volume} {5}},\ \bibinfo {pages} {eaav0582} (\bibinfo {year} {2019})}\BibitemShut {NoStop}%
\bibitem [{\citenamefont {Black}\ \emph {et~al.}(2022)\citenamefont {Black}, \citenamefont {Brodnik}, \citenamefont {Liu}, \citenamefont {Yu}, \citenamefont {Carlson}, \citenamefont {Zang}, \citenamefont {Briles},\ and\ \citenamefont {Papp}}]{blackOpticalparametricOscillationPhotoniccrystal2022}%
  \BibitemOpen
  \bibfield  {author} {\bibinfo {author} {\bibfnamefont {J.~A.}\ \bibnamefont {Black}}, \bibinfo {author} {\bibfnamefont {G.}~\bibnamefont {Brodnik}}, \bibinfo {author} {\bibfnamefont {H.}~\bibnamefont {Liu}}, \bibinfo {author} {\bibfnamefont {S.-P.}\ \bibnamefont {Yu}}, \bibinfo {author} {\bibfnamefont {D.~R.}\ \bibnamefont {Carlson}}, \bibinfo {author} {\bibfnamefont {J.}~\bibnamefont {Zang}}, \bibinfo {author} {\bibfnamefont {T.~C.}\ \bibnamefont {Briles}},\ and\ \bibinfo {author} {\bibfnamefont {S.~B.}\ \bibnamefont {Papp}},\ }\bibfield  {title} {\bibinfo {title} {Optical-parametric oscillation in photonic-crystal ring resonators},\ }\href {https://doi.org/10.1364/OPTICA.469210} {\bibfield  {journal} {\bibinfo  {journal} {Optica}\ }\textbf {\bibinfo {volume} {9}},\ \bibinfo {pages} {1183} (\bibinfo {year} {2022})}\BibitemShut {NoStop}%
\bibitem [{\citenamefont {Liu}\ \emph {et~al.}(2024)\citenamefont {Liu}, \citenamefont {Wang}, \citenamefont {Chauhan}, \citenamefont {Harrington}, \citenamefont {Nelson},\ and\ \citenamefont {Blumenthal}}]{liuIntegratedPhotonicMolecule2024}%
  \BibitemOpen
  \bibfield  {author} {\bibinfo {author} {\bibfnamefont {K.}~\bibnamefont {Liu}}, \bibinfo {author} {\bibfnamefont {J.}~\bibnamefont {Wang}}, \bibinfo {author} {\bibfnamefont {N.}~\bibnamefont {Chauhan}}, \bibinfo {author} {\bibfnamefont {M.~W.}\ \bibnamefont {Harrington}}, \bibinfo {author} {\bibfnamefont {K.~D.}\ \bibnamefont {Nelson}},\ and\ \bibinfo {author} {\bibfnamefont {D.~J.}\ \bibnamefont {Blumenthal}},\ }\bibfield  {title} {\bibinfo {title} {Integrated photonic molecule {{Brillouin}} laser with a high-power sub-100-{{mHz}} fundamental linewidth},\ }\href {https://doi.org/10.1364/OL.503126} {\bibfield  {journal} {\bibinfo  {journal} {Optics Letters}\ }\textbf {\bibinfo {volume} {49}},\ \bibinfo {pages} {45} (\bibinfo {year} {2024})}\BibitemShut {NoStop}%
\bibitem [{\citenamefont {Kittlaus}\ \emph {et~al.}(2018)\citenamefont {Kittlaus}, \citenamefont {Otterstrom}, \citenamefont {Kharel}, \citenamefont {Gertler},\ and\ \citenamefont {Rakich}}]{kittlausNonreciprocalInterbandBrillouin2018}%
  \BibitemOpen
  \bibfield  {author} {\bibinfo {author} {\bibfnamefont {E.~A.}\ \bibnamefont {Kittlaus}}, \bibinfo {author} {\bibfnamefont {N.~T.}\ \bibnamefont {Otterstrom}}, \bibinfo {author} {\bibfnamefont {P.}~\bibnamefont {Kharel}}, \bibinfo {author} {\bibfnamefont {S.}~\bibnamefont {Gertler}},\ and\ \bibinfo {author} {\bibfnamefont {P.~T.}\ \bibnamefont {Rakich}},\ }\bibfield  {title} {\bibinfo {title} {Non-reciprocal interband {{Brillouin}} modulation},\ }\href {https://doi.org/10.1038/s41566-018-0254-9} {\bibfield  {journal} {\bibinfo  {journal} {Nature Photonics}\ }\textbf {\bibinfo {volume} {12}},\ \bibinfo {pages} {613} (\bibinfo {year} {2018})}\BibitemShut {NoStop}%
\bibitem [{\citenamefont {Chen}\ \emph {et~al.}(2023)\citenamefont {Chen}, \citenamefont {Li}, \citenamefont {Lee}, \citenamefont {Chakravarthi}, \citenamefont {Fu},\ and\ \citenamefont {Li}}]{chenOptomechanicalRingResonator2023}%
  \BibitemOpen
  \bibfield  {author} {\bibinfo {author} {\bibfnamefont {I.-T.}\ \bibnamefont {Chen}}, \bibinfo {author} {\bibfnamefont {B.}~\bibnamefont {Li}}, \bibinfo {author} {\bibfnamefont {S.}~\bibnamefont {Lee}}, \bibinfo {author} {\bibfnamefont {S.}~\bibnamefont {Chakravarthi}}, \bibinfo {author} {\bibfnamefont {K.-M.}\ \bibnamefont {Fu}},\ and\ \bibinfo {author} {\bibfnamefont {M.}~\bibnamefont {Li}},\ }\bibfield  {title} {\bibinfo {title} {Optomechanical ring resonator for efficient microwave-optical frequency conversion},\ }\href {https://doi.org/10.1038/s41467-023-43393-x} {\bibfield  {journal} {\bibinfo  {journal} {Nature Communications}\ }\textbf {\bibinfo {volume} {14}},\ \bibinfo {pages} {7594} (\bibinfo {year} {2023})}\BibitemShut {NoStop}%
\bibitem [{\citenamefont {Meenehan}\ \emph {et~al.}(2015)\citenamefont {Meenehan}, \citenamefont {Cohen}, \citenamefont {MacCabe}, \citenamefont {Marsili}, \citenamefont {Shaw},\ and\ \citenamefont {Painter}}]{meenehanPulsedExcitationDynamics2015}%
  \BibitemOpen
  \bibfield  {author} {\bibinfo {author} {\bibfnamefont {S.~M.}\ \bibnamefont {Meenehan}}, \bibinfo {author} {\bibfnamefont {J.~D.}\ \bibnamefont {Cohen}}, \bibinfo {author} {\bibfnamefont {G.~S.}\ \bibnamefont {MacCabe}}, \bibinfo {author} {\bibfnamefont {F.}~\bibnamefont {Marsili}}, \bibinfo {author} {\bibfnamefont {M.~D.}\ \bibnamefont {Shaw}},\ and\ \bibinfo {author} {\bibfnamefont {O.}~\bibnamefont {Painter}},\ }\bibfield  {title} {\bibinfo {title} {Pulsed {{Excitation Dynamics}} of an {{Optomechanical Crystal Resonator}} near {{Its Quantum Ground State}} of {{Motion}}},\ }\href {https://doi.org/10.1103/PhysRevX.5.041002} {\bibfield  {journal} {\bibinfo  {journal} {Physical Review X}\ }\textbf {\bibinfo {volume} {5}},\ \bibinfo {pages} {041002} (\bibinfo {year} {2015})}\BibitemShut {NoStop}%
\bibitem [{\citenamefont {MacCabe}\ \emph {et~al.}(2020)\citenamefont {MacCabe}, \citenamefont {Ren}, \citenamefont {Luo}, \citenamefont {Cohen}, \citenamefont {Zhou}, \citenamefont {Sipahigil}, \citenamefont {Mirhosseini},\ and\ \citenamefont {Painter}}]{maccabeNanoacousticResonatorUltralong2020}%
  \BibitemOpen
  \bibfield  {author} {\bibinfo {author} {\bibfnamefont {G.~S.}\ \bibnamefont {MacCabe}}, \bibinfo {author} {\bibfnamefont {H.}~\bibnamefont {Ren}}, \bibinfo {author} {\bibfnamefont {J.}~\bibnamefont {Luo}}, \bibinfo {author} {\bibfnamefont {J.~D.}\ \bibnamefont {Cohen}}, \bibinfo {author} {\bibfnamefont {H.}~\bibnamefont {Zhou}}, \bibinfo {author} {\bibfnamefont {A.}~\bibnamefont {Sipahigil}}, \bibinfo {author} {\bibfnamefont {M.}~\bibnamefont {Mirhosseini}},\ and\ \bibinfo {author} {\bibfnamefont {O.}~\bibnamefont {Painter}},\ }\bibfield  {title} {\bibinfo {title} {Nano-acoustic resonator with ultralong phonon lifetime},\ }\href {https://doi.org/10.1126/science.abc7312} {\bibfield  {journal} {\bibinfo  {journal} {Science}\ }\textbf {\bibinfo {volume} {370}},\ \bibinfo {pages} {840} (\bibinfo {year} {2020})}\BibitemShut {NoStop}%
\bibitem [{\citenamefont {Fiaschi}\ \emph {et~al.}(2021)\citenamefont {Fiaschi}, \citenamefont {Hensen}, \citenamefont {Wallucks}, \citenamefont {Benevides}, \citenamefont {Li}, \citenamefont {Alegre},\ and\ \citenamefont {Gr{\"o}blacher}}]{fiaschiOptomechanicalQuantumTeleportation2021}%
  \BibitemOpen
  \bibfield  {author} {\bibinfo {author} {\bibfnamefont {N.}~\bibnamefont {Fiaschi}}, \bibinfo {author} {\bibfnamefont {B.}~\bibnamefont {Hensen}}, \bibinfo {author} {\bibfnamefont {A.}~\bibnamefont {Wallucks}}, \bibinfo {author} {\bibfnamefont {R.}~\bibnamefont {Benevides}}, \bibinfo {author} {\bibfnamefont {J.}~\bibnamefont {Li}}, \bibinfo {author} {\bibfnamefont {T.~P.~M.}\ \bibnamefont {Alegre}},\ and\ \bibinfo {author} {\bibfnamefont {S.}~\bibnamefont {Gr{\"o}blacher}},\ }\bibfield  {title} {\bibinfo {title} {Optomechanical quantum teleportation},\ }\href {https://doi.org/10.1038/s41566-021-00866-z} {\bibfield  {journal} {\bibinfo  {journal} {Nature Photonics}\ }\textbf {\bibinfo {volume} {15}},\ \bibinfo {pages} {817} (\bibinfo {year} {2021})}\BibitemShut {NoStop}%
\bibitem [{\citenamefont {Cleland}\ \emph {et~al.}(2023)\citenamefont {Cleland}, \citenamefont {Wollack},\ and\ \citenamefont {{Safavi-Naeini}}}]{clelandStudyingPhononCoherence2023}%
  \BibitemOpen
  \bibfield  {author} {\bibinfo {author} {\bibfnamefont {A.~Y.}\ \bibnamefont {Cleland}}, \bibinfo {author} {\bibfnamefont {E.~A.}\ \bibnamefont {Wollack}},\ and\ \bibinfo {author} {\bibfnamefont {A.~H.}\ \bibnamefont {{Safavi-Naeini}}},\ }\href {https://doi.org/10.48550/arXiv.2302.00221} {\bibinfo {title} {Studying phonon coherence with a quantum sensor}} (\bibinfo {year} {2023}),\ \Eprint {https://arxiv.org/abs/2302.00221} {arXiv:2302.00221} \BibitemShut {NoStop}%
\bibitem [{\citenamefont {Mayor}\ \emph {et~al.}(2024)\citenamefont {Mayor}, \citenamefont {Malik}, \citenamefont {Primo}, \citenamefont {Gyger}, \citenamefont {Jiang}, \citenamefont {Alegre},\ and\ \citenamefont {{Safavi-Naeini}}}]{mayorTwodimensionalOptomechanicalCrystal2024}%
  \BibitemOpen
  \bibfield  {author} {\bibinfo {author} {\bibfnamefont {F.~M.}\ \bibnamefont {Mayor}}, \bibinfo {author} {\bibfnamefont {S.}~\bibnamefont {Malik}}, \bibinfo {author} {\bibfnamefont {A.~G.}\ \bibnamefont {Primo}}, \bibinfo {author} {\bibfnamefont {S.}~\bibnamefont {Gyger}}, \bibinfo {author} {\bibfnamefont {W.}~\bibnamefont {Jiang}}, \bibinfo {author} {\bibfnamefont {T.~P.~M.}\ \bibnamefont {Alegre}},\ and\ \bibinfo {author} {\bibfnamefont {A.~H.}\ \bibnamefont {{Safavi-Naeini}}},\ }\href {https://doi.org/10.48550/arXiv.2406.14484} {\bibinfo {title} {A two-dimensional optomechanical crystal for quantum transduction}} (\bibinfo {year} {2024}),\ \Eprint {https://arxiv.org/abs/2406.14484} {arXiv:2406.14484} \BibitemShut {NoStop}%
\bibitem [{\citenamefont {Doeleman}\ \emph {et~al.}(2023)\citenamefont {Doeleman}, \citenamefont {Schatteburg}, \citenamefont {Benevides}, \citenamefont {Vollenweider}, \citenamefont {Macri},\ and\ \citenamefont {Chu}}]{doelemanBrillouinOptomechanicsQuantum2023}%
  \BibitemOpen
  \bibfield  {author} {\bibinfo {author} {\bibfnamefont {H.~M.}\ \bibnamefont {Doeleman}}, \bibinfo {author} {\bibfnamefont {T.}~\bibnamefont {Schatteburg}}, \bibinfo {author} {\bibfnamefont {R.}~\bibnamefont {Benevides}}, \bibinfo {author} {\bibfnamefont {S.}~\bibnamefont {Vollenweider}}, \bibinfo {author} {\bibfnamefont {D.}~\bibnamefont {Macri}},\ and\ \bibinfo {author} {\bibfnamefont {Y.}~\bibnamefont {Chu}},\ }\bibfield  {title} {\bibinfo {title} {Brillouin optomechanics in the quantum ground state},\ }\href {https://doi.org/10.1103/PhysRevResearch.5.043140} {\bibfield  {journal} {\bibinfo  {journal} {Physical Review Research}\ }\textbf {\bibinfo {volume} {5}},\ \bibinfo {pages} {043140} (\bibinfo {year} {2023})}\BibitemShut {NoStop}%
\bibitem [{\citenamefont {Marquardt}\ \emph {et~al.}(2007)\citenamefont {Marquardt}, \citenamefont {Chen}, \citenamefont {Clerk},\ and\ \citenamefont {Girvin}}]{marquardtQuantumTheoryCavityAssisted2007}%
  \BibitemOpen
  \bibfield  {author} {\bibinfo {author} {\bibfnamefont {F.}~\bibnamefont {Marquardt}}, \bibinfo {author} {\bibfnamefont {J.~P.}\ \bibnamefont {Chen}}, \bibinfo {author} {\bibfnamefont {A.~A.}\ \bibnamefont {Clerk}},\ and\ \bibinfo {author} {\bibfnamefont {S.~M.}\ \bibnamefont {Girvin}},\ }\bibfield  {title} {\bibinfo {title} {Quantum {{Theory}} of {{Cavity-Assisted Sideband Cooling}} of {{Mechanical Motion}}},\ }\href {https://doi.org/10.1103/PhysRevLett.99.093902} {\bibfield  {journal} {\bibinfo  {journal} {Physical Review Letters}\ }\textbf {\bibinfo {volume} {99}},\ \bibinfo {pages} {093902} (\bibinfo {year} {2007})}\BibitemShut {NoStop}%
\bibitem [{\citenamefont {{ID Quantique}}(2024)}]{IDQSNSPD}%
  \BibitemOpen
  \bibfield  {author} {\bibinfo {author} {\bibnamefont {{ID Quantique}}},\ }\href@noop {} {\bibinfo {title} {{{ID281 SNSPD System}}}} (\bibinfo {year} {2024})\BibitemShut {NoStop}%
\bibitem [{\citenamefont {Johansson}\ \emph {et~al.}(2012)\citenamefont {Johansson}, \citenamefont {Nation},\ and\ \citenamefont {Nori}}]{qutip}%
  \BibitemOpen
  \bibfield  {author} {\bibinfo {author} {\bibfnamefont {J.~R.}\ \bibnamefont {Johansson}}, \bibinfo {author} {\bibfnamefont {P.~D.}\ \bibnamefont {Nation}},\ and\ \bibinfo {author} {\bibfnamefont {F.}~\bibnamefont {Nori}},\ }\bibfield  {title} {\bibinfo {title} {{{QuTiP}}: {{An}} open-source {{Python}} framework for the dynamics of open quantum systems},\ }\href {https://doi.org/10.1016/j.cpc.2012.02.021} {\bibfield  {journal} {\bibinfo  {journal} {Computer Physics Communications}\ }\textbf {\bibinfo {volume} {183}},\ \bibinfo {pages} {1760} (\bibinfo {year} {2012})}\BibitemShut {NoStop}%
\bibitem [{\citenamefont {Johansson}\ \emph {et~al.}(2013)\citenamefont {Johansson}, \citenamefont {Nation},\ and\ \citenamefont {Nori}}]{qutip2}%
  \BibitemOpen
  \bibfield  {author} {\bibinfo {author} {\bibfnamefont {J.~R.}\ \bibnamefont {Johansson}}, \bibinfo {author} {\bibfnamefont {P.~D.}\ \bibnamefont {Nation}},\ and\ \bibinfo {author} {\bibfnamefont {F.}~\bibnamefont {Nori}},\ }\bibfield  {title} {\bibinfo {title} {{{QuTiP}} 2: {{A Python}} framework for the dynamics of open quantum systems},\ }\href {https://doi.org/10.1016/j.cpc.2012.11.019} {\bibfield  {journal} {\bibinfo  {journal} {Computer Physics Communications}\ }\textbf {\bibinfo {volume} {184}},\ \bibinfo {pages} {1234} (\bibinfo {year} {2013})}\BibitemShut {NoStop}%
\bibitem [{\citenamefont {Sahay}\ \emph {et~al.}(2023)\citenamefont {Sahay}, \citenamefont {Claes},\ and\ \citenamefont {Puri}}]{sahayTailoringFusionbasedError2023}%
  \BibitemOpen
  \bibfield  {author} {\bibinfo {author} {\bibfnamefont {K.}~\bibnamefont {Sahay}}, \bibinfo {author} {\bibfnamefont {J.}~\bibnamefont {Claes}},\ and\ \bibinfo {author} {\bibfnamefont {S.}~\bibnamefont {Puri}},\ }\bibfield  {title} {\bibinfo {title} {Tailoring fusion-based error correction for high thresholds to biased fusion failures},\ }\href {https://doi.org/10.1103/PhysRevLett.131.120604} {\bibfield  {journal} {\bibinfo  {journal} {Physical Review Letters}\ }\textbf {\bibinfo {volume} {131}},\ \bibinfo {pages} {120604} (\bibinfo {year} {2023})}\BibitemShut {NoStop}%
\bibitem [{\citenamefont {Song}\ \emph {et~al.}(2024)\citenamefont {Song}, \citenamefont {Kang}, \citenamefont {Kim},\ and\ \citenamefont {Lee}}]{songEncodedFusionBasedQuantumComputation2024}%
  \BibitemOpen
  \bibfield  {author} {\bibinfo {author} {\bibfnamefont {W.}~\bibnamefont {Song}}, \bibinfo {author} {\bibfnamefont {N.}~\bibnamefont {Kang}}, \bibinfo {author} {\bibfnamefont {Y.-S.}\ \bibnamefont {Kim}},\ and\ \bibinfo {author} {\bibfnamefont {S.-W.}\ \bibnamefont {Lee}},\ }\bibfield  {title} {\bibinfo {title} {Encoded-{{Fusion-Based Quantum Computation}} for {{High Thresholds}} with {{Linear Optics}}},\ }\href {https://doi.org/10.1103/PhysRevLett.133.050605} {\bibfield  {journal} {\bibinfo  {journal} {Physical Review Letters}\ }\textbf {\bibinfo {volume} {133}},\ \bibinfo {pages} {050605} (\bibinfo {year} {2024})}\BibitemShut {NoStop}%
\bibitem [{\citenamefont {Kok}\ \emph {et~al.}(2007)\citenamefont {Kok}, \citenamefont {Munro}, \citenamefont {Nemoto}, \citenamefont {Ralph}, \citenamefont {Dowling},\ and\ \citenamefont {Milburn}}]{kokLinearOpticalQuantum2007}%
  \BibitemOpen
  \bibfield  {author} {\bibinfo {author} {\bibfnamefont {P.}~\bibnamefont {Kok}}, \bibinfo {author} {\bibfnamefont {W.~J.}\ \bibnamefont {Munro}}, \bibinfo {author} {\bibfnamefont {K.}~\bibnamefont {Nemoto}}, \bibinfo {author} {\bibfnamefont {T.~C.}\ \bibnamefont {Ralph}}, \bibinfo {author} {\bibfnamefont {J.~P.}\ \bibnamefont {Dowling}},\ and\ \bibinfo {author} {\bibfnamefont {G.~J.}\ \bibnamefont {Milburn}},\ }\bibfield  {title} {\bibinfo {title} {Linear optical quantum computing with photonic qubits},\ }\href {https://doi.org/10.1103/RevModPhys.79.135} {\bibfield  {journal} {\bibinfo  {journal} {Reviews of Modern Physics}\ }\textbf {\bibinfo {volume} {79}},\ \bibinfo {pages} {135} (\bibinfo {year} {2007})}\BibitemShut {NoStop}%
\bibitem [{\citenamefont {Grassl}\ \emph {et~al.}(1997)\citenamefont {Grassl}, \citenamefont {Beth},\ and\ \citenamefont {Pellizzari}}]{grasslCodesQuantumErasure1997}%
  \BibitemOpen
  \bibfield  {author} {\bibinfo {author} {\bibfnamefont {M.}~\bibnamefont {Grassl}}, \bibinfo {author} {\bibfnamefont {{\relax Th}.}~\bibnamefont {Beth}},\ and\ \bibinfo {author} {\bibfnamefont {T.}~\bibnamefont {Pellizzari}},\ }\bibfield  {title} {\bibinfo {title} {Codes for the quantum erasure channel},\ }\href {https://doi.org/10.1103/PhysRevA.56.33} {\bibfield  {journal} {\bibinfo  {journal} {Physical Review A}\ }\textbf {\bibinfo {volume} {56}},\ \bibinfo {pages} {33} (\bibinfo {year} {1997})}\BibitemShut {NoStop}%
\bibitem [{\citenamefont {Gottesman}(1997)}]{gottesmanStabilizerCodesQuantum1997}%
  \BibitemOpen
  \bibfield  {author} {\bibinfo {author} {\bibfnamefont {D.~E.}\ \bibnamefont {Gottesman}},\ }\emph {\bibinfo {title} {Stabilizer {{Codes}} and {{Quantum Error Correction}}}},\ \href {https://doi.org/10.7907/rzr7-dt72} {Ph.D. thesis},\ \bibinfo  {school} {California Institute of Technology} (\bibinfo {year} {1997})\BibitemShut {NoStop}%
\bibitem [{\citenamefont {Bartolucci}\ \emph {et~al.}(2021)\citenamefont {Bartolucci}, \citenamefont {Birchall}, \citenamefont {{Gimeno-Segovia}}, \citenamefont {Johnston}, \citenamefont {Kieling}, \citenamefont {Pant}, \citenamefont {Rudolph}, \citenamefont {Smith}, \citenamefont {Sparrow},\ and\ \citenamefont {Vidrighin}}]{bartolucciCreationEntangledPhotonic2021}%
  \BibitemOpen
  \bibfield  {author} {\bibinfo {author} {\bibfnamefont {S.}~\bibnamefont {Bartolucci}}, \bibinfo {author} {\bibfnamefont {P.~M.}\ \bibnamefont {Birchall}}, \bibinfo {author} {\bibfnamefont {M.}~\bibnamefont {{Gimeno-Segovia}}}, \bibinfo {author} {\bibfnamefont {E.}~\bibnamefont {Johnston}}, \bibinfo {author} {\bibfnamefont {K.}~\bibnamefont {Kieling}}, \bibinfo {author} {\bibfnamefont {M.}~\bibnamefont {Pant}}, \bibinfo {author} {\bibfnamefont {T.}~\bibnamefont {Rudolph}}, \bibinfo {author} {\bibfnamefont {J.}~\bibnamefont {Smith}}, \bibinfo {author} {\bibfnamefont {C.}~\bibnamefont {Sparrow}},\ and\ \bibinfo {author} {\bibfnamefont {M.~D.}\ \bibnamefont {Vidrighin}},\ }\href {https://doi.org/10.48550/arXiv.2106.13825} {\bibinfo {title} {Creation of {{Entangled Photonic States Using Linear Optics}}}} (\bibinfo {year} {2021}),\ \Eprint {https://arxiv.org/abs/2106.13825} {arXiv:2106.13825} \BibitemShut {NoStop}%
\bibitem [{\citenamefont {Brubaker}\ \emph {et~al.}(2022)\citenamefont {Brubaker}, \citenamefont {Kindem}, \citenamefont {Urmey}, \citenamefont {Mittal}, \citenamefont {Delaney}, \citenamefont {Burns}, \citenamefont {Vissers}, \citenamefont {Lehnert},\ and\ \citenamefont {Regal}}]{brubakerOptomechanicalGroundStateCooling2022}%
  \BibitemOpen
  \bibfield  {author} {\bibinfo {author} {\bibfnamefont {B.~M.}\ \bibnamefont {Brubaker}}, \bibinfo {author} {\bibfnamefont {J.~M.}\ \bibnamefont {Kindem}}, \bibinfo {author} {\bibfnamefont {M.~D.}\ \bibnamefont {Urmey}}, \bibinfo {author} {\bibfnamefont {S.}~\bibnamefont {Mittal}}, \bibinfo {author} {\bibfnamefont {R.~D.}\ \bibnamefont {Delaney}}, \bibinfo {author} {\bibfnamefont {P.~S.}\ \bibnamefont {Burns}}, \bibinfo {author} {\bibfnamefont {M.~R.}\ \bibnamefont {Vissers}}, \bibinfo {author} {\bibfnamefont {K.~W.}\ \bibnamefont {Lehnert}},\ and\ \bibinfo {author} {\bibfnamefont {C.~A.}\ \bibnamefont {Regal}},\ }\bibfield  {title} {\bibinfo {title} {Optomechanical {{Ground-State Cooling}} in a {{Continuous}} and {{Efficient Electro-Optic Transducer}}},\ }\href {https://doi.org/10.1103/PhysRevX.12.021062} {\bibfield  {journal} {\bibinfo  {journal} {Physical Review X}\ }\textbf {\bibinfo {volume} {12}},\ \bibinfo {pages} {021062} (\bibinfo {year} {2022})}\BibitemShut {NoStop}%
\bibitem [{\citenamefont {Mayor}\ \emph {et~al.}(2025)\citenamefont {Mayor}, \citenamefont {Malik}, \citenamefont {Primo}, \citenamefont {Gyger}, \citenamefont {Jiang}, \citenamefont {Alegre},\ and\ \citenamefont {{Safavi-Naeini}}}]{mayorHighPhotonphononPair2025}%
  \BibitemOpen
  \bibfield  {author} {\bibinfo {author} {\bibfnamefont {F.~M.}\ \bibnamefont {Mayor}}, \bibinfo {author} {\bibfnamefont {S.}~\bibnamefont {Malik}}, \bibinfo {author} {\bibfnamefont {A.~G.}\ \bibnamefont {Primo}}, \bibinfo {author} {\bibfnamefont {S.}~\bibnamefont {Gyger}}, \bibinfo {author} {\bibfnamefont {W.}~\bibnamefont {Jiang}}, \bibinfo {author} {\bibfnamefont {T.~P.~M.}\ \bibnamefont {Alegre}},\ and\ \bibinfo {author} {\bibfnamefont {A.~H.}\ \bibnamefont {{Safavi-Naeini}}},\ }\bibfield  {title} {\bibinfo {title} {High photon-phonon pair generation rate in a two-dimensional optomechanical crystal},\ }\href {https://doi.org/10.1038/s41467-025-57948-7} {\bibfield  {journal} {\bibinfo  {journal} {Nature Communications}\ }\textbf {\bibinfo {volume} {16}},\ \bibinfo {pages} {2576} (\bibinfo {year} {2025})}\BibitemShut {NoStop}%
\bibitem [{\citenamefont {Barrett}\ and\ \citenamefont {Kok}(2005)}]{barrettEfficientHighfidelityQuantum2005}%
  \BibitemOpen
  \bibfield  {author} {\bibinfo {author} {\bibfnamefont {S.~D.}\ \bibnamefont {Barrett}}\ and\ \bibinfo {author} {\bibfnamefont {P.}~\bibnamefont {Kok}},\ }\bibfield  {title} {\bibinfo {title} {Efficient high-fidelity quantum computation using matter qubits and linear optics},\ }\href {https://doi.org/10.1103/PhysRevA.71.060310} {\bibfield  {journal} {\bibinfo  {journal} {Physical Review A}\ }\textbf {\bibinfo {volume} {71}},\ \bibinfo {pages} {060310} (\bibinfo {year} {2005})}\BibitemShut {NoStop}%
\bibitem [{\citenamefont {Lei}\ \emph {et~al.}(2023)\citenamefont {Lei}, \citenamefont {Asadi}, \citenamefont {Zhong}, \citenamefont {Kuzmich}, \citenamefont {Simon},\ and\ \citenamefont {Hosseini}}]{leiQuantumOpticalMemory2023}%
  \BibitemOpen
  \bibfield  {author} {\bibinfo {author} {\bibfnamefont {Y.}~\bibnamefont {Lei}}, \bibinfo {author} {\bibfnamefont {F.~K.}\ \bibnamefont {Asadi}}, \bibinfo {author} {\bibfnamefont {T.}~\bibnamefont {Zhong}}, \bibinfo {author} {\bibfnamefont {A.}~\bibnamefont {Kuzmich}}, \bibinfo {author} {\bibfnamefont {C.}~\bibnamefont {Simon}},\ and\ \bibinfo {author} {\bibfnamefont {M.}~\bibnamefont {Hosseini}},\ }\bibfield  {title} {\bibinfo {title} {Quantum optical memory for entanglement distribution},\ }\href {https://doi.org/10.1364/OPTICA.493732} {\bibfield  {journal} {\bibinfo  {journal} {Optica}\ }\textbf {\bibinfo {volume} {10}},\ \bibinfo {pages} {1511} (\bibinfo {year} {2023})}\BibitemShut {NoStop}%
\bibitem [{\citenamefont {Dodonov}\ \emph {et~al.}(1974)\citenamefont {Dodonov}, \citenamefont {Malkin},\ and\ \citenamefont {Man'ko}}]{dodonovEvenOddCoherent1974}%
  \BibitemOpen
  \bibfield  {author} {\bibinfo {author} {\bibfnamefont {V.~V.}\ \bibnamefont {Dodonov}}, \bibinfo {author} {\bibfnamefont {I.~A.}\ \bibnamefont {Malkin}},\ and\ \bibinfo {author} {\bibfnamefont {V.~I.}\ \bibnamefont {Man'ko}},\ }\bibfield  {title} {\bibinfo {title} {Even and odd coherent states and excitations of a singular oscillator},\ }\href {https://doi.org/10.1016/0031-8914(74)90215-8} {\bibfield  {journal} {\bibinfo  {journal} {Physica}\ }\textbf {\bibinfo {volume} {72}},\ \bibinfo {pages} {597} (\bibinfo {year} {1974})}\BibitemShut {NoStop}%
\bibitem [{\citenamefont {Lund}\ \emph {et~al.}(2004)\citenamefont {Lund}, \citenamefont {Jeong}, \citenamefont {Ralph},\ and\ \citenamefont {Kim}}]{lundConditionalProductionSuperpositions2004}%
  \BibitemOpen
  \bibfield  {author} {\bibinfo {author} {\bibfnamefont {A.~P.}\ \bibnamefont {Lund}}, \bibinfo {author} {\bibfnamefont {H.}~\bibnamefont {Jeong}}, \bibinfo {author} {\bibfnamefont {T.~C.}\ \bibnamefont {Ralph}},\ and\ \bibinfo {author} {\bibfnamefont {M.~S.}\ \bibnamefont {Kim}},\ }\bibfield  {title} {\bibinfo {title} {Conditional production of superpositions of coherent states with inefficient photon detection},\ }\href {https://doi.org/10.1103/PhysRevA.70.020101} {\bibfield  {journal} {\bibinfo  {journal} {Physical Review A}\ }\textbf {\bibinfo {volume} {70}},\ \bibinfo {pages} {020101} (\bibinfo {year} {2004})}\BibitemShut {NoStop}%
\bibitem [{\citenamefont {Ourjoumtsev}\ \emph {et~al.}(2007)\citenamefont {Ourjoumtsev}, \citenamefont {Jeong}, \citenamefont {{Tualle-Brouri}},\ and\ \citenamefont {Grangier}}]{ourjoumtsevGenerationOpticalSchrodinger2007}%
  \BibitemOpen
  \bibfield  {author} {\bibinfo {author} {\bibfnamefont {A.}~\bibnamefont {Ourjoumtsev}}, \bibinfo {author} {\bibfnamefont {H.}~\bibnamefont {Jeong}}, \bibinfo {author} {\bibfnamefont {R.}~\bibnamefont {{Tualle-Brouri}}},\ and\ \bibinfo {author} {\bibfnamefont {P.}~\bibnamefont {Grangier}},\ }\bibfield  {title} {\bibinfo {title} {Generation of optical `{{Schr{\"o}dinger}} cats' from photon number states},\ }\href {https://doi.org/10.1038/nature06054} {\bibfield  {journal} {\bibinfo  {journal} {Nature}\ }\textbf {\bibinfo {volume} {448}},\ \bibinfo {pages} {784} (\bibinfo {year} {2007})}\BibitemShut {NoStop}%
\bibitem [{\citenamefont {Chen}\ \emph {et~al.}(2024{\natexlab{b}})\citenamefont {Chen}, \citenamefont {Hsieh}, \citenamefont {Ning}, \citenamefont {Wu}, \citenamefont {Chen}, \citenamefont {Shi}, \citenamefont {Yang}, \citenamefont {Steuernagel}, \citenamefont {Wu},\ and\ \citenamefont {Lee}}]{chenGenerationHeraldedOptical2024}%
  \BibitemOpen
  \bibfield  {author} {\bibinfo {author} {\bibfnamefont {Y.-R.}\ \bibnamefont {Chen}}, \bibinfo {author} {\bibfnamefont {H.-Y.}\ \bibnamefont {Hsieh}}, \bibinfo {author} {\bibfnamefont {J.}~\bibnamefont {Ning}}, \bibinfo {author} {\bibfnamefont {H.-C.}\ \bibnamefont {Wu}}, \bibinfo {author} {\bibfnamefont {H.~L.}\ \bibnamefont {Chen}}, \bibinfo {author} {\bibfnamefont {Z.-H.}\ \bibnamefont {Shi}}, \bibinfo {author} {\bibfnamefont {P.}~\bibnamefont {Yang}}, \bibinfo {author} {\bibfnamefont {O.}~\bibnamefont {Steuernagel}}, \bibinfo {author} {\bibfnamefont {C.-M.}\ \bibnamefont {Wu}},\ and\ \bibinfo {author} {\bibfnamefont {R.-K.}\ \bibnamefont {Lee}},\ }\bibfield  {title} {\bibinfo {title} {Generation of heralded optical cat states by photon addition},\ }\href {https://doi.org/10.1103/PhysRevA.110.023703} {\bibfield  {journal} {\bibinfo  {journal} {Physical Review A}\ }\textbf {\bibinfo {volume} {110}},\ \bibinfo {pages} {023703} (\bibinfo {year} {2024}{\natexlab{b}})}\BibitemShut {NoStop}%
\bibitem [{\citenamefont {Zhao}\ \emph {et~al.}(2023)\citenamefont {Zhao}, \citenamefont {Krauss}, \citenamefont {Bienaime}, \citenamefont {Whitlock}, \citenamefont {Koch}, \citenamefont {Qvarfort},\ and\ \citenamefont {Metelmann}}]{zhaoFastRobustCat2023}%
  \BibitemOpen
  \bibfield  {author} {\bibinfo {author} {\bibfnamefont {S.}~\bibnamefont {Zhao}}, \bibinfo {author} {\bibfnamefont {M.~G.}\ \bibnamefont {Krauss}}, \bibinfo {author} {\bibfnamefont {T.}~\bibnamefont {Bienaime}}, \bibinfo {author} {\bibfnamefont {S.}~\bibnamefont {Whitlock}}, \bibinfo {author} {\bibfnamefont {C.~P.}\ \bibnamefont {Koch}}, \bibinfo {author} {\bibfnamefont {S.}~\bibnamefont {Qvarfort}},\ and\ \bibinfo {author} {\bibfnamefont {A.}~\bibnamefont {Metelmann}},\ }\href {https://doi.org/10.48550/arXiv.2312.05218} {\bibinfo {title} {Fast and robust cat state preparation utilizing higher order nonlinearities}} (\bibinfo {year} {2023}),\ \Eprint {https://arxiv.org/abs/2312.05218} {arXiv:2312.05218} \BibitemShut {NoStop}%
\bibitem [{\citenamefont {Gottesman}\ \emph {et~al.}(2001)\citenamefont {Gottesman}, \citenamefont {Kitaev},\ and\ \citenamefont {Preskill}}]{gottesmanEncodingQubitOscillator2001}%
  \BibitemOpen
  \bibfield  {author} {\bibinfo {author} {\bibfnamefont {D.}~\bibnamefont {Gottesman}}, \bibinfo {author} {\bibfnamefont {A.}~\bibnamefont {Kitaev}},\ and\ \bibinfo {author} {\bibfnamefont {J.}~\bibnamefont {Preskill}},\ }\bibfield  {title} {\bibinfo {title} {Encoding a qubit in an oscillator},\ }\href {https://doi.org/10.1103/PhysRevA.64.012310} {\bibfield  {journal} {\bibinfo  {journal} {Physical Review A}\ }\textbf {\bibinfo {volume} {64}},\ \bibinfo {pages} {012310} (\bibinfo {year} {2001})}\BibitemShut {NoStop}%
\bibitem [{\citenamefont {Konno}\ \emph {et~al.}(2024)\citenamefont {Konno}, \citenamefont {Asavanant}, \citenamefont {Hanamura}, \citenamefont {Nagayoshi}, \citenamefont {Fukui}, \citenamefont {Sakaguchi}, \citenamefont {Ide}, \citenamefont {China}, \citenamefont {Yabuno}, \citenamefont {Miki}, \citenamefont {Terai}, \citenamefont {Takase}, \citenamefont {Endo}, \citenamefont {Marek}, \citenamefont {Filip}, \citenamefont {{van Loock}},\ and\ \citenamefont {Furusawa}}]{konnoLogicalStatesFaulttolerant2024}%
  \BibitemOpen
  \bibfield  {author} {\bibinfo {author} {\bibfnamefont {S.}~\bibnamefont {Konno}}, \bibinfo {author} {\bibfnamefont {W.}~\bibnamefont {Asavanant}}, \bibinfo {author} {\bibfnamefont {F.}~\bibnamefont {Hanamura}}, \bibinfo {author} {\bibfnamefont {H.}~\bibnamefont {Nagayoshi}}, \bibinfo {author} {\bibfnamefont {K.}~\bibnamefont {Fukui}}, \bibinfo {author} {\bibfnamefont {A.}~\bibnamefont {Sakaguchi}}, \bibinfo {author} {\bibfnamefont {R.}~\bibnamefont {Ide}}, \bibinfo {author} {\bibfnamefont {F.}~\bibnamefont {China}}, \bibinfo {author} {\bibfnamefont {M.}~\bibnamefont {Yabuno}}, \bibinfo {author} {\bibfnamefont {S.}~\bibnamefont {Miki}}, \bibinfo {author} {\bibfnamefont {H.}~\bibnamefont {Terai}}, \bibinfo {author} {\bibfnamefont {K.}~\bibnamefont {Takase}}, \bibinfo {author} {\bibfnamefont {M.}~\bibnamefont {Endo}}, \bibinfo {author} {\bibfnamefont {P.}~\bibnamefont {Marek}}, \bibinfo {author} {\bibfnamefont {R.}~\bibnamefont {Filip}}, \bibinfo {author} {\bibfnamefont {P.}~\bibnamefont {{van Loock}}},\ and\ \bibinfo {author} {\bibfnamefont {A.}~\bibnamefont {Furusawa}},\ }\bibfield  {title} {\bibinfo {title} {Logical states for fault-tolerant quantum computation with propagating light},\ }\href {https://doi.org/10.1126/science.adk7560} {\bibfield  {journal} {\bibinfo  {journal} {Science}\ }\textbf {\bibinfo {volume} {383}},\ \bibinfo {pages} {289} (\bibinfo {year} {2024})}\BibitemShut {NoStop}%
\end{thebibliography}%

\end{document}